\pdfoutput=1
\documentclass[11pt,twoside,a4paper,dn,final]{cms-tdr}
\def\cmsCernNoTag{}
\def\cmsCernDate{1 August 2023}
\def\svnDate{}
\def\cmsMessage{Published in the European Physical Journal C as \\\DOI{10.1140/epjc/s10052-023-11713-6}.}
\begin{document}\cmsNoteHeader{NOTE-2022/007}

\newlength\cmsTabSkip\setlength{\cmsTabSkip}{1ex}
\newcommand{\MHz}{\unit{MHz}}
\newcommand{\mmsq}{\unit{mm$^2$}}
\newcommand{\sigmavis}{\ensuremath{\sigma_{\text{vis}}}\xspace}
\newcommand{\sigmab}{\ensuremath{\sigma_{\text{b}}}\xspace}
\newcommand{\hzub}{\ensuremath{\text{Hz}/\mu\text{b}}\xspace}
\newcommand{\Linst}{\ensuremath{\mathcal{L}_{\text{inst}}}\xspace}
\newcommand{\Linsti}{\ensuremath{\mathcal{L}_{\text{inst}}^{i}}\xspace}
\newcommand{\frev}{\ensuremath{f_{\text{rev}}}\xspace}
\newcommand{\Vmax}{\ensuremath{V_{\text{maxEff}}}\xspace}
\newlength\cmsFigureWidth\ifthenelse{\boolean{cms@external}}{\setlength{\cmsFigureWidth}{\columnwidth}}{\setlength{\cmsFigureWidth}{0.7\textwidth}}
\newlength\cmsFigureWidthSmaller\ifthenelse{\boolean{cms@external}}{\setlength{\cmsFigureWidthSmaller}{\columnwidth}}{\setlength{\cmsFigureWidthSmaller}{0.55\textwidth}}
\ifthenelse{\boolean{cms@external}}{\providecommand{\cmsLeft}{top\xspace}}{\providecommand{\cmsLeft}{left\xspace}}
\ifthenelse{\boolean{cms@external}}{\providecommand{\cmsRight}{bottom\xspace}}{\providecommand{\cmsRight}{right\xspace}} 
\ifthenelse{\boolean{cms@external}}{\providecommand{\cmsLeftCap}{Top\xspace}}{\providecommand{\cmsLeftCap}{Left\xspace}}
\ifthenelse{\boolean{cms@external}}{\providecommand{\cmsRightCap}{Bottom\xspace}}{\providecommand{\cmsRightCap}{Right\xspace}} 
\ifthenelse{\boolean{cms@external}}{\providecommand{\cmsUL}{top\xspace}}{\providecommand{\cmsUL}{top left\xspace}} 
\ifthenelse{\boolean{cms@external}}{\providecommand{\cmsUR}{center\xspace}}{\providecommand{\cmsUR}{top right\xspace}} 
\ifthenelse{\boolean{cms@external}}{\providecommand{\cmsTable}[1]{\resizebox{\columnwidth}{!}{#1}}}{\providecommand{\cmsTable}[1]{#1}}
\hyphenation{firm-ware}

\cmsNoteHeader{DN-21-008}
\title{The Pixel Luminosity Telescope: A detector for luminosity measurement at CMS using silicon pixel sensors}

\author[ca]{The CMS BRIL Collaboration\footnote{Corresponding author: Paul Lujan, \texttt{paul.lujan@cern.ch}}}

\date{\today}

\abstract{The Pixel Luminosity Telescope is a silicon pixel detector dedicated to luminosity measurement at the CMS experiment at the LHC. It is located approximately 1.75\unit{m} from the interaction point and arranged into 16 ``telescopes'', with eight telescopes installed around the beam pipe at either end of the detector and each telescope composed of three individual silicon sensor planes. The per-bunch instantaneous luminosity is measured by counting events where all three planes in the telescope register a hit, using a special readout at the full LHC bunch-crossing rate of 40\MHz. The full pixel information is read out at a lower rate and can be used to determine calibrations, corrections, and systematic uncertainties for the online and offline measurements. This paper details the commissioning, operational history, and performance of the detector during Run~2 (2015--18) of the LHC, as well as preparations for Run 3, which will begin in 2022.}

\hypersetup{pdfauthor={The CMS BRIL Collaboration},pdftitle={The Pixel Luminosity Telescope: A detector for luminosity measurement at CMS using silicon pixel sensors},pdfsubject={CMS},pdfkeywords={CMS, luminosity, pixel detectors}}

\maketitle 

\section{Introduction}
\label{sec:intro}

Precise determination of the luminosity at the CERN LHC is a critical component of any experiment, as the value of the integrated luminosity is an input to all cross-section measurements and many searches for new physics; in addition, real-time (``online'') feedback of instantaneous luminosity is important to optimize the performance of the LHC accelerator and the data taking of the experiment. To this end, the CMS Beam Radiation, Instrumentation, and Luminosity (BRIL) project operates several luminosity measurement subsystems in order to provide precision online and offline luminosity measurements.

\begin{figure*}[htbp]
\centering
  \includegraphics[width=0.49\textwidth]{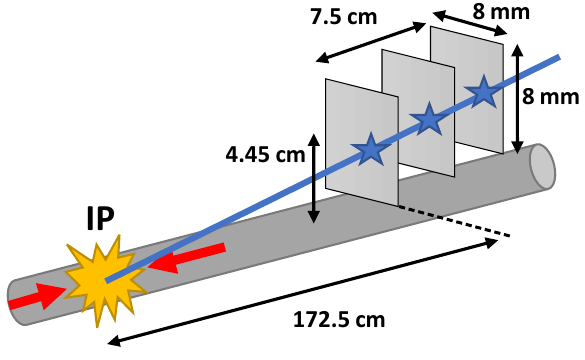}
  \includegraphics[width=0.49\textwidth]{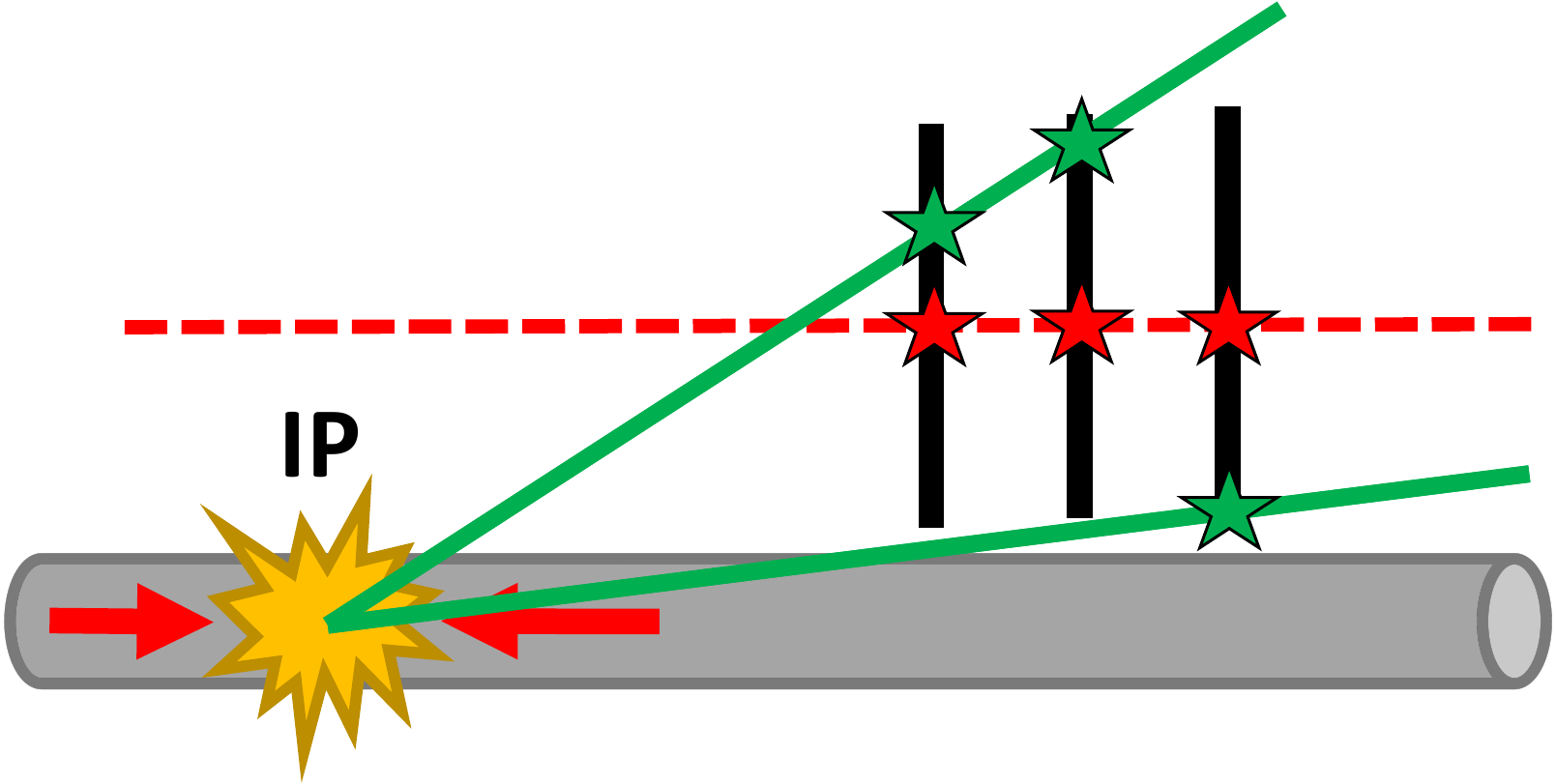}
  \caption{On the left is a sketch (not to scale) illustrating the basic operating principle of the PLT: a track originating from the CMS interaction point passing through a single PLT telescope will produce a triple coincidence. The center of the first plane is 4.45\cm from the beam axis, with the other two planes slightly farther away in the radial direction to match the slope of tracks coming from the IP. This produces a pointing angle of 1.15$^{\circ}$ between the beam axis and the line connecting the centers of each plane. On the right is a sketch illustrating two possible sources of accidentals in the PLT: the solid green lines show a combinatorial background, where hits from two tracks that do not individually pass through all three planes produce a triple coincidence together, while the dashed red line shows a track from a noncollision source, in this case beam-induced background, passing through all three planes of a PLT telescope.}
  \label{fig:plt_sketches}
\end{figure*}

The Pixel Luminosity Telescope (PLT)~\cite{Kornmayer:2016wkz,Lujan:2017kvh,eps2021} is a dedicated luminosity monitor (``luminometer'') using silicon pixel sensors. It was installed in January 2015 as part of the Run 2 upgrades for the BRIL project~\cite{Dabrowski:2016rqm}, and was operated successfully throughout Run 2 of the LHC from 2015 to 2018. The PLT consists of 48 silicon sensors arranged into ``telescopes'', where each telescope contains three sensors separated along the $z$ axis (parallel to the beam line), such that particles originating from the CMS interaction point (IP) will pass through all three planes in the telescope, as shown in Fig.~\ref{fig:plt_sketches} (left). Of the 16 PLT telescopes, eight are positioned on either side of the pixel endcaps (approximately 1.75\unit{m} from the IP), arranged in a circle around the beam pipe, at a pseudorapidity $\abs{\eta} \approx 4.2$.

The PLT uses much of the same technology as the CMS phase-0 pixel detector~\cite{Chatrchyan:2008zzk} (which operated in CMS up to the end of 2016), including the sensors~\cite{Bolla:2002my,Allkofer:2007ek} and readout chips (ROCs)~\cite{Kastli:2005jj,Barbero:467141}, but takes advantage of a separate ``fast-or'' readout mode in the readout chips, which was not used in the main CMS pixel detector. In this fast-or mode, if any pixels in a sensor register a hit over threshold during a single 25\unit{ns} time interval, a single pulse is produced. By its nature, this signal does not contain any detailed information on the hit, but can be read out at the full bunch crossing frequency of 40\MHz. The readout hardware then counts the number of ``triple coincidences'', \ie, events where all three planes in a telescope register a signal, to determine the instantaneous luminosity. This fast-or readout thus allows the PLT to provide online per-bunch luminosity with excellent statistical precision, with the triple coincidence requirement providing a strong suppression of background from noise and activated material in the detector. The full pixel data can also be read out from the ROCs, as in the CMS pixel detector, upon receipt of a trigger signal, which in the PLT is provided by a dedicated generator at a rate of typically a few kHz; this data can be used for additional studies to validate and correct the fast-or luminosity measurement.

The instantaneous luminosity should be proportional to $\mu$, the mean number of triple coincidences. The proportionality constant, referred to as the visible cross section \sigmavis, is determined using the Van der Meer (VdM) scan method described in Section~\ref{sec:VdM}. In practice, there are effects which can cause a nonlinear response in the PLT and which need to be corrected. The primary source of nonlinearity in the PLT is ``accidentals'', where a triple coincidence is registered from three hits that do not actually come from a single particle track originating from the IP. This can be due to combinatorial sources, where hits from multiple tracks (or other sources, such as detector noise) combine to form a triple coincidence when none of the individual tracks passes through all three planes. Accidentals can also occur when particles not originating from the IP pass through the PLT, such as beam-induced background (BIB) traveling parallel to the LHC beam, or particles produced in secondary interactions with the detector or by activated material, as illustrated in Fig.~\ref{fig:plt_sketches} (right). These are discussed further in Section~\ref{sec:accidentals}. In addition, the value of the calibration constant \sigmavis may vary over the data-taking period due to changes in the operating conditions of the PLT, which in the course of Run 2 was principally due to radiation damage in the sensors. This variation also needs to be measured and corrected for in the final luminosity measurement, as discussed in Section~\ref{sec:efficiency}.

Because of the accumulated radiation damage in the PLT sensors and other components, a new copy of the PLT was constructed during the Long Shutdown 2 (LS2) period (2019--22) and installed in July 2021 for the beginning of LHC Run 3 (2022--24). A second copy is currently under construction to be made available as a ``hot spare'', as it is expected that radiation damage will make a replacement during Run~3 necessary to maintain the best performance.

This paper is structured as follows: Section~\ref{sec:cms} describes the CMS detector and the relevant parts of the LHC. Section~\ref{sec:technical_description} gives a technical description of the PLT components, and Section~\ref{sec:detector_calibration} describes the various calibration procedures used for the PLT. Section~\ref{sec:performance} describes additional studies used for monitoring detector performance and other quantities of interest. In Section~\ref{sec:lumi}, the procedure for obtaining and calibrating the luminosity measurement of the PLT is described. Finally, Section~\ref{sec:run3} discusses preparations for Run~3, with Section~\ref{sec:summary} summarizing the results.

\section{The CMS detector and the LHC}
\label{sec:cms}

The central feature of the CMS apparatus is a superconducting solenoid of 6\unit{m} internal diameter, providing a magnetic field of 3.8\unit{T}. Within the solenoid volume are a silicon pixel and strip tracker, a lead tungstate crystal electromagnetic calorimeter, and a brass and scintillator hadron calorimeter, each composed of a barrel and two endcap sections. Forward calorimeters extend the pseudorapidity coverage provided by the barrel and endcap detectors. Muons are detected in gas-ionization chambers embedded in the steel flux-return yoke outside the solenoid. A more detailed description of the CMS detector, together with a definition of the coordinate system used and the relevant kinematic variables, can be found in Ref.~\cite{Chatrchyan:2008zzk}. 

In addition to the PLT, the CMS BRIL group produces luminosity measurements using several other methods. These include two methods using the CMS hadronic forward (HF) calorimeter, one based on occupancy counting (HFOC) and one using the energy sum (HFET); a rate measurement with the Fast Beam Conditions Monitor (BCM1F)~\cite{Leonard:2014gpa,Hempel:2017nvn}; Pixel Cluster Counting (PCC), measuring the rate of clusters in the main CMS pixel detector; a measurement of the ambient dose equivalent rate with the RAMSES (Radiation Monitoring System for the Environment and Safety) detectors~\cite{Forkel-Wirth:687619} mounted in the CMS cavern; and one using the rate of muon stubs in the CMS muon drift tubes (DT).

The LHC orbit is divided into 3564 BXs, where a BX is a time interval of 25\unit{ns}. A single orbit is defined by the time it takes for a single bunch to completely circle the LHC ring, or, equivalently, for each bunch to pass by a single point, such as the CMS IP, once. The length of an orbit is thus 89\mus, and the corresponding revolution frequency \frev is 11.246\unit{kHz}~\cite{Evans:2008zzb}. An individual BX can contain proton bunches in both beams (a ``colliding'' bunch pair), a proton bunch in only one beam (a ``noncolliding'' or ``unpaired'' bunch), or be empty in both beams (an ``empty'' BX). Because of limitations imposed by the LHC cryogenic, injection, and abort systems, the maximum number of colliding bunches in a given fill during Run 2 was approximately 2500; these bunches are typically arranged into ``trains'', long sequences of filled bunches with intervals of empty BXs separating them. We refer to the first bunch in a train as the ``leading'' bunch and following bunches as ``train'' bunches; the fill pattern often includes a few isolated colliding bunches not part of any train. In addition, the last 120 BXs (3\mus) of the orbit are guaranteed to remain empty, in order to ensure a safe interval in case an LHC beam abort is necessary; this is called the ``abort gap''. Each BX is numbered with a bunch crossing ID (BCID) in the range 1--3564.

\section{Technical description}
\label{sec:technical_description}

The PLT is constructed from 16 individual telescopes, where each telescope consists of three sensors, with each sensor mounted in the $x$-$y$ plane (\ie, perpendicular to the beam axis) and separated from the other two planes along the $z$ axis, with a total length of approximately 7.5\unit{cm}. The planes are also slightly offset in the radial direction, producing a pointing angle of 1.15$^\circ$ towards the IP, so that a track produced by a particle originating at the IP will pass through the same relative point on each sensor.

\begin{figure}[htbp]
\centering
  \includegraphics[width=0.9\columnwidth]{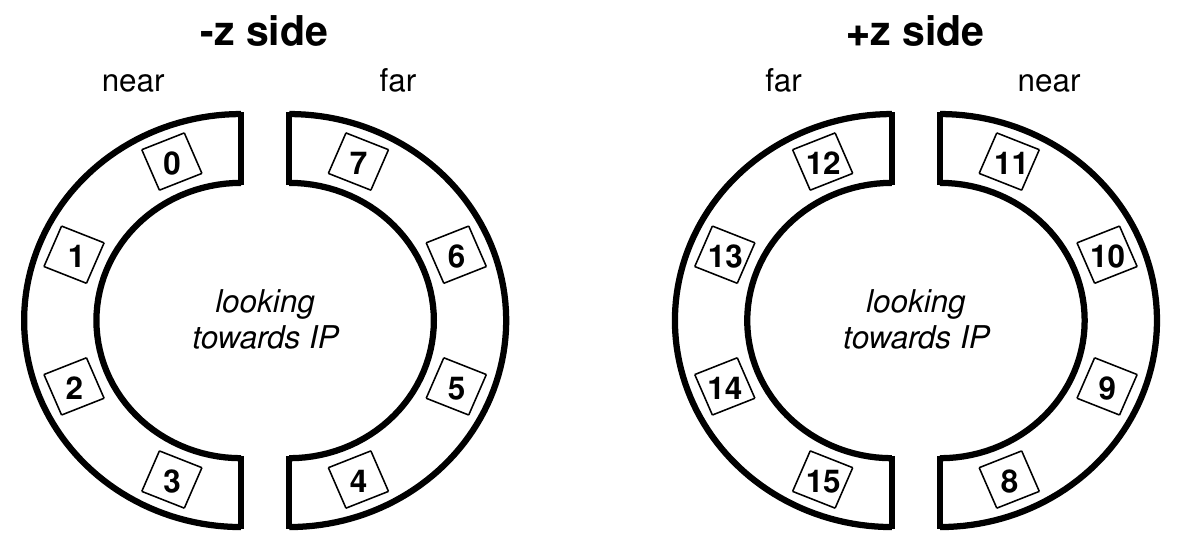}
  \caption{Schematic of the arrangement of the PLT telescopes, numbered by position, for the $-z$ side (left) and the $+z$ side (right), viewed looking towards the IP. The ``near'' side is the side closer to the center of the LHC ring.}
  \label{fig:readout_channels}
\end{figure}

The 16 telescopes are arranged into four quadrants, each quadrant containing four telescopes arranged in a semicircle. The quadrants are labelled as either $-z$ or $+z$, depending on which end of the CMS detector they are located, and ``near'' (closer to the center of the LHC ring) or ``far'', depending on which side of the beam pipe they are on. The telescopes are numbered 0--3 in the $-z$ near quadrant, 4--7 in the $-z$ far quadrant, 8--11 in the $+z$ near quadrant, and 12--15 in the $+z$ far quadrant. When looking from outside the PLT towards the IP, the numbers increase in the counterclockwise direction. Figure~\ref{fig:readout_channels} shows the location of the individual telescopes, looking from outside the pixel bulkhead towards the IP.

\subsection{Front-end hardware}

The silicon sensors used in the PLT are the same as those used in the CMS phase-0 pixel detector~\cite{Bolla:2002my,Allkofer:2007ek}, using an ``n-in-n'' technology with a silicon thickness of 285\mum. They are divided into 52 columns and 80 rows of pixels, with each pixel $150{\times}100\mum$, for a total active area of $8{\times}8\mmsq$. However, to decrease the contribution from accidentals, as discussed in Section~\ref{sec:accidentals}, only a smaller active area is used. In 2015, the active area was $4.2{\times}4.1\mmsq$ (28 columns${\times}$41 rows) in the central plane of a telescope and $5.1{\times}5.0\mmsq$ (34 columns${\times}$50 rows) in the outer (first and third) planes, with the larger area in the outer planes to ensure that tracks are not lost even if the alignment of the three planes is slightly imperfect. In 2016, this was reduced to $3.6{\times}3.6\mmsq$ (24 columns${\times}$36 rows) in the center plane and $3.9{\times}3.8\mmsq$ (26 columns${\times}$38 rows) in the outer planes, and this setting was used for the rest of Run 2.

The sensors are read out by the PSI46v2 ROC~\cite{Kastli:2005jj,Barbero:467141}, which was also developed for the CMS phase-0 pixel detector. It features an array of $52{\times}80$ readout cells, each bump-bonded to the corresponding pixel on the sensor, with readout, calibration, and control circuitry located on the periphery of the chip. For readout purposes, the columns are grouped into 26 pairs, as ``double columns'', and each double column has its own readout buffer and timestamp buffer in the periphery. The sensors and ROCs are mounted to a ``hybrid'' board, a small circuit board for providing the connections to the other parts of the detector.

\begin{figure*}[htbp]
\centering
  \includegraphics[width=0.95\textwidth]{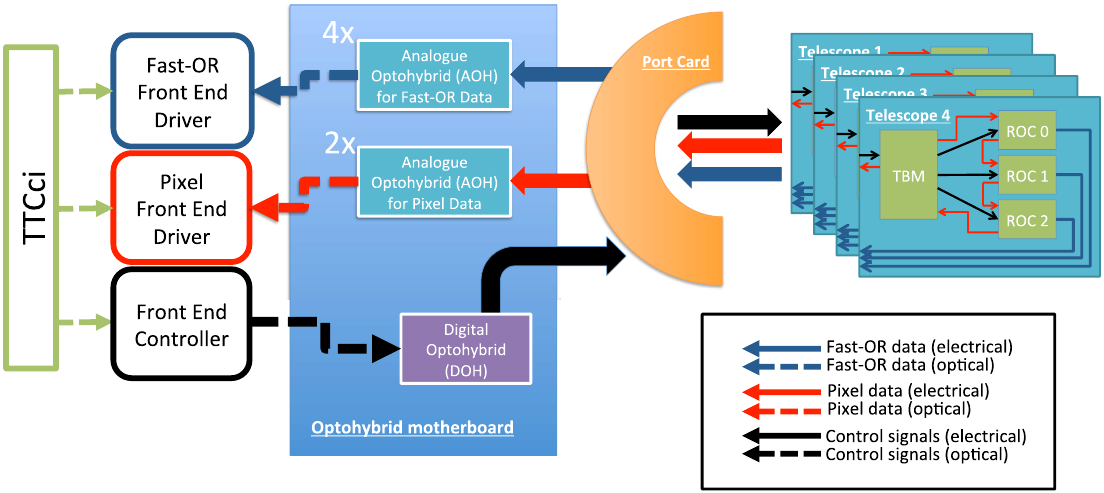}
  \caption{A schematic of the readout scheme for a single PLT quadrant, showing the data flow from the individual ROCs through the port card and OMB to the FEDs and FEC on the back end. The FEC and pixel FED are shared among all four quadrants, while one fast-or FED serves two quadrants.}
  \label{fig:readout_schematic}
\end{figure*}

A schematic of the connections of the ROCs to the rest of the PLT, and the overall flow of data and control signals, is shown for a single PLT quadrant in Fig.~\ref{fig:readout_schematic}. The three ROCs for a single telescope are connected to a high-density interconnect (HDI) card, which contains a Token Bit Manager (TBM) chip~\cite{BartzTBM}. The TBM chip distributes clock and trigger signals, coordinates the readout of the three individual ROCs, and produces a single readout for each telescope. The TBM is only used to manage the pixel data, as the fast-or ROC data follow an independent data path, managed by a fast-or driver chip (also located on the HDI).

Four telescopes are connected to a port card, which manages the communication and control signals for a single quadrant of the detector. The port card is in turn connected to the opto-motherboard (OMB). The OMB contains six analog optohybrids (AOHs)~\cite{Friedl:2004ji}, which convert the analog signals from the detector into optical signals. These signals are then sent over fibers from the CMS experimental cavern, where the front-end electronics are located, into the CMS service cavern, where the PLT back-end readout electronics are located. Four AOHs are used for the fast-or signals, one for each telescope, and two are used for the pixel data. The OMB also contains a digital optohybrid (DOH)~\cite{Benotto:2008zz}, which receives the optical clock, trigger, and control signals from the back-end hardware and distributes them to the detector. Several other support chips are on the OMB, including a tracker PLL chip, which decodes the clock and trigger signals and ensures the clock stability, a Delay25 chip~\cite{Furtado:920425} used for fine timing adjustments of the clock and trigger signals, a Gatekeeper chip for translating signal levels between the PLL and the other chips on the port card, and a slow hub chip and adapter chip, which distribute control signals via I$^2$C connections.

\begin{figure}[htbp]
\centering
  \includegraphics[width=0.8\columnwidth]{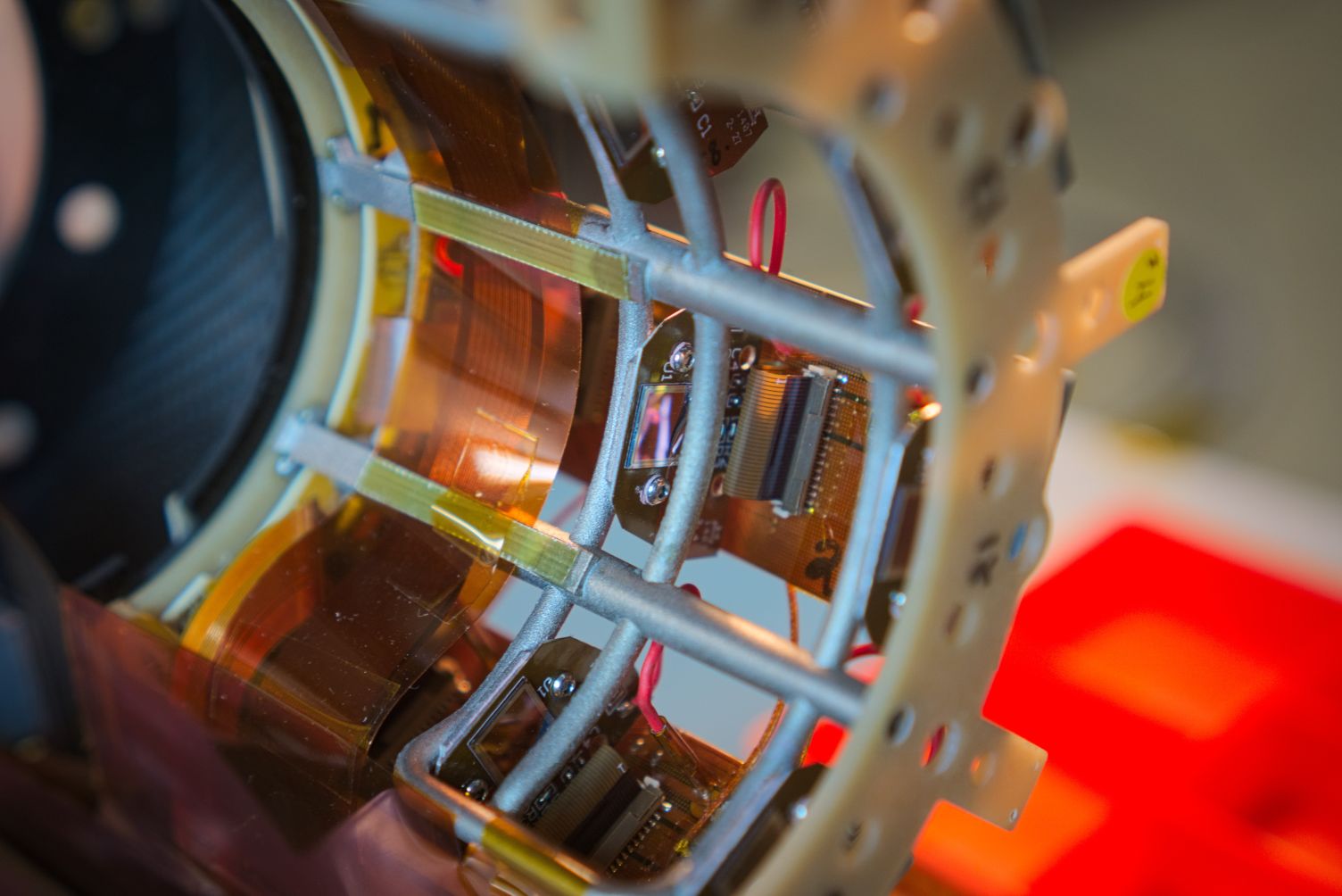}
  \caption{Closeup of an assembled cassette, with the gray cooling tubes visible in the foreground, the hybrid boards behind (one is visible at the center of the picture, carrying the silicon sensor visible as the silver rectangle), and the HDIs running horizontally in the background. The port card is just visible to the left of the ribbon in the foreground.}
  \label{fig:cassette}
\end{figure}

The hybrid boards, HDIs, and port card making up a single quadrant are mounted on a ``cassette'', a lightweight structure providing mechanical support to the PLT components. The cassette also supports the cooling tubes. Cooling of the silicon sensors is necessary to mitigate radiation damage effects and minimize leakage currents. The cooling tubes must be as small as possible, be capable of withstanding the high pressures used in the CMS tracker cooling system, and feature many small-radius bends in order to distribute cooling to the whole PLT. To meet these requirements, a 3-D printing process using selective laser melting was used to fabricate the cooling structure from titanium powder. The resulting cooling tube has a diameter of 2.8\unit{mm}. It is connected to the plant that provides the cooling for the CMS strip tracker, which uses a working fluid of C$_6$F$_{14}$ at a temperature of $-15^\circ\unit{C}$ (decreased to $-20^\circ\unit{C}$ in 2018). Figure~\ref{fig:cassette} shows an assembled PLT cassette.

The cassette is in turn mounted inside a ``carriage'', which is a mechanical structure carrying the cassette and OMB, as well as the sensors for two other BRIL detectors, BCM1F and the Beam Conditions Monitor for Losses (BCML1)~\cite{Mueller:1319599,Guthoff:1977429,Kassel:2271160}, and their support electronics. The carriage is then inserted into the pixel bulkhead, surrounding the CMS beam pipe.

\subsection{Back-end hardware}

The back-end readout electronics comprise a front-end controller (FEC) card in the Versa Module Eurocard (VME) format, which issues commands to the ROCs, TBMs, and OMB, and three front-end driver (FED) VME cards. Two of the FEDs are the ``fast-or FEDs'', one for each side  ($-z$ and $+z$) of the PLT. These read out the fast-or data from the ROCs, using custom firmware developed for the PLT, and look for events where a triple coincidence is observed in a telescope. The number of triple coincidences is then histogrammed for each telescope and BX. The histogram data are accumulated for an integration period of 4096 orbits (approximately 0.36\unit{s}), and then read out over an optical bridge to a dedicated readout PC. The fast-or FEDs feature two histogram buffers, so that data can be accumulated in one buffer while the other is being read out. The fast-or FED can also read out other information, such as the number of hits per individual plane in an integration period aggregated over all BXes, which can be used for additional studies or diagnostics. One readout channel of the fast-or FED corresponds to one PLT telescope, numbered as shown in Fig.~\ref{fig:readout_channels}.

The other FED, the ``pixel FED'', reads out, digitizes, and decodes the pixel data from the ROCs, and is identical to the FEDs used by the phase-0 pixel detector~\cite{Pernicka:1091743}. These data are then read out over an Slink~\cite{vanderBij:1997rc} connection and saved to a dedicated Slink PC; some additional data used for diagnostics and calibration can also be read out over the optical bridge to the main readout PC.

The back-end electronics also include a CAEN model SY1527 mainframe containing the low voltage (LV) and high voltage (HV) power supplies for the detector, and a programmable logic controller which automatically shuts down the PLT under circumstances where the detector cannot be operated safely, such as a loss of cooling.

The PLT back end receives clock and orbit signals from the main CMS timing and control distribution system (TCDS)~\cite{Hegeman:2016hlt}, but does not use the main CMS trigger system, instead using a CMS trigger card (TTCci) to generate its own triggers that are sent to the front end for reading out pixel data. For most of Run 2 operation, a simple zero-bias trigger was used that equally sampled all BXs in the LHC orbit at a rate of approximately 3.3\unit{kHz}. During some special fills, such as those used for VdM scans, special triggers were used with an overall higher rate (since the collision rate, and hence the amount of data, is significantly less in these fills) and with the trigger optimized to select zero-bias events primarily from the colliding bunches in the fill, since most of the BXs were empty in these fills. The overall rate for the VdM fills was approximately 70\unit{kHz} for 2016 and 10\unit{kHz} in 2017--18.

\subsection{Data acquisition and processing}
\label{sec:daq}

The triple-coincidence data received from the fast-or FEDs are published to the BRILDAQ, a dedicated data acquisition (DAQ) system for BRIL data, which operates independently of the main CMS DAQ to ensure that luminosity information is available to the LHC regardless of the status of CMS. The BRILDAQ processor reads the raw PLT data, aggregates it into integration intervals of $2^{14}$ orbits (approximately 1.4\unit{s}), calculates the average number of triple coincidences $\mu$, and applies the calibration constants to obtain the instantaneous luminosity value. The resulting PLT luminosity data are published to CMS and LHC in real time via the CERN DIP protocol~\cite{bib:DIP}, made available for online monitoring through the BRIL web monitoring system, and saved to the luminosity database for further offline analysis. For use in offline physics analysis, the data are further aggregated into time intervals of $2^{18}$ orbits (approximately 23.3\unit{s}), known as ``lumi sections'' (LS). The PLT background measurement described in Section~\ref{sec:background} is also published to BRILDAQ and DIP. The raw data files are also saved to disk.

The pixel data are similarly both saved to disk and published to BRILDAQ, where some quantities of interest (such as the online measurement of the rate of accidentals, as described in Section~\ref{sec:accidentals}) can be viewed with the BRIL web monitoring tools.

The PLT is designed to operate at all times, regardless of the LHC beam conditions, to ensure that luminosity is always available for machine operations. The only exceptions are when the cooling or dry air supply to the PLT are interrupted, or when the LHC is operating in unusual conditions (\eg, in certain machine development studies) that cause significantly higher losses than normal.

\subsection{Pilot PLT}

The PLT was originally developed during Run 1 of the LHC (2010--12)~\cite{HallWilton:2011zz,Bugg:2011zz}. This version of the PLT used similar technology to the final version, but with different readout sensors, consisting of single-crystal diamond sensors with an area of $4{\times}4\mmsq$. It was envisioned that the use of diamond sensors would provide increased resistance to radiation damage and eliminate the need for a separate cooling system for the sensors. A pilot project was installed in CMS on the table used for the CASTOR detector~\cite{Chatrchyan:2008zzk} (behind HF) in 2012, at a distance of 14.5\unit{m} from the IP. However, the results from this run showed some undesirable features of the luminosity performance. In particular, it was observed that the efficiency of the charge collection varied with time during a fill, believed to be caused by a polarization effect in the diamond~\cite{Gan:2015zya}. As a consequence, it was decided to use silicon sensors for the final Run 2 PLT, as this was a well-understood and developed technology, although this necessitated the addition of the cooling system described above.

\subsection{Operational experience}
\label{sec:operations}

Over the course of Run 2, the PLT experienced several hardware failures. In 2015, there was a complete failure of two telescopes (channels 14 and 15) over the course of the year. These losses were traced to a failure of the low-current differential signaling (LCDS) chips in the port cards, which are responsible for sending signals to and from the TBMs and ROCs; the failures in these chips appeared to be linked to thermal cycles caused by interruptions in the CMS cooling. In 2016, new port cards were assembled and subjected to an extensive program of thermal cycling, and were installed in the 2016--17 year-end technical stop, successfully recovering those telescopes.

During 2016, two other telescopes (channels 0 and 4) failed in a different way, although also apparently triggered by thermal cycles in the PLT environment. In these telescopes, the pixel data disappeared entirely, but the telescope still had a fully functioning fast-or readout. These problems were traced to a failure in the analog level translator (ALT) chips on the OMB, which translated signal levels from those used by the TBM to those needed by the AOH. Because of the difficulty of replacing these components and the lack of available replacements, these were not fixed during Run 2. After their failures, although these telescopes were still producing fast-or data, they were excluded from the primary luminosity calculation because the lack of pixel data meant that they could not be calibrated and monitored as effectively.

Being in a high-radiation environment, the PLT hardware can be affected by occasional single-event upsets (SEUs), where an incident particle on the chip can cause data corruption. If the SEU happens to affect the configuration registers of the chip, this can cause partial or complete data loss from this chip. As a result, automated algorithms to detect these dropouts and reconfigure the PLT as quickly as possible (typically within a few seconds) were implemented in early 2016, allowing the PLT to continue providing good luminosity with minimal downtime and no manual intervention. These automatic recoveries were typically performed on the order of a few times per fill during Run 2.

The principal operational challenge in the PLT over the course of Run 2 was radiation damage in the sensors and the other front-end detector components, the former of which resulted in a decreased efficiency in the triple-coincidence measurement. This was partially compensated for by increasing the bias HV applied to the sensors, but continuous monitoring and correction was necessary. The initial bias voltage of 150\unit{V} was gradually increased over the course of Run 2 to a maximum of 800\unit{V} during 2018 running. In addition, the ROC thresholds for detecting a hit were occasionally recalibrated in an attempt to retain good efficiency even with decreased signal amplitude.

\section{Detector calibration}
\label{sec:detector_calibration}

Several calibration steps must be performed with the detector in order to ensure the best quality data. The relative positions of the detector planes must be measured and aligned in order to be able to properly reconstruct tracks from the pixel data; the effect of accidentals should be measured in order to correct for their effects in both the fast-or luminosity measurement and studies using full track reconstruction; the active area of the pixel sensors must be selected in order to retain good statistical uncertainty in the fast-or luminosity measurement while minimizing systematic effects; and the sensor efficiency should be measured over time in order to understand and correct for the time-dependent effect of radiation damage in the sensors. These studies are described in this section.

\subsection{Alignment}
\label{sec:alignment}

The intended position of the planes in the PLT is such that a line passing through the center of all three planes in a telescope will also pass through the nominal interaction point of CMS. However, the positions of the planes may vary slightly from their intended design values because of uncertainties introduced in the installation process.  The alignment process consists of two parts; first, aligning the active areas of each plane so that tracks will pass through the active area in all three planes, and second, measuring the difference of the plane positions from their nominal values so that tracks can be correctly reconstructed.

\begin{figure*}[hbtp]
  \centering
    \includegraphics[width=0.9\textwidth]{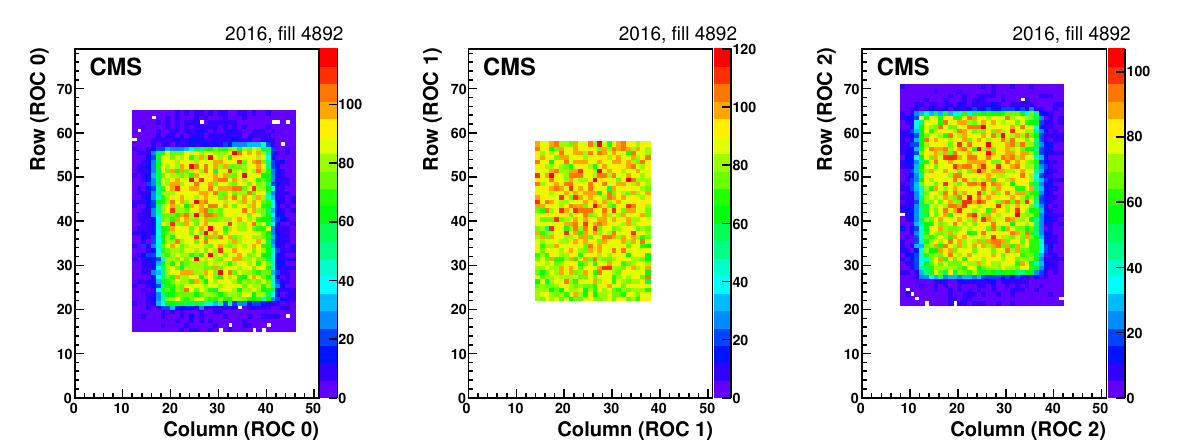}
    \caption{A demonstration of the active area alignment procedure in 2016, using data from LHC fill 4892. The color scale indicates the occupancy (number of hits) in each pixel. The center plane (center) uses the normal active area, while the outer planes (left and right) use a larger active area, allowing the image of the center plane to be clearly visible.}
    \label{fig:mask_alignment}
\end{figure*}

The alignment of the active areas is performed by using a special PLT configuration, in which the central plane had the normal active area (chosen to be at the center of the plane), but the two outer planes had a much larger active area. Tracks were then reconstructed from the collected data, using a pure sample of tracks in which only one cluster (a group of one or more contiguous hits, including diagonally touching pixels) was present on each plane. Each hit belonging to a reconstructed track is then plotted in a two-dimensional histogram of the sensor columns and rows. Figure~\ref{fig:mask_alignment} shows the results for fill 4892 in 2016. The center plane active area is visible, and in the outer planes, we see a central area with high occupancy, corresponding to tracks passing through all three planes, and a fringe with much lower occupancy from hits from other sources. The active area is then set to include this high-occupancy region on the outer planes. The procedure was repeated during early commissioning runs for each year, since the reinstallation of the PLT after the year-end technical stop could result in a change in the alignment.

Note that the track reconstruction procedure treats the tracks as straight lines; because particles passing through the PLT from the IP are travelling nearly parallel to the magnetic field, the deflection due to the magnetic field is negligibly small. No constraint to the nominal IP is applied in track reconstruction, and the linear fit uses only the position of the hits (for clusters containing more than one pixel, the hit position is taken as the average of the individual pixel positions, weighted by the charge in each pixel), without considering the resolution of the hit.

Second, in order to correctly reconstruct tracks in the PLT, the positions of the planes of the telescope must be determined, so that hits can be properly aligned. This requires determining both the relative rotation and displacement of the planes; we measure these quantities relative to the innermost plane in $z$, for a total of six alignment constants (one angular and two linear displacements for each of the other two planes).  To measure the alignment, a fill is selected with normal physics conditions and with no known operational issues for PLT. We then select a pure track sample consisting of tracks with exactly one cluster on each plane of a telescope, to avoid any problems with multiple track reconstruction. Each set of three hits is then fit with a linear function, and the slopes of the line in the $x$ and $y$ directions, as well as the fit residuals for each hit in the $x$ and $y$ directions, are recorded. All positions are expressed in local telescope coordinates, where $y$ is the radial direction in CMS coordinates and $x$ is the perpendicular direction.

The first step in the alignment is to measure the relative rotation of the second and third planes with respect to the first plane. To perform this measurement, a so-called ``XdY'' plot is constructed by taking, for each hit on a plane, the residual distance in the $y$ direction from the hit to the best-fit line as a function of its $x$ position; a ``YdX'' plot is constructed similarly. The XdY plot is then fit with a linear function, and the rotation of the plane is taken as the arctangent of the slope of the line.

Once the rotational alignment has been determined, the tracks are then refitted using the new alignment constants, and new XdY and YdX plots generated. The average of the XdY plot is used to determine the amount that the plane needs to be translated in the $y$ direction to be correctly aligned, and similarly, the YdX plot is used to determine the $x$ alignment.

\begin{figure*}[hbtp]
  \centering
    \includegraphics[width=\textwidth]{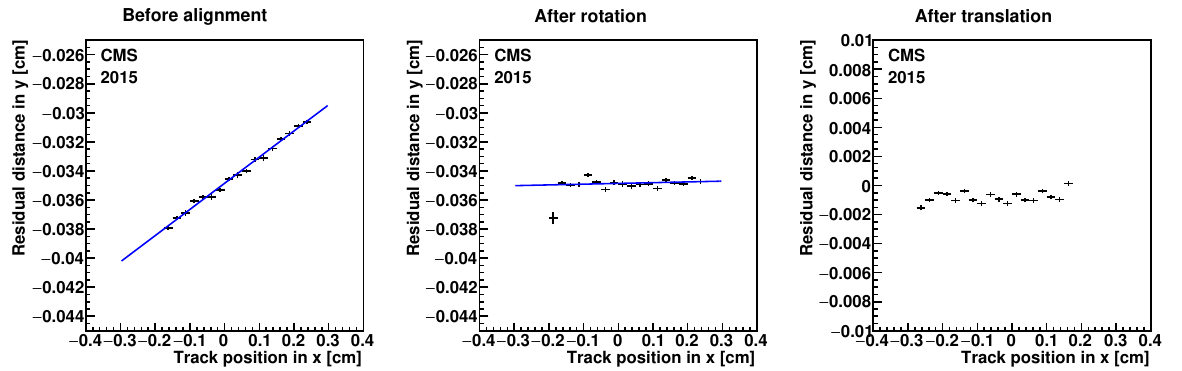}
    \caption{The XdY plots for the second plane (ROC 1) in telescope 7 for fill 4444 in 2015, showing the profiled distribution of the $y$ residual distance between the hit position and the fitted track as a function of the $x$ coordinate of the track at the plane. The three plots show the three stages of the alignment: (left) before alignment, assuming that the second and third planes are in exactly the design position relative to the first plane; (center) after the rotational correction, where the position of the plane has been rotated using the slope of the fitted line in the first plot (and similarly for the third plane); (right) after the translational correction, when the alignment procedure is complete. The blue line shows the fit used to determine the slope in the first plot, and the offset in the second plot.}
    \label{fig:alignment_xdy}
\end{figure*}

Finally, once the translational and rotational alignments have been determined, the tracks are refitted a third time using the final alignment to check that the average of the slope and residual distributions are 0 and that the XdY and YdX plots are flat. Figure~\ref{fig:alignment_xdy} shows the XdY plots for a single plane at the various steps in the calibration for the alignment performed in fill 4444 in 2015.

Note that this alignment procedure only aligns the three planes of a telescope with respect to each other; it does not change the global coordinates of the first plane of the telescope. For this, an analysis combining the data from multiple telescopes and using the CMS interaction point is necessary, as discussed in Section~\ref{sec:beamspot}.

\begin{figure*}[hbtp]
  \centering
    \includegraphics[width=\textwidth]{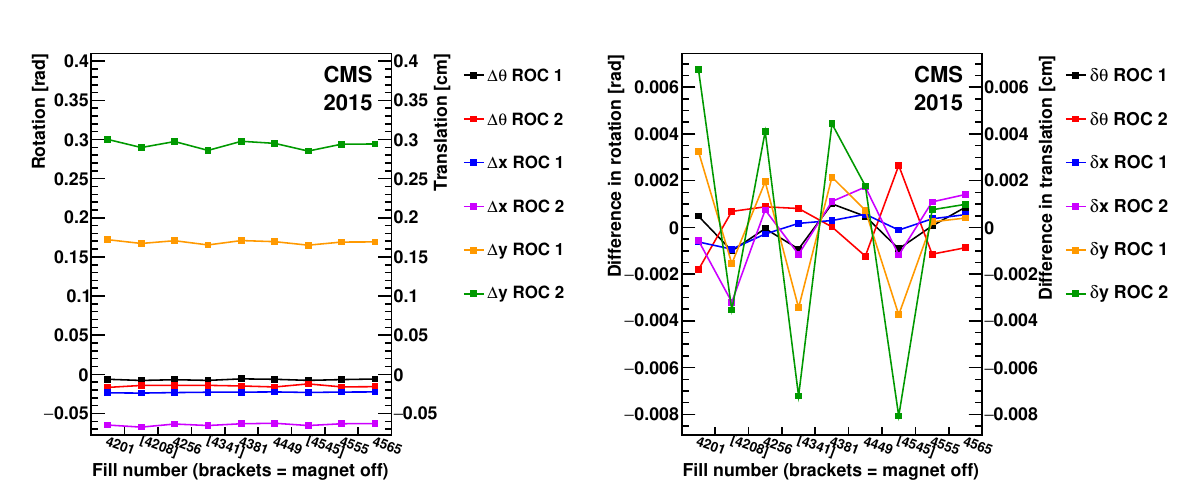}
    \caption{Alignment vs. time for a single PLT telescope (channel 10) for nine different fills in 2015, where brackets denote fills in which the CMS magnet was off. The alignment is described by six parameters: rotation ($\Delta\theta$), translation in $x$ ($\Delta x$), and translation in $y$ ($\Delta y$) of ROC 1 relative to ROC 0, and the same three quantities for ROC 2 relative to ROC 0. The nominal value for each of these is zero, except for $\Delta y$, which has a physical nonzero value, because of the 1.15$^\circ$ angle between the beam axis and the center line of a telescope. The left plot shows the absolute alignment values, and the right shows the difference of these values (designated by $\delta$) compared to their average.}
    \label{fig:alignvstime}
\end{figure*}

During the 2015 run, the CMS magnet was not on for all fills, because of operational issues with the magnet. To check the stability of the alignment over time, we selected nine fills in 2015, with each pair of fills separated by at least one magnet ramp. In some cases there were no physics fills taken when the magnet was off, so we have two consecutive magnet-on fills. Figure~\ref{fig:alignvstime} shows the results for a single PLT telescope (channel 10). The absolute alignment constants are shown on the left, while the right shows the difference in the value for each individual fill from the average value. In general the alignment is quite stable over the course of the year, but we do see that there is a difference in the $y$ translation constants (and possibly a small difference in $x$) between the magnet-on and magnet-off fills. However, since in general the physics data of interest only uses fills with the magnet on, we can treat the alignment as constant over the course of a given year.

\subsection{Accidentals}
\label{sec:accidentals}

As discussed previously, accidentals are one of the most significant contributions to the systematic uncertainty in the fast-or luminosity. We can use the pixel data to estimate the contribution to the fast-or rate from accidentals as follows. First, the alignment procedure is carried out, as described in Section~\ref{sec:alignment}. Once the detector has been aligned, histograms are built of the distributions of the $x$ and $y$ track slopes in local sensor coordinates, and the $x$ and $y$ residuals of each hit relative to the fitted track on each plane of the telescope. The mean and standard deviation $\sigma$ of each distribution are then recorded. The observed distributions are Gaussian in shape. The mean for each of these distributions should be 0, except for the $y$ slope, which should have a mean value of approximately 0.027 from the natural slope of the PLT (\ie, the fact that the second and third planes are located slightly farther away from the beam line in the radial direction). We then define a candidate track as an accidental if either of the slopes, or any of the residual values, is more than $5\sigma$ away from the mean value for that distribution; otherwise, the track is considered to be good. In the case of multiple candidate tracks in a single telescope, we consider all possible combinations of hits, and we take the event as good if any combination forms a good track, since in the zero-counting method used for luminosity determination (as described in Section~\ref{sec:lumi}), it only matters if the number of triple coincidences is zero or nonzero.

\begin{figure}[tbhp]
\centering
  \includegraphics[width=0.49\textwidth]{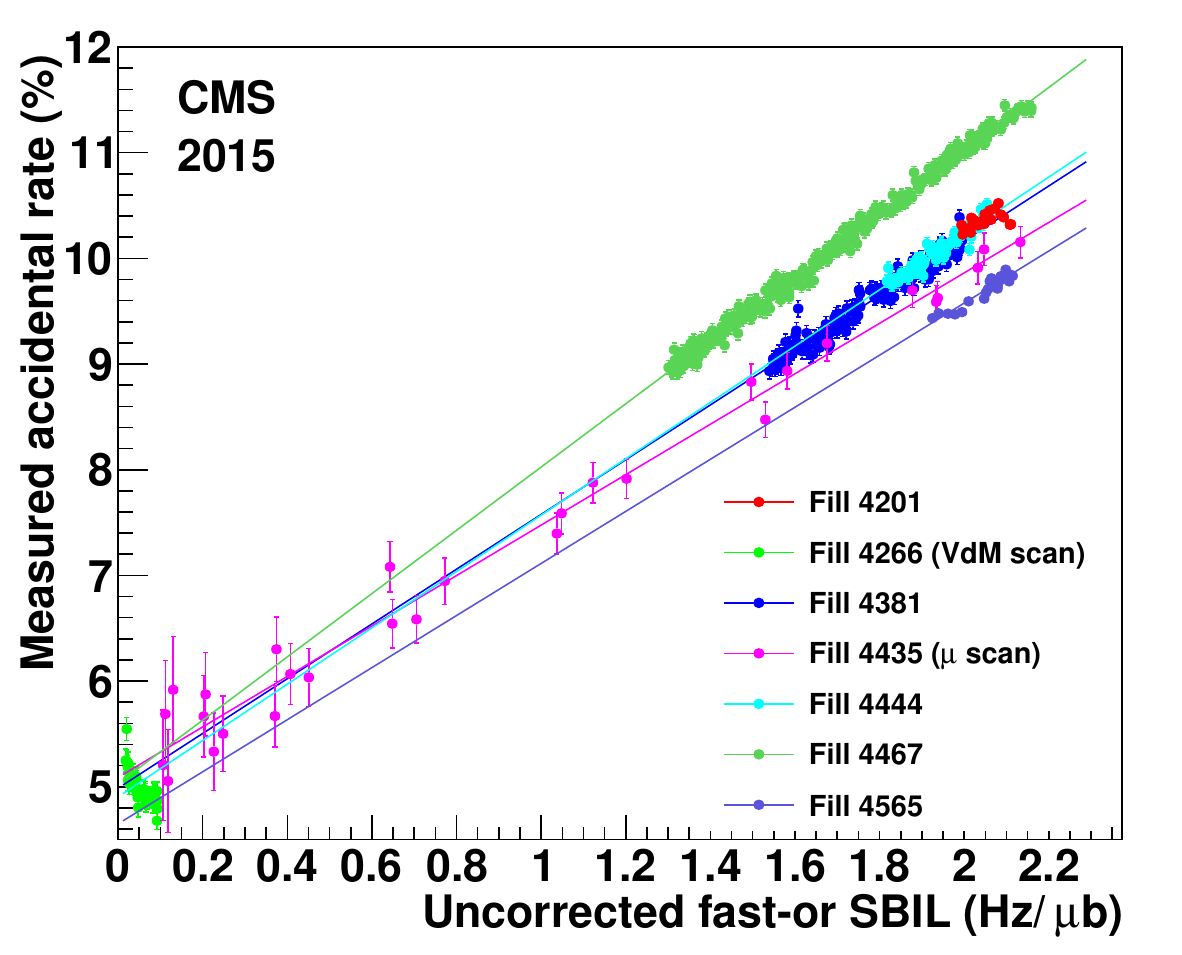}
  \includegraphics[width=0.49\textwidth]{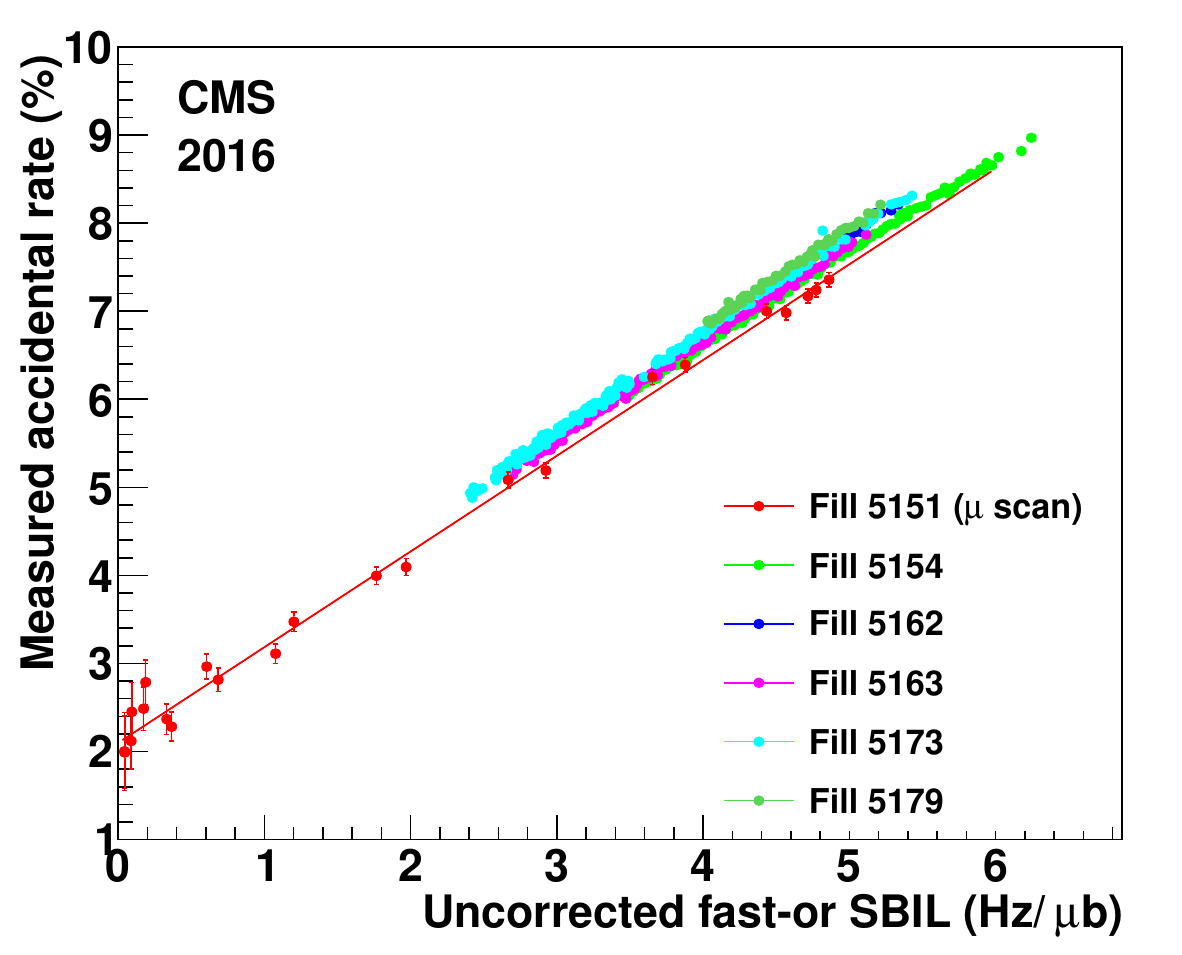}
  \includegraphics[width=0.44\textwidth]{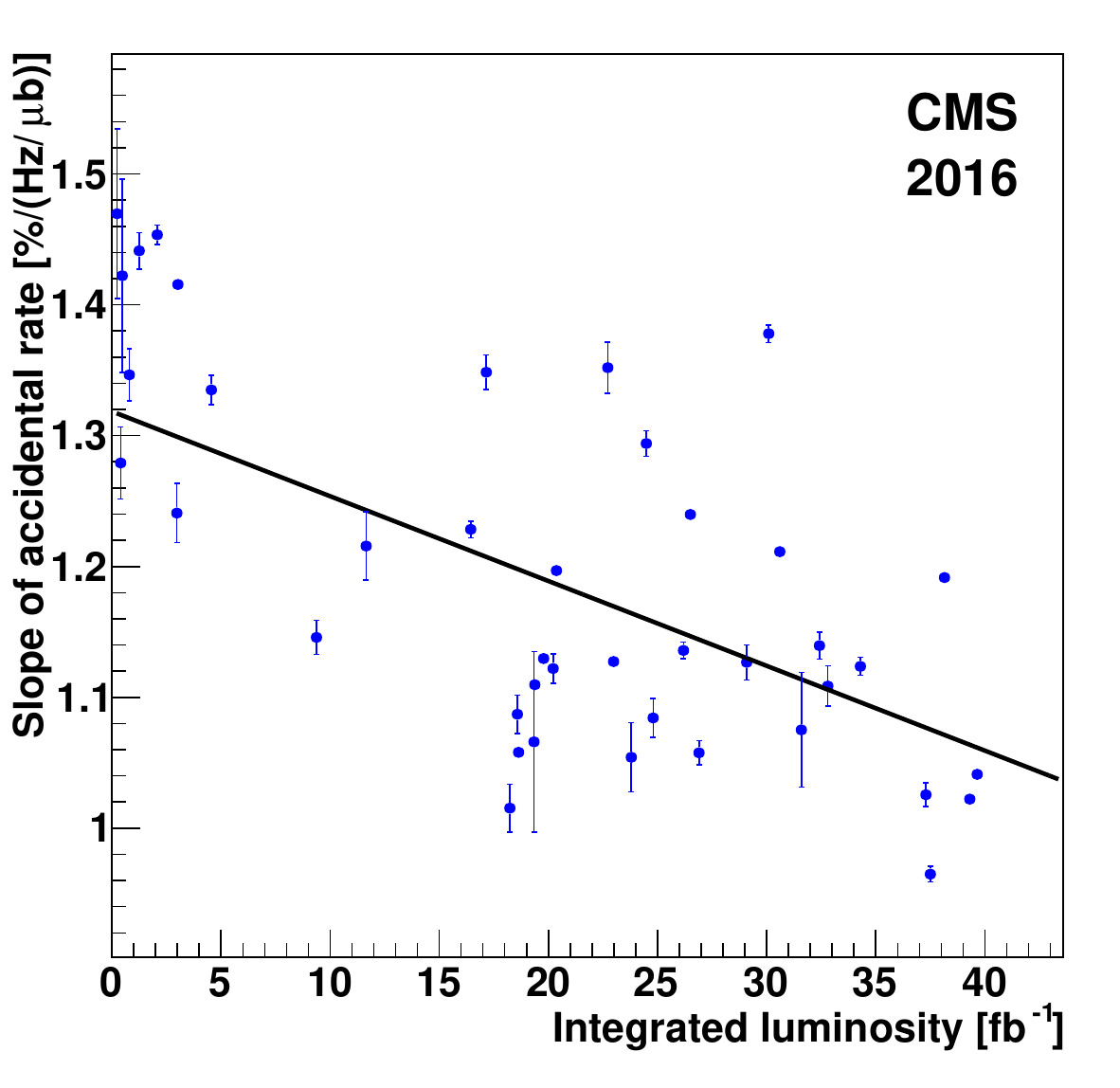}
  \caption{Measured PLT accidental rate, as a function of SBIL, for selected fills in 2015 (\cmsUL) and 2016 (\cmsUR). For 2015, a linear fit for each fill is shown; in 2016, for clarity, only the linear fit for fill 5151 is shown. The apparent increase in the accidental rate at very low luminosities in 2015 is because of the larger relative contribution from beam-induced background, as discussed in the text. The bottom plot shows the evolution of the slope of the per-fill accidental rate fit over the course of 2016, where the black line shows a linear fit to the results.}
  \label{fig:accidental_rates}
\end{figure}

The top plots in Fig.~\ref{fig:accidental_rates} show the measured accidental rate as a function of the single-bunch instantaneous luminosity (SBIL) for a variety of 2015 and 2016 fills, including fills for regular physics, VdM calibration, and ``$\mu$ scans''. In the $\mu$ scan, the fill starts with normal physics conditions, but then the beams are separated, so that the behavior of the luminometers can be probed over a wide range of instantaneous luminosity. The overall observed accidental rate is generally linear as a function of SBIL. In 2015, the rate is reasonably consistent across fills, although there is some fill-to-fill variation, which is taken as a systematic uncertainty in the correction. In 2016, the slope of the per-fill accidental rate is also observed to change over time, as illustrated in Fig.~\ref{fig:accidental_rates} (bottom), which is accounted for in the correction. This is presumably caused by the overall decrease in efficiency from radiation damage in the sensors. Note that the measured accidental rate is significantly lower in 2016, despite the much higher SBIL range in the 2016 data; this is because of the optimization of the sensor active area described in Section~\ref{sec:activearea}, which improves the rejection of tracks originating from locations other than the IP.

Note that an accidental rate that is a constant fraction of the luminosity simply effectively increases the acceptance of the detector, so it can automatically be accounted for in the VdM calibration. Thus, the constant term of the fits shown in Fig.~\ref{fig:accidental_rates} does not affect the overall calibration as long as it remains constant over fills; the only part that matters is the luminosity-dependent accidental contribution (the slopes of the lines in Fig.~\ref{fig:accidental_rates}).

In general, the significant majority of triple coincidences (approximately 85--90\%) that are classified as accidentals fail the residual requirements; that is, they do not actually form a straight line and thus are likely due to random combinations of hits from two or more sources. The remainder do form straight lines, but do not have slope values within $5\sigma$ of the average slope. This suggests that they are tracks not originating from the IP, but from sources such as beam-induced background (BIB) or activated material in the detector; some may also be combinatorial background that happen to form a line by chance.

We can confirm this hypothesis by examining the accidental rates in a VdM scan; as the luminosity in these fills is very low, we would expect a constant fraction of accidental tracks, consistent with the $y$-intercept of the lines shown in Fig.~\ref{fig:accidental_rates}. When the beam separations are small, this is indeed the case. However, at larger beam separations, the beam-induced background remains the same, while the luminosity decreases. As a result, the overall accidental rate becomes constant (rather than the accidental fraction being constant), resulting in a higher apparent accidental fraction.

To investigate a possible dependence of the measured accidental rate on the particular value of the selection used to reject accidentals, we measured the accidental rate in a representative 2015 fill where the nominal $5\sigma$ requirement was varied to $4.5\sigma$ and $5.5\sigma$. Decreasing the requirement results in no distinguishable change in the measured accidental rate, but increasing to $5.5\sigma$ results in a noticeable decrease in the accidental rate (by $\approx$5\%). This is because a $5.5\sigma$ interval is large enough that tracks with a $y$ slope (in local sensor coordinates) near 0 can pass this criterion, so tracks parallel to the beam from beam-induced background are no longer rejected. This suggests that a slightly smaller value of the accidental rejection threshold than $5\sigma$ may actually be optimal, to ensure that we are safely away from this region.

\begin{figure}[tbhp]
\centering
  \includegraphics[width=\cmsFigureWidth]{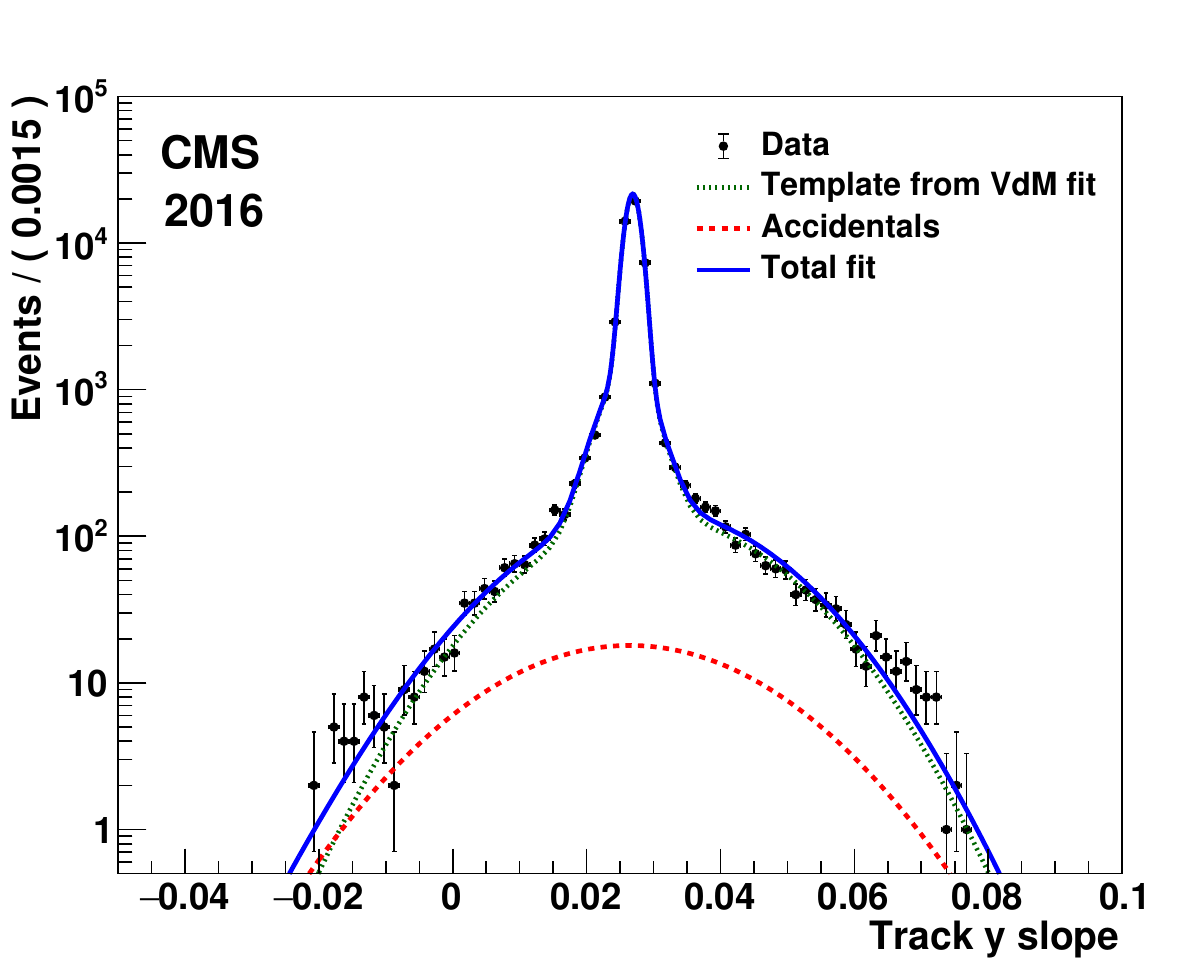}
  \caption{Maximum likelihood fit to the slope distribution from fill 4979 in 2016 with an instantaneous luminosity of approximately 6\ten{33}\percms. The dotted green curve represents the component from the distribution at VdM luminosity, the dashed red curve represents the additional accidental contribution at higher luminosities, and the solid blue curve is their sum.}
  \label{fig:accidental_likelihood}
\end{figure}

A new algorithm for measuring the accidental rates was developed in 2017 and used to analyze the 2016 data. In this algorithm, the distribution of the track slopes is fit with a maximum likelihood, containing two components, one representing the slope distribution at VdM luminosity (obtained from a fit to that distribution), and one representing the additional accidental component at higher luminosity. This method thus accounts for the fact that, assuming the slope value of accidental tracks is mostly randomly distributed, some accidental tracks will pass the slope requirements in the $5\sigma$ method by chance. The function used for the fit at VdM luminosity is the sum of three Gaussian terms with independent means and standard deviations, and the function used for the accidental component is a single Gaussian term. An example of such a fit is shown in Fig.~\ref{fig:accidental_likelihood}. In general, the results from the two methods are broadly consistent, although the likelihood fit method yields a lower accidental rate than the $5\sigma$ method.

\subsection{Optimization of active area}
\label{sec:activearea}

The accidental rate measured in Section~\ref{sec:accidentals} depends strongly on the active area of the sensors. As the accidental rates measured in 2015 were substantial, we conducted a study prior to the start of 2016 running in order to determine the optimal active area, balancing the need to reduce the accidental rate with the need to maintain good statistical precision in the PLT measurement.

\begin{figure}[tbhp]
\centering
  \includegraphics[width=\cmsFigureWidth]{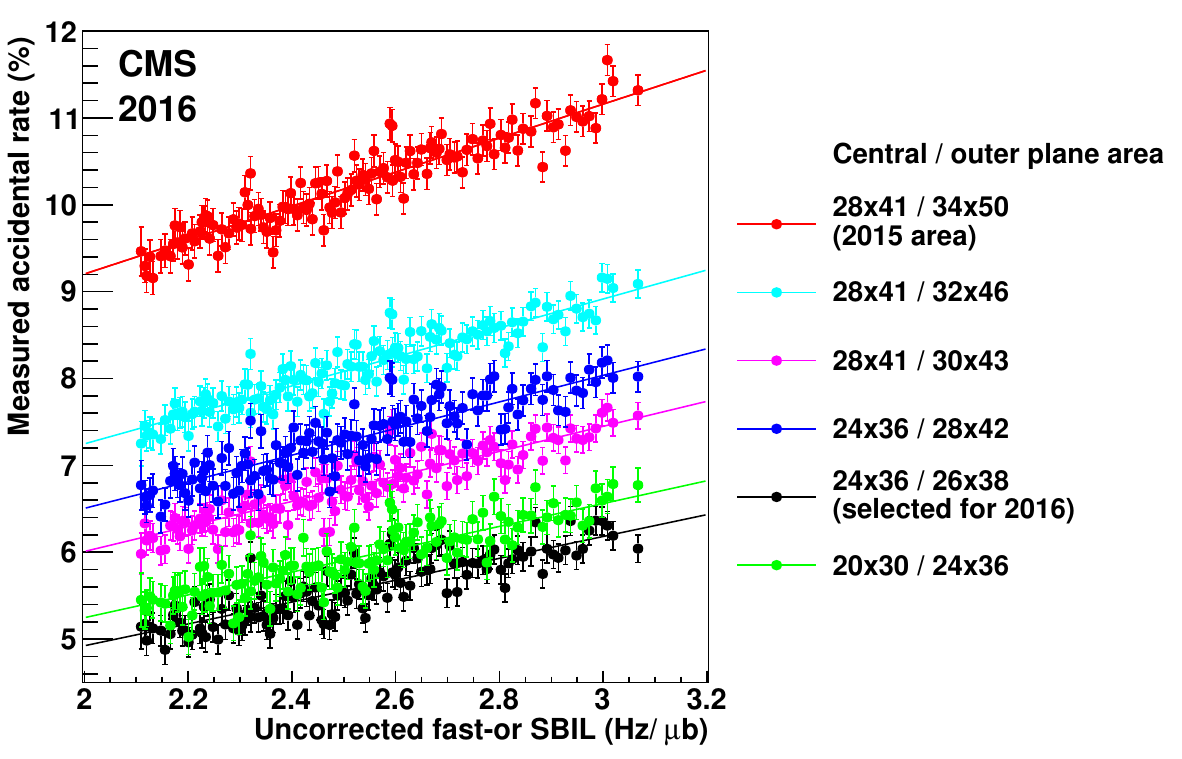}
  \caption{Measured PLT accidental rate for a typical LHC fill, as a function of SBIL, for different active areas. The 2015 active area was 28 columns${\times}$41 rows in the central plane and 34 columns${\times}$50 rows in the outer planes, and the selected active area for 2016 was 24 columns${\times}$36 rows in the central plane and 26 columns${\times}$38 rows in the outer planes. We observe that the size of the ``fringe'' area (the extra area in the outer planes compared to the central plane) has a substantial effect on the accidental rate.}
  \label{fig:accidental_vs_area}
\end{figure}

To measure the effect of reducing the active area, the accidental rate was measured using the procedure described in Section~\ref{sec:accidentals} on fill 4892 in 2015. A variety of smaller active areas were then considered by redoing the accidental analysis but excluding pixels that would not fall within the new active area. The results are shown in Fig.~\ref{fig:accidental_vs_area}. We can observe that even relatively small changes in the active area can result in a significant change in the measured accidental rate. In particular, the ``fringe'' region (the area in the outer planes beyond the area in the central plane) contributes significantly to the measured accidental rate; while some fringe area is necessary in case of misalignment, these results suggest that the size of the fringe area could be reduced from the 2015 value. The expected loss in statistical precision was measured in simulation and compared to the effect on the accidental rate.

As a result of these studies, an active area size of 24 columns${\times}$36 rows ($3.6{\times}3.6\mmsq$) in the center plane and 26 columns${\times}$38 rows ($3.9{\times}3.8\mmsq$) in the outer planes was adopted and used throughout the rest of Run 2. This resulted in an approximately 40\% decrease in the accidental rate while incurring only a modest loss in statistical precision (approximately 10\% in simulation).

\subsection{Efficiency measurement with track reconstruction}
\label{sec:efficiency}

Because of radiation damage in the sensors and ROCs, the efficiency of reconstructing a hit gradually decreases over time, and because of the triple coincidence requirement, the PLT luminosity measurement is particularly sensitive to these effects. The loss of efficiency can be measured using the pixel data.

Let us designate the three planes in a PLT telescope as 0, 1, and 2. We can measure the efficiency of plane 0 by using the reconstructed track data as follows~\cite{Pompeo:2017kig}. First, we consider the number of events where we find a stub consisting of a hit in each of planes 1 and 2. We then take this stub and extrapolate it to the $z$ location of plane 0, and find the resulting point of intersection. Let $n_{12}$ be the number of such stubs where the extrapolated track lies in the active region of plane 0. Then, we consider the number of events $n_{0|12}$ where a hit is found in plane 0 that matches the extrapolated stub. Specifically, we require there to be a hit in plane 0 within 5 rows and 5 columns of the location of the extrapolated stub on plane 0; this area is chosen to be significantly larger than the expected uncertainties from the extrapolation and hit resolution. The efficiency of plane 0, $e_0$, is then defined as $n_{0|12}/n_{12}.$ We can define efficiencies for planes 1 and 2 by using two-hit stubs in the other two planes in a similar fashion. This efficiency corresponds to the efficiency of plane 0 with respect to planes 1 and 2 in the same telescope, and we refer to it as ``track-hit'' efficiency.

We expect that some fraction of the two-hit stubs $n_{12}$ will be due to accidentals rather than from a genuine track, and that this fraction will be higher than the accidental rate for triple coincidences. In this case, of course, no matching hit will be found in plane 0 and so the efficiency will be systematically lower than the true value. To reduce the contribution from accidentals, we thus require that the track slopes in the $x$-$z$ and $y$-$z$ planes are constrained to be consistent with tracks originating from the IP. Specifically, we require that the slope in $x$ is within a certain value $\Delta$ of the nominal value of 0, and similarly that the slope in $y$ is within $\Delta$ of the nominal value of 0.027. We choose a relatively small window of $\Delta = 0.005$ to minimize contributions from accidentals.

\begin{figure}[htbp]
\centering
  \includegraphics[width=0.32\textwidth]{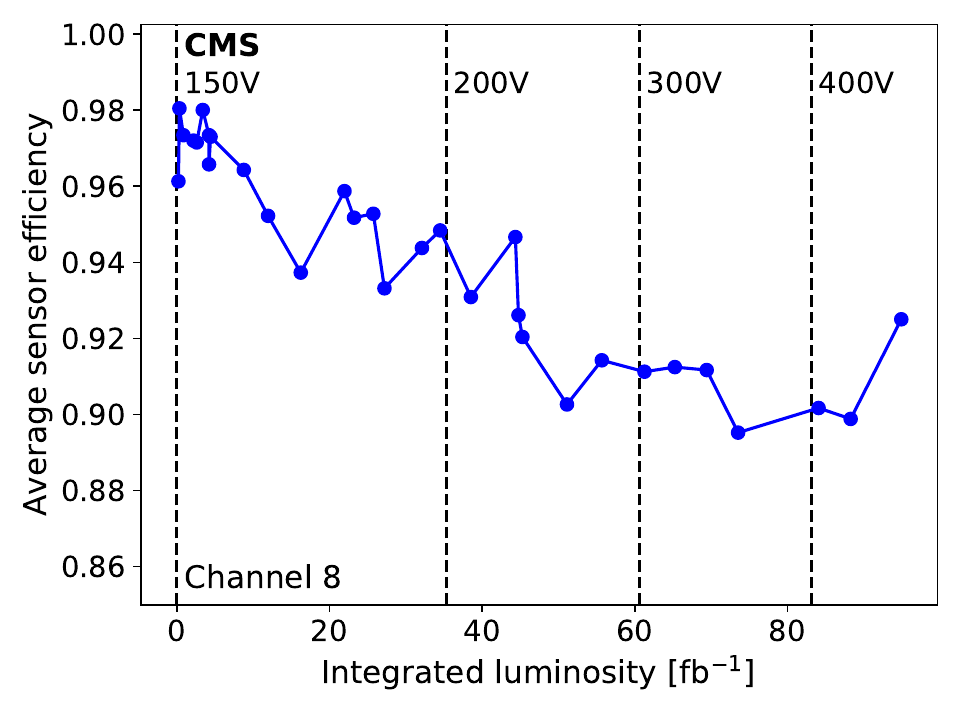}
  \includegraphics[width=0.32\textwidth]{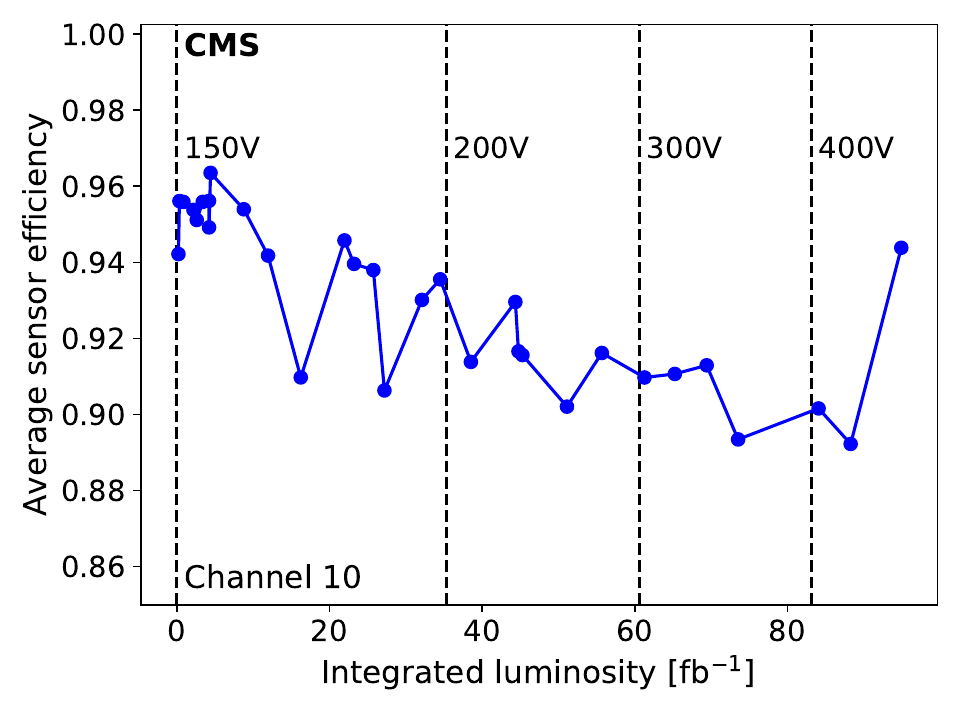}
  \includegraphics[width=0.32\textwidth]{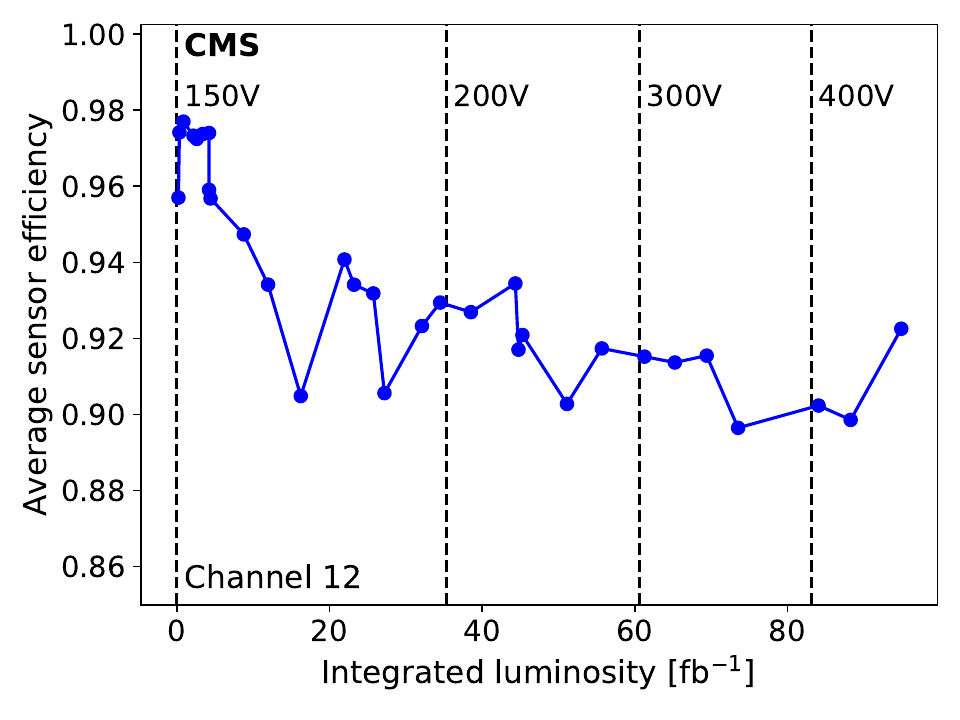}
  \caption{Average sensor efficiencies for three telescopes as a function of integrated luminosity in 2015--17, for channels 8 (\cmsLeft), 10 (center), and 12 (\cmsRight). The dashed lines indicate points at which the bias voltage used for the sensors was changed. The uncertainties in the efficiency values are too small to be visible in this plot.}
  \label{fig:ROCEfficiency_3Telescopes}
\end{figure}

However, some contribution from accidental stubs will remain, and this means that the measured efficiency will always be underestimated. As a consequence, we use this measurement primarily as a relative measure rather than an absolute value. Similarly, in principle, the efficiency for a telescope should simply be the product of the efficiencies of the three individual planes. However, this measurement may be affected by correlations among the relative efficiencies of the different planes in the telescope, such as from a constant rate of accidental stubs present. By looking at the correlation coefficient, we can observe that there is often significant correlation (from 0.2 up to 0.95) among the different sensors. This is presumably due to the fact that the efficiency loss is driven by the radiation damage that all the sensors are exposed to, although the effect is not necessarily exactly the same across all sensors. As a result, for measuring the efficiency of a telescope, we use the average of the three individual sensor efficiencies in a telescope, which we refer to as the ``average sensor efficiency''. Results are shown for three telescopes, corresponding to channels 8, 10, and 12, as a function of integrated luminosity in 2015--17, in Fig.~\ref{fig:ROCEfficiency_3Telescopes}. The changes in HV appear to slow down the loss in efficiency somewhat, although they do not increase it. We note that the final efficiency corrections described in Section~\ref{sec:lin_eff} are derived using different methods, although they generally agree with the measurements here.

\section{Measurements of detector properties, performance, and beam conditions}
\label{sec:performance}

While the primary deliverable of the PLT is the luminosity measurement using triple coincidences, the PLT is also capable of measuring other quantities of interest, both for internal monitoring of the detector performance and of the LHC machine conditions. These measurements are described in this section.

\subsection{Pulse heights}
\label{pulse_heights}

In addition to measuring the position of hits, the ROC is also capable of measuring the charge deposited by a particle traversing the depletion region of a sensor. Analysis of these data over time can also provide a measure of the effect of radiation damage in the PLT. For this analysis, we measure the charge deposited in each cluster of hits, where a cluster is defined as a contiguous group of hits.

First, the gain must be calibrated, to translate the raw values produced by the ROC into a charge (measured in number of electrons). This is performed by injecting a known amount of calibration charge into each pixel, measuring the resulting response, and repeating the process for a variety of input charge values over all pixels. The resulting curve is then fitted and used to derive the calibration. As the fit does not always converge or have good quality, only pixels with good calibrations are selected.

\begin{figure}[htbp]
    \centering
    \includegraphics[width=\cmsFigureWidth]{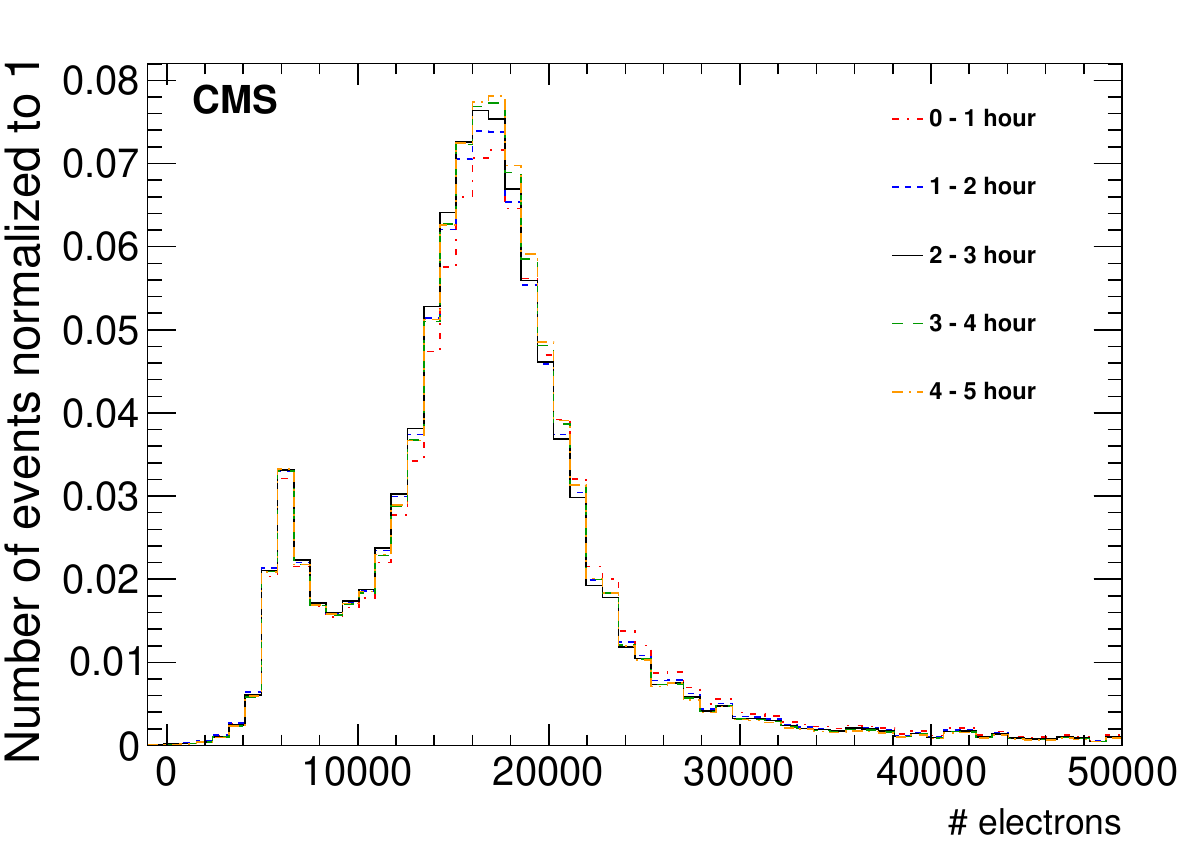}
    \caption{Pulse height distributions for fill 6035 in 2016, divided into 1-hour intervals, where each color represents a separate interval. Since each interval may not contain the same number of events, each individual histogram is normalized to a total of 1.}
    \label{fig:phintervals}
\end{figure}

For these studies, we select pixels which have good fits for all gain calibrations taken during the course of Run 2, and ROCs which have a consistently high number of pixels with a good fit. First, we can verify that the pulse height is stable with time during a single fill. To check this, we split the data from the fill into 1-hour intervals, and examine the pulse height distribution for each interval. Figure~\ref{fig:phintervals} shows the results for fill 6035 in 2016. We observe that the pulse heights are indeed stable over the course of a single fill.

Since the charge distribution represents the collected charge from an individual PLT sensor, it is expected that the distribution should be a Landau distribution convolved with a Gaussian distribution; as the radiation damage to the sensor increases, the Gaussian component becomes dominant. However, as we can see in Fig.~\ref{fig:phintervals}, there is often a second lower peak in the distribution, which can reach a significant amplitude. This peak at $\approx$4500 electrons could be produced by a number of causes such as radiation damage, the quality of the gain calibration, or time walk effects in which a signal is distributed across multiple BXs. The second part of this study was focused on the examination of the factors that contribute to this peak.

Two possibilities are considered for the production of the secondary peak. The first is that the secondary peak results from hits from other sources (noise or other detector background), which would have a different energy distribution. To test this hypothesis, we look at hits only from events where a triple coincidence is produced in the telescope, thus significantly reducing the contribution from noncollision sources. The second is that the secondary peak is produced by time walk in the ROC, and actually is the result of a signal from a collision source spilling over into the next BX. To test this hypothesis, we look at events from leading bunches (where there is no collision in the preceding BX) and compare to events from empty BXs immediately following a colliding bunch.

\begin{figure}[htbp]
    \centering
    \includegraphics[width=0.45\textwidth]{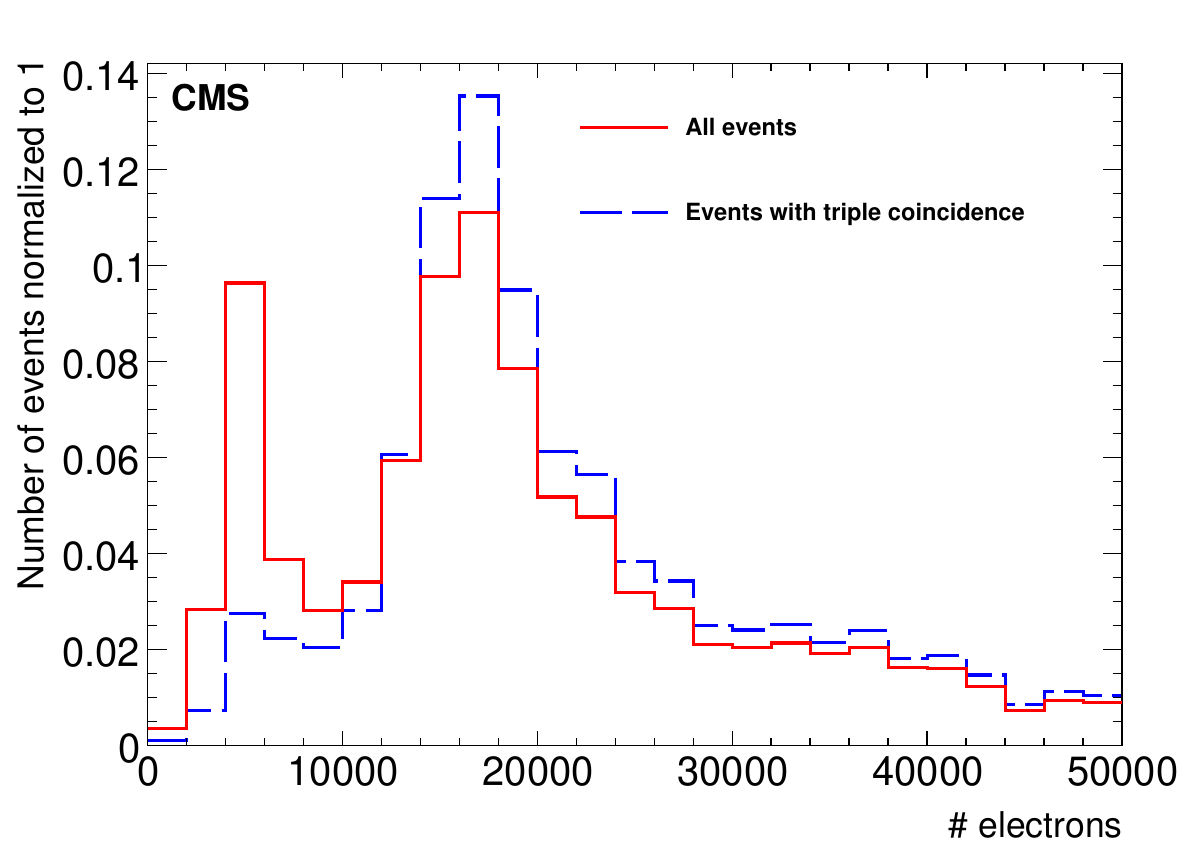}
    \includegraphics[width=0.45\textwidth]{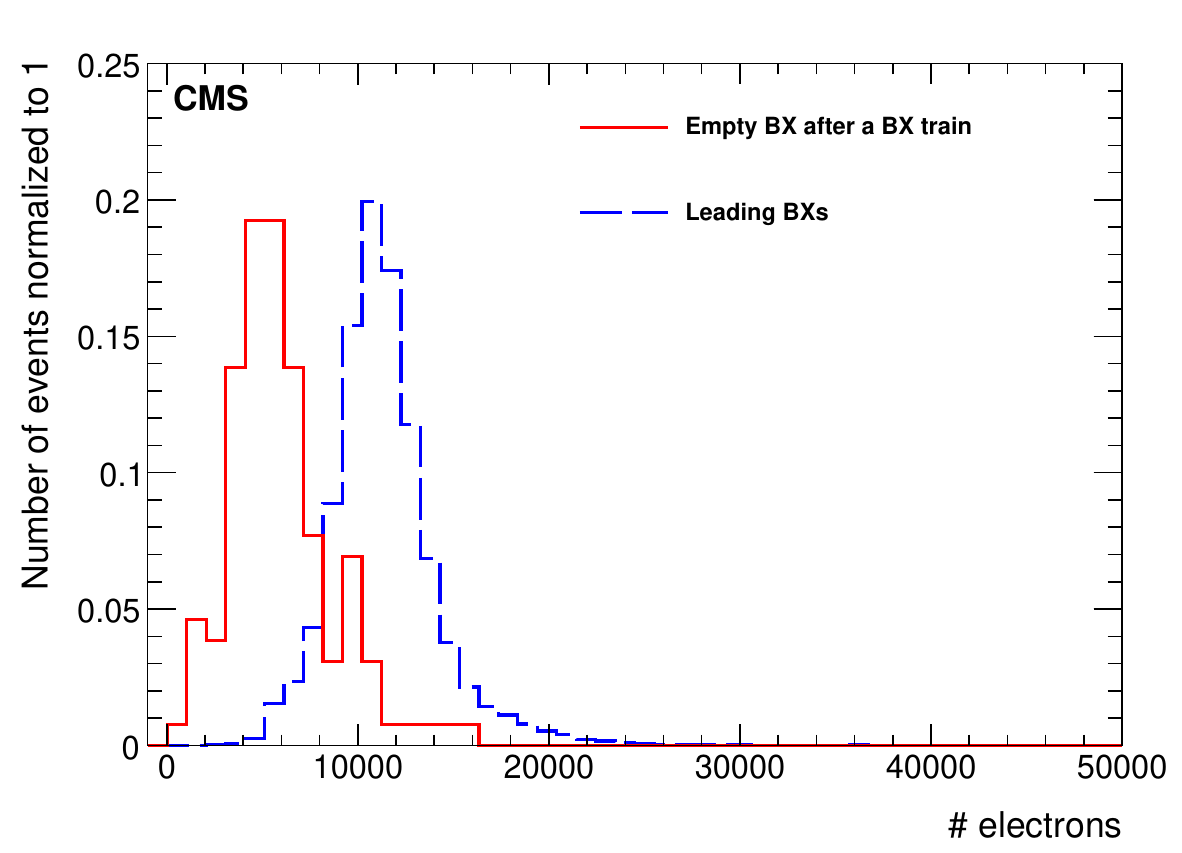}
    \caption{\cmsLeftCap: Pulse height distribution for a single ROC in a single LHC fill before (solid red line) and after (dashed blue line) the triple coincidence requirement is applied. \cmsRightCap: Pulse heights only for events in the leading bunch of a train (dashed blue line), and only for events corresponding to the first empty BX after a train (solid red line). In both plots, all histograms are normalized to unit area. These distributions are from a 2018 fill, where the hit thresholds on the ROCs were lower than in 2016 (Fig.~\ref{fig:phintervals}).}
    \label{fig:ph_secondary}
\end{figure}

The results of these tests are shown in Fig.~\ref{fig:ph_secondary}. We observe that applying the triple coincidence requirement significantly reduces the second peak, although it does not eliminate it entirely. However, the test with leading bunches strongly supports the hypothesis of the secondary peak being caused by timewalk effects, as the secondary peak is not visible at all in the leading bunches, while the empty BXs following colliding bunches, which should contain only timewalk signals, are peaked much lower, corresponding to the secondary peak visible on the left of the first plot in Fig.~\ref{fig:ph_secondary}.

The next step is to examine the pulse height distribution over time (or, more precisely, as a function of integrated luminosity). This measurement is planned for Run~3, as a way to provide an additional monitor of the effect of radiation damage on the sensors.

\subsection{Measurement of the bias voltage for full hit efficiency}
\label{sec:depletion_voltage}

In order to maximize the signals from each sensor, sufficient bias voltage must be applied to create a depletion layer across the p-n junction. As the sensor suffers radiation damage, the voltage required to maintain a high hit efficiency will increase over time. In this section, a measurement of the voltage necessary for maximum hit efficiency, which we designate \Vmax, as a function of integrated luminosity is discussed.

The measurement is based on a series of HV bias scans performed during LHC fills. The resulting triple-coincidence rate corresponding to each point in the scan is measured for each PLT channel. A plateau in the rate is expected as the HV set point is increased, and the minimum set point to reach the plateau is designated as \Vmax for that channel and scan. This point is defined as the lowest HV point such that the difference in rate between the point and the next point is less than 1\%, and between the point and the second following point is less than 2\%. This is the point where each sensor is sufficiently depleted to yield enough signal, although it does not necessarily correspond to a fully depleted sensor. These scans were performed by hand occasionally at the beginning of Run~2. Towards the end of Run~2, an automated scan program was introduced which allowed scans to be run regularly (approximately once every month). 

\begin{figure}[htbp]
\centering
  \includegraphics[width=\cmsFigureWidth]{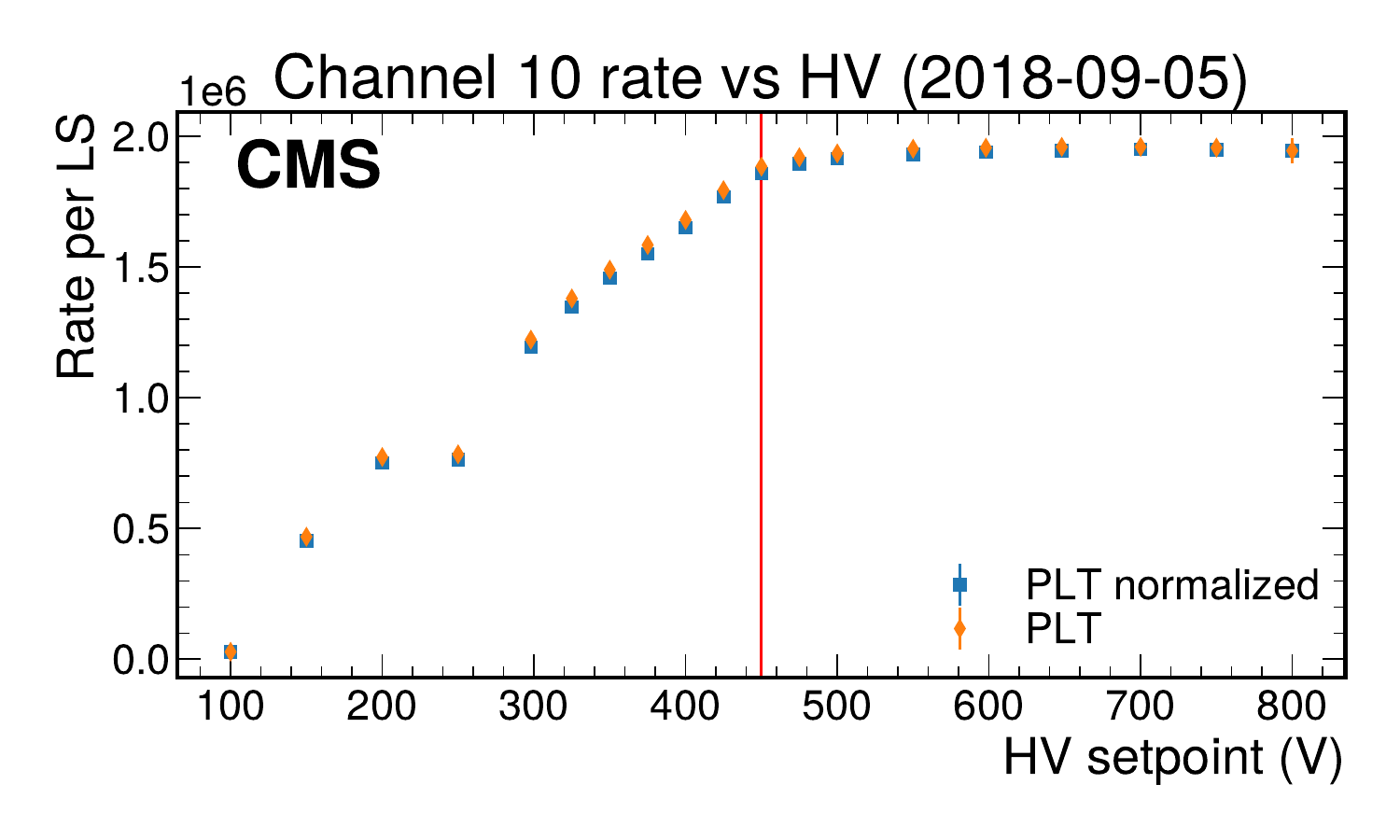}
  \caption{Raw (orange diamond) and normalized (blue square) rates at each HV set point for channel 10 in a 2018 scan. The vertical line indicates the calculated \Vmax. We observe some nonuniform behavior at low HV values, likely due to time walk effects.}
  \label{fig:biasscan_rates}
\end{figure}

For each scan, the observed fast-or rate is plotted against the HV applied at each step. The beginning of the step is excluded to allow the rate to stabilize after the HV change, and to account for the fact that the luminosity is generally naturally decreasing over the course of time, the PLT rate is normalized to a reference luminometer (HFET if it is available, or BCM1F otherwise). The normalized rate is obtained by taking the ratio of the PLT rate to the reference luminometer at each scan point, and then scaling by an overall arbitrary factor to match the scale of the raw rate. Figure~\ref{fig:biasscan_rates} shows an example of the resulting scan curves and the extracted \Vmax.

\begin{figure*}[htbp]
\centering
  \includegraphics[width=0.4\textwidth]{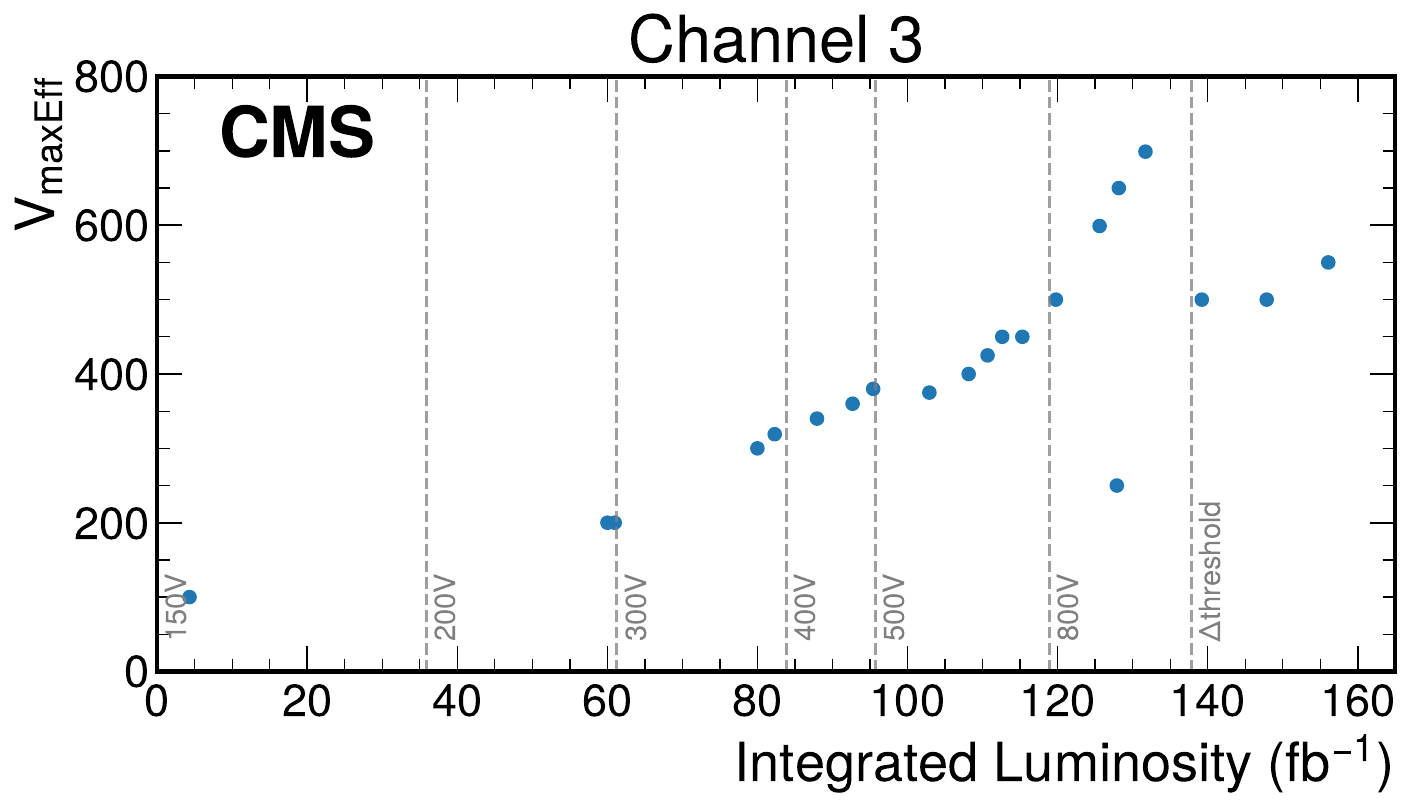}
  \includegraphics[width=0.4\textwidth]{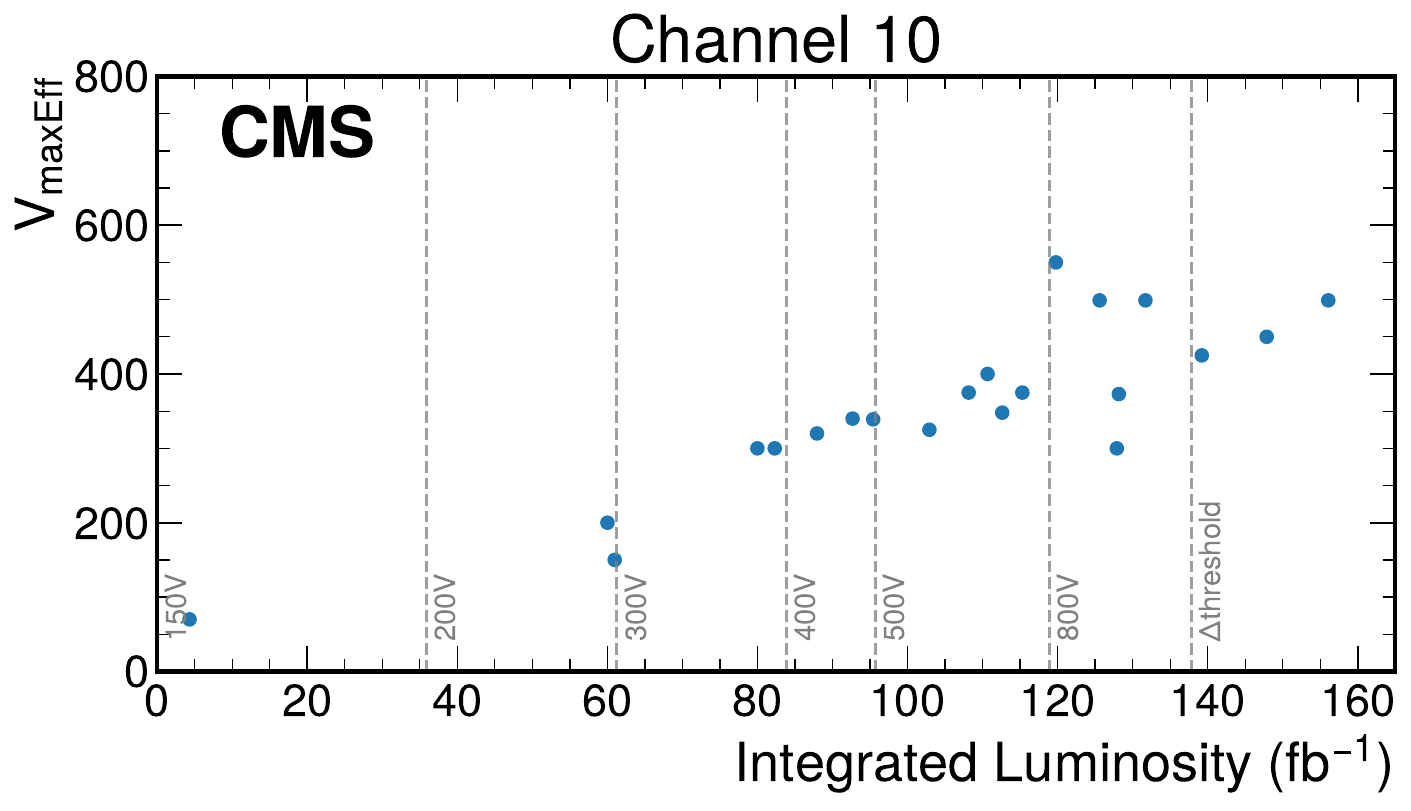}
  \includegraphics[width=0.4\textwidth]{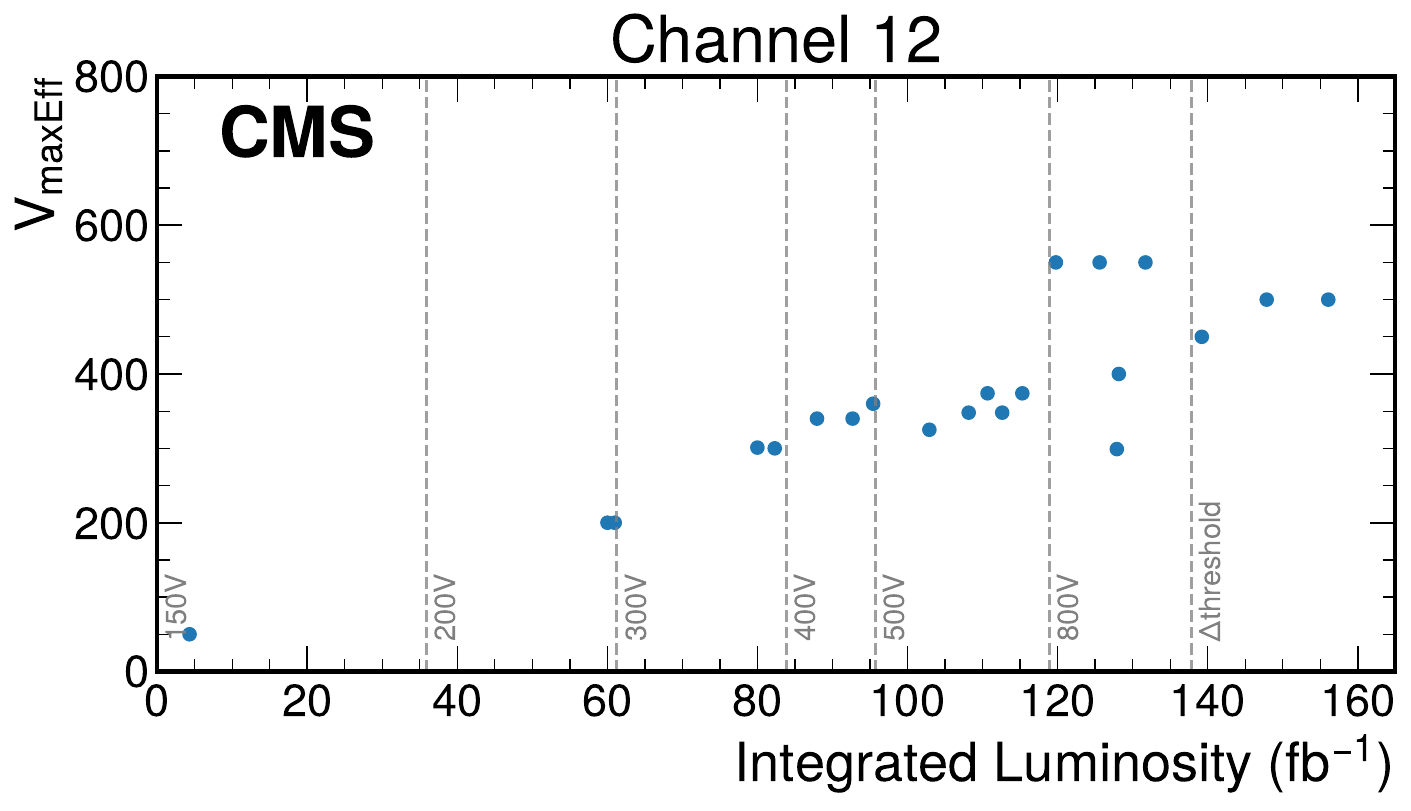}
  \includegraphics[width=0.4\textwidth]{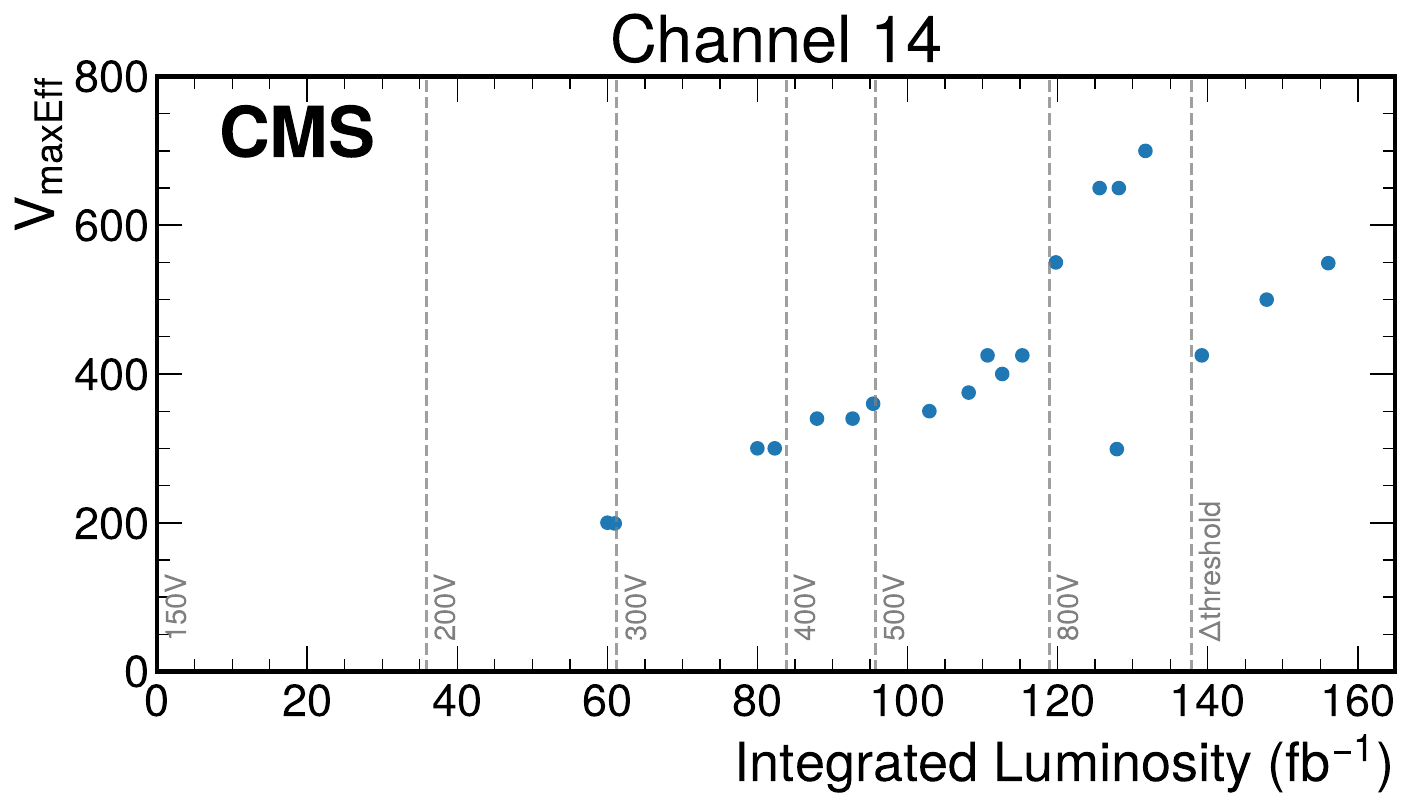}
  \caption{Calculated \Vmax derived from HV scans as a function of integrated luminosity for four selected PLT channels: channel 3 (upper left), channel 10 (upper right), channel 12 (lower left), and channel 14 (lower right). The dashed vertical lines indicate changes in operating conditions, with the rightmost denoting the change in the ROC thresholds ($\Delta$threshold) and the rest denoting a change in the applied HV. In general, an upward trend is visible. The change in the ROC thresholds resulted in a smaller voltage being necessary.}
  \label{fig:depletion_voltage}
\end{figure*}

Figure~\ref{fig:depletion_voltage} shows the resulting calculated \Vmax as a function of integrated luminosity for several PLT channels. Changes to the operational HV set point and the global ROC thresholds are indicated as vertical dotted lines. Note that, at times, the \Vmax for certain channels approaches the operational HV set point (which must always be higher to maintain sensor efficiency). A decrease in the thresholds should increase the overall amount of signal, thereby requiring a lower applied HV to obtain maximum hit efficiency.

These observations suggest that the per-channel setting of \Vmax must be measured uniformly and regularly during operations. In addition, the thresholds should be closely monitored and adjusted. Work is underway to automate this process for Run 3, as is further discussed in Section~\ref{sec:run3}.

\subsection{Data quality monitoring using machine learning}
\label{sec:dqm}

While operational issues affecting the fast-or luminosity are immediately obvious, it is possible for there to be problems in the pixel data without the fast-or data being affected, thus causing difficulties with use of the pixel data in later analyses. One potential cause is drift of the analog output levels from a ROC; if the pixel FED is not properly recalibrated to account for this change, the pixel data may be incorrectly decoded. This can be easily visualized on an occupancy map; in normal operation, the occupancy of a single ROC should be relatively uniform, increasing slightly towards the edge closest to the beam. However, when these errors occur, some rows or columns will have decreased occupancy, while others will increase correspondingly. While these effects are obvious visually, the large amount of data makes individual inspection of these maps impractical, and so an automated algorithm was developed to detect potential problems in these occupancy maps.

The algorithm uses occupancy maps for each ROC integrated over five-minute intervals, resulting in a total of more than three million maps in the full Run~2 data set. The occupancy maps are then preprocessed to compensate for the average trends, and a set of 31 variables describing the maps is then defined, such as the average and standard deviation of the number of hits per pixel, the standard deviation within and among rows and columns, and the number of pixels with a significantly outlying number of hits. The variables are normalized to remove any dependence on the overall average occupancy. An unsupervised machine learning technique, the $k$-means clustering algorithm~\cite{kmeans}, is then used to divide the occupancy maps into different sets, with one set corresponding to good maps and the other sets corresponding to different types of problems visible in the data.

\begin{figure}[htbp]
\centering
\includegraphics[width=.45\textwidth]{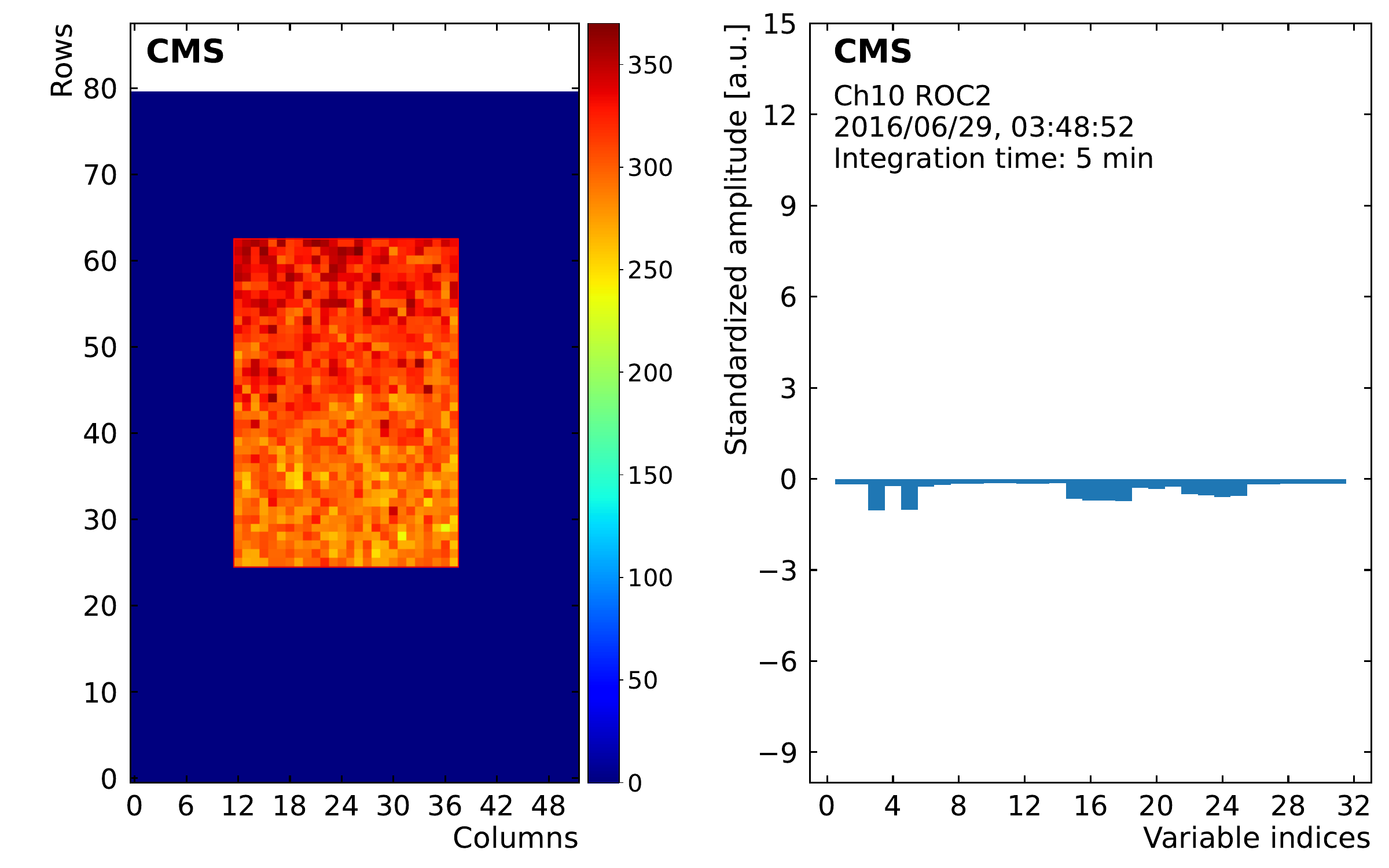}
\includegraphics[width=.45\textwidth]{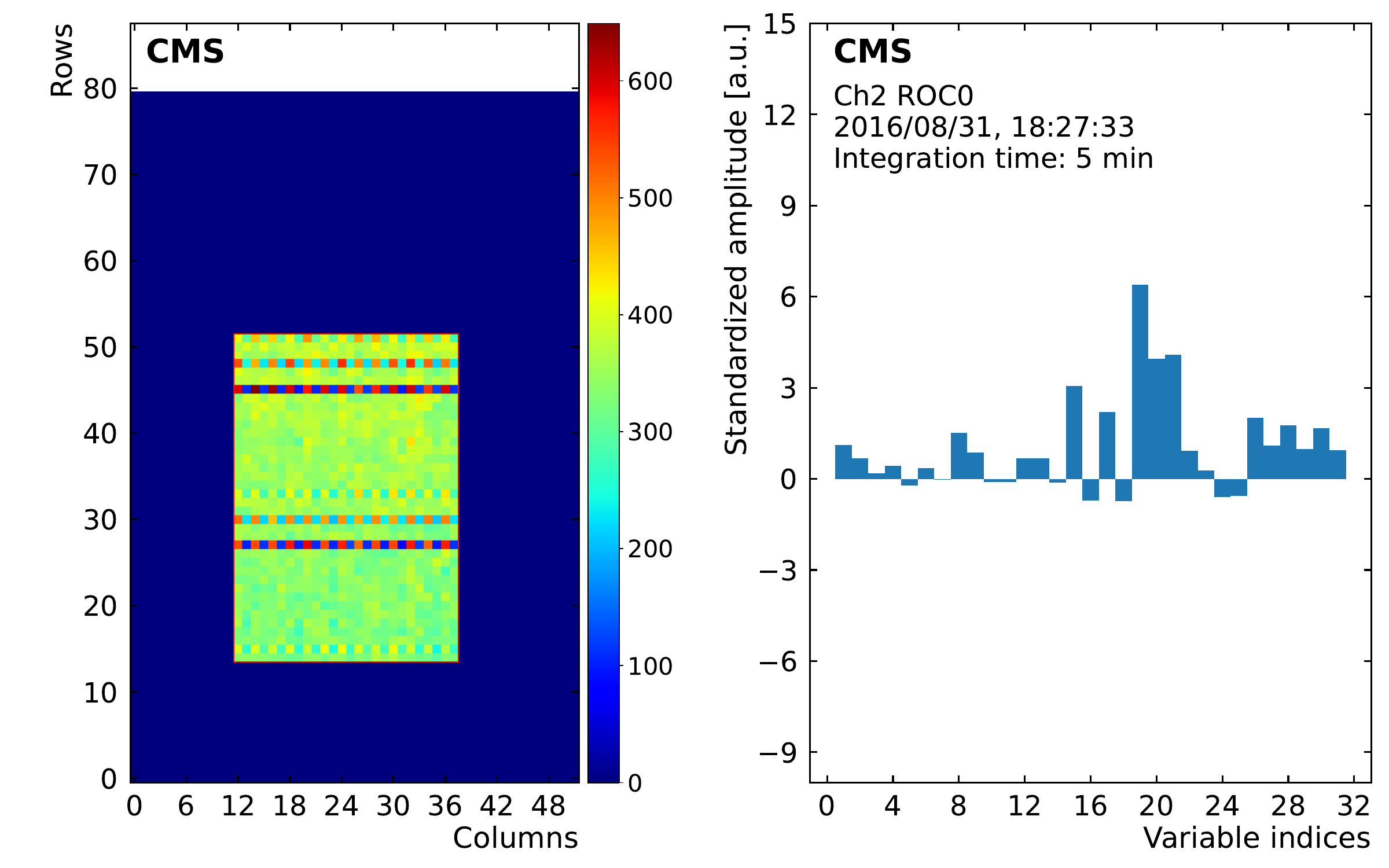}
\caption{The \cmsLeft two plots show an occupancy map for a single ROC during a period of good operation and the corresponding values of the 31 variables used as input to the $k$-means clustering. The \cmsRight two plots show similar plots for a period when the pixel data was not correctly decoded, resulting in line errors in the occupancy plot.}
\label{fig:dqm_sample}
\end{figure}

Figure~\ref{fig:dqm_sample} shows a sample of the occupancy map and the 31 discriminating variables for a period of good operation and a period with the decoding problem described above. When applied to the full Run~2 data set, the $k$-means algorithm successfully identified good maps with a greater than 95\% purity, and divided the bad maps into categories such as one or a few pixels with very low occupancy, row or column errors, and other types of issues. This allows for the possibility to develop this into an automated recovery algorithm for Run~3, allowing the PLT to quickly recover from issues that could significantly affect the data quality; however, issues with only a small effect (such as a single temporarily dead pixel) could be safely ignored.

\subsection{Background measurement with fast-or data}
\label{sec:background}

Measurement of the beam-induced background (BIB) is, along with luminosity measurement, one of the primary responsibilities of the CMS BRIL group. There are several potential sources of BIB, such as interactions of the beam particles with residual gas in the LHC beam pipe, or beam halo particles produced by interactions of off-axis beam particles with the LHC collimators. The BCM1F detector is the primary BRIL detector responsible for BIB measurements; however, in 2016, we investigated the possibility of a background measurement using PLT data as a backup to the BCM1F measurement.

There are two algorithms considered for making a background measurement using the PLT fast-or data. The first relies on the fact that the LHC filling scheme usually includes one or more noncolliding bunches, where a filled bunch is present in one beam but not in the other. In this case, the observed rate in the PLT can be taken to be due to the BIB from the filled beam, since the triple-coincidence rate from non-beam background is negligibly small compared to the BIB rate (as can be observed by looking at BXs far away from any filled bunches). The second method takes advantage of the 1.75\unit{m} distance between the PLT and the IP. This means that BIB from the incoming beam will arrive at the PLT approximately 6\unit{ns} prior to the collision, and thus approximately 12\unit{ns} before the collision products arrive. Thus, it should be possible to observe the BIB rate in the empty BX prior to a colliding bunch train (a ``precolliding'' BX), since the LHC timing places the collisions in the first 2.5\unit{ns} of the BX.

These algorithms were implemented into the PLT processor in October 2016 and the calculated background values published to BRILDAQ and DIP. In preliminary studies, it was found that the two algorithms gave very similar results, although the precolliding BX method has the advantage that it does not require the LHC filling scheme to contain any noncolliding bunches. The agreement of these methods serves to validate the assumptions made in constructing the measurement.

\begin{figure*}[tbhp]
\centering
  \includegraphics[width=0.9\textwidth]{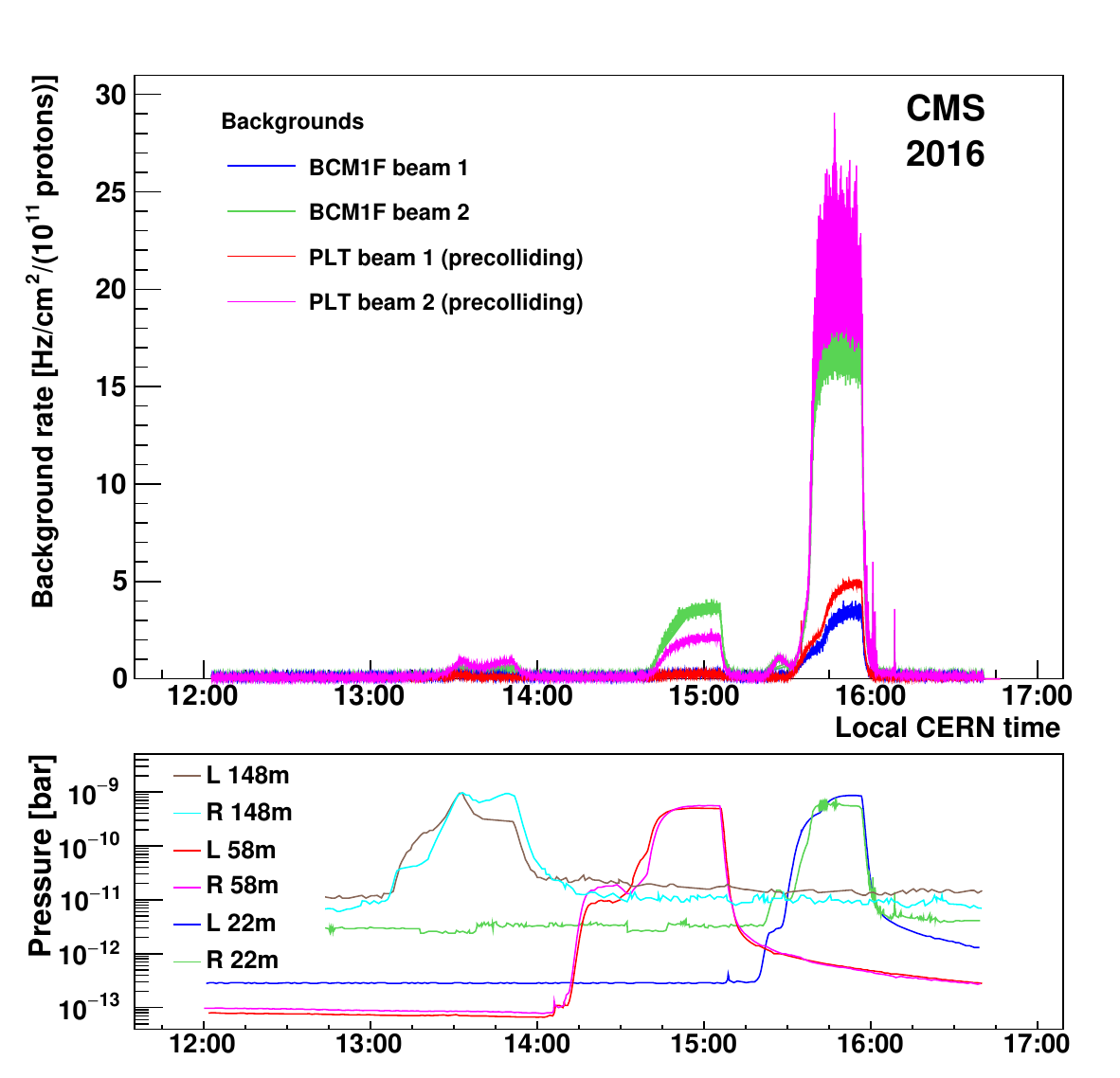}
  \caption{Measured PLT background rate as a function of time, using the precolliding BX method, in beams 1 (red) and 2 (purple) as compared to the BCM1F background in beams 1 (blue) and 2 (green). This study was carried out in fill 5005, a special LHC fill in which background levels were deliberately increased by injecting gas into the beam pipe. The lower panel shows the vacuum pressure, as measured by three pairs of gauges located where the gas was injected, the first pair 148\unit{m} left (L) and right (R) of the CMS interaction point, and the other two pairs 58 and 22\unit{m} on either side.}
  \label{fig:pltbackground}
\end{figure*}

Figure~\ref{fig:pltbackground} shows the measured PLT background, using the precolliding BX method, compared to the BCM1F background in fill 5005, a special LHC fill in 2016. In this fill, the vacuum conditions were intentionally degraded in order to cause increased BIB by injecting gas into the beam pipe at three separate pairs of locations, first 148\unit{m} in both directions away from the CMS interaction point, then 58 and 22\unit{m}. These produce distinct visible spikes in the background rates, with the closer injections having a much larger effect on the rates at CMS. The PLT measurement is normalized to the BCM1F measurement, and we observe good qualitative agreement between the PLT and BCM1F measurements, indicating the general validity of the PLT background measurement method. However, it appears that the PLT beam 1 measurement has a somewhat nonlinear response compared to the BCM1F measurement; this may be due to different timing properties of the PLT and BCM1F. However, since the background measurement is primarily needed to assess whether the beam conditions are safe for CMS operation, precision measurement is not necessary, showing that the PLT background measurement could serve as a viable backup in Run~3 if necessary.

\subsection{Performance in high-pileup conditions}
\label{sec:high_pileup}

In most of Run 2, the typical SBIL at the beginning of a fill was approximately 6.5--8\unit{\hzub}, corresponding to a pileup of approximately 40--50, sometimes going up to as high as 10\unit{\hzub} (pileup 60). However, for machine development studies during Run 2, the LHC had a few fills with significantly higher pileup ($>$100). This gives us an excellent opportunity to study the linearity behavior of the PLT at very high instantaneous luminosities, especially since these will be more common in Run 3 of the LHC.

This analysis uses data from the special high-pileup fill 7358, which was recorded by CMS at the end of the 2018 $\Pp\Pp$ run. The fill featured two bunch trains, each with 10 colliding bunches, as well as two isolated colliding bunches. The average pileup at the beginning of the fill was approximately 100, and BX number 1648 had the highest individual pileup at $\approx$130. The fill also featured a $\mu$ scan, which scanned the range from the maximum to a pileup of $\approx$30. For comparison, a more typical physics fill with a maximum pileup of $\approx$50, fill 6854, is used as a reference.

\begin{figure*}[ht!]
\centering
\includegraphics[width=0.32\textwidth]{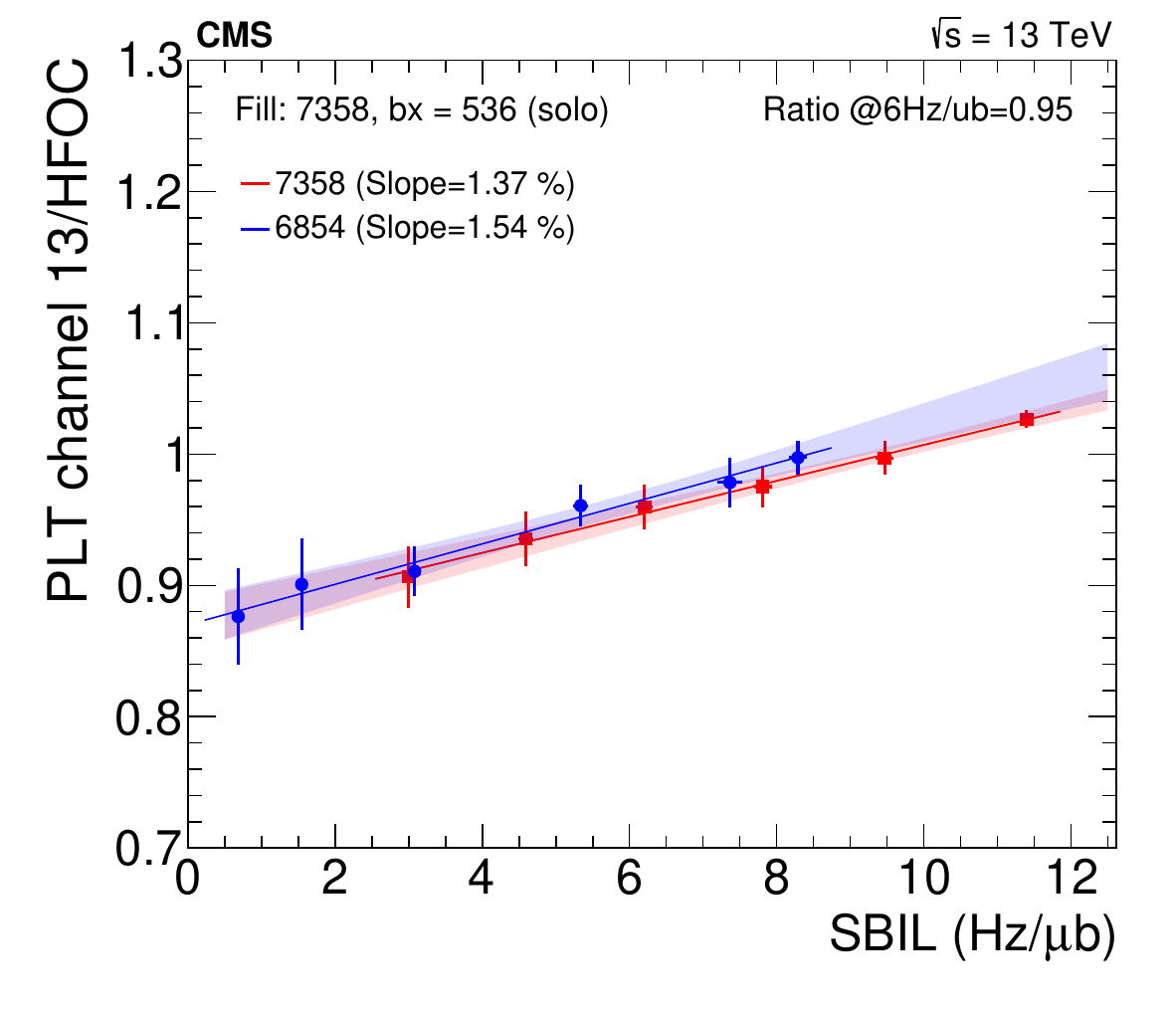}
\includegraphics[width=0.32\textwidth]{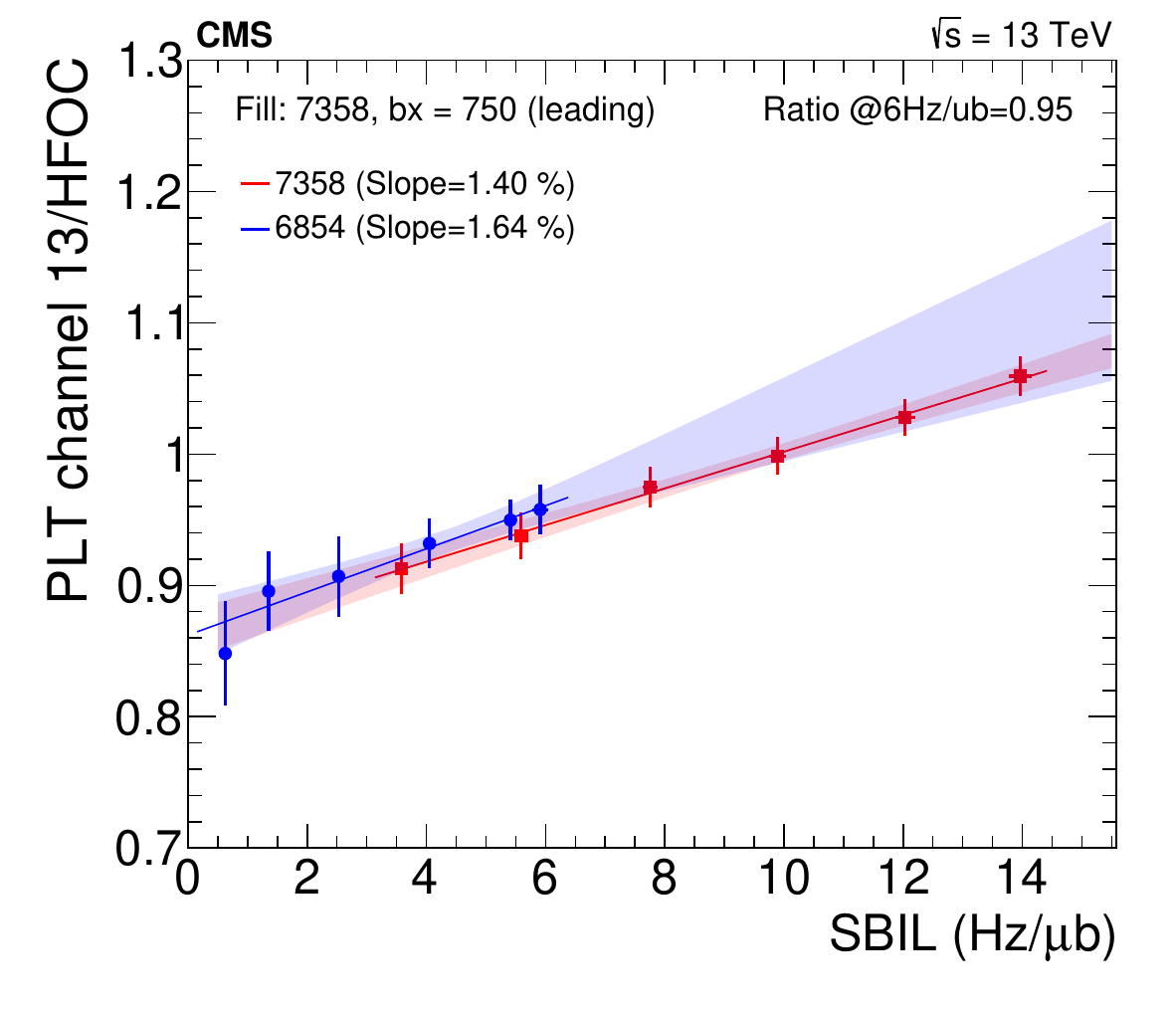}
\includegraphics[width=0.32\textwidth]{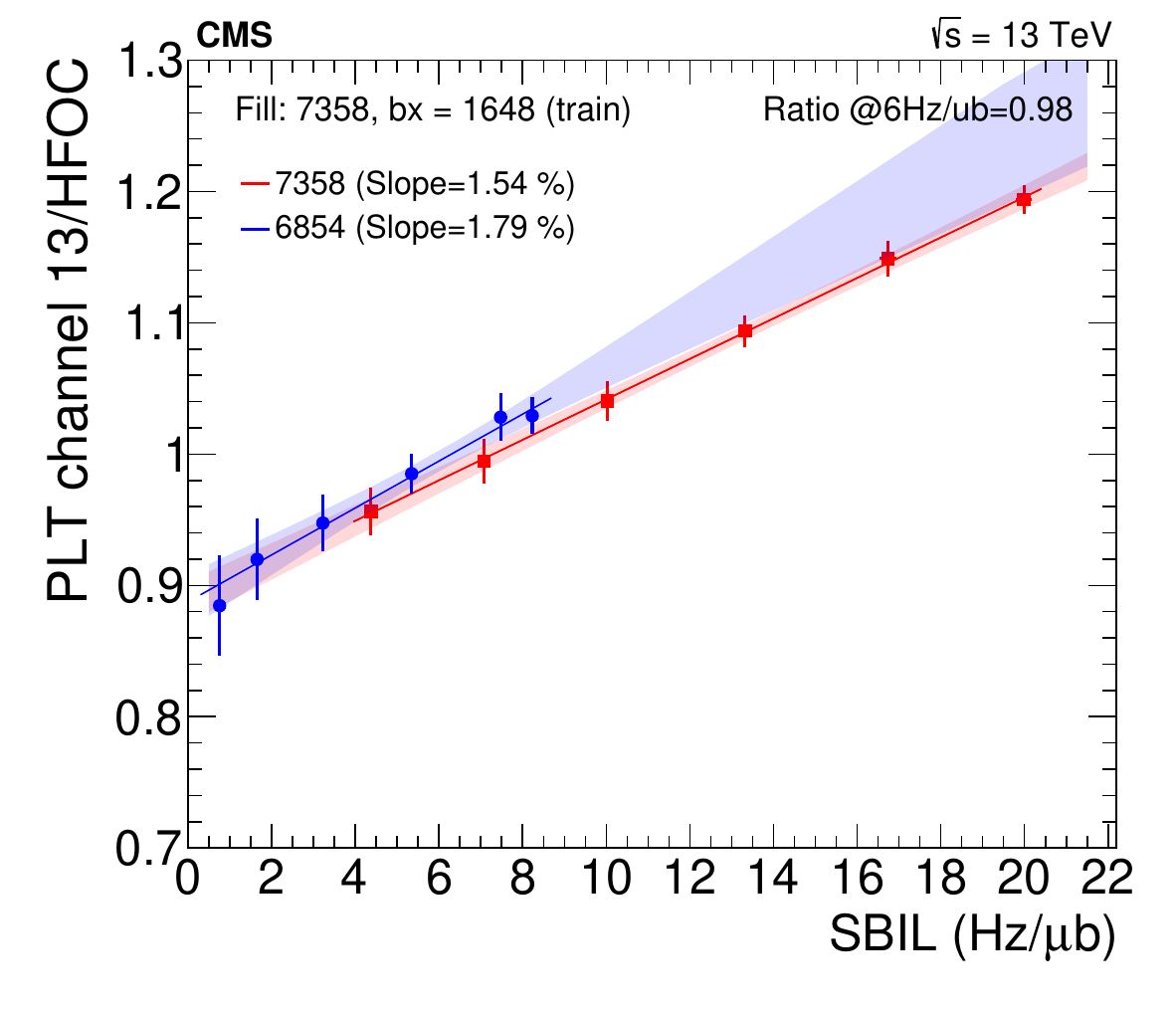}\\
\caption{Ratios of PLT to HFOC instantaneous luminosity, as a function of SBIL, in the high-pileup fill 7358 (red line) and the reference fill 6854 (blue line), for a single PLT channel (channel 13). The left plot shows a single isolated bunch (BCID 536 in fill 7358 and BCID 823 in fill 6854), the center plot shows a leading bunch in a bunch train (BCID 750 in fill 7358 and BCID 62 in fill 6854), and the right plot shows the train bunch with the highest luminosity (BCID 1648 in fill 7358 and BCID 63 in fill 6854). The shaded bands indicate the uncertainty in the linear fit for each fill.}
\label{fig:highpu_slopes}
\end{figure*}

Figure~\ref{fig:highpu_slopes} shows the ratio of the luminosity from a single PLT channel (channel 13) to the HFOC luminosity, which is used as a reference luminometer. The PLT luminosity is measured as described in Section~\ref{sec:lumi}, but no nonlinearity corrections are applied for this study, while the HFOC measurement includes all corrections described in Ref.~\cite{CMS-PAS-LUM-18-002}. The ratio is shown for three different types of colliding bunches: a single isolated bunch, a leading bunch in a bunch train, and the bunch with the highest instantaneous luminosity (which is inside a bunch train). The trends are fit with a linear function, with the 1$\sigma$ uncertainty in the fit shown as a shaded band. In general, the trends observed in the standard fill agree well with the data in high-pileup conditions. Some other PLT channels show a more pronounced difference, possibly due to changes in efficiency between the reference fill and the high-pileup fill.

\begin{figure}[ht!]
\centering
\includegraphics[width=\cmsFigureWidthSmaller]{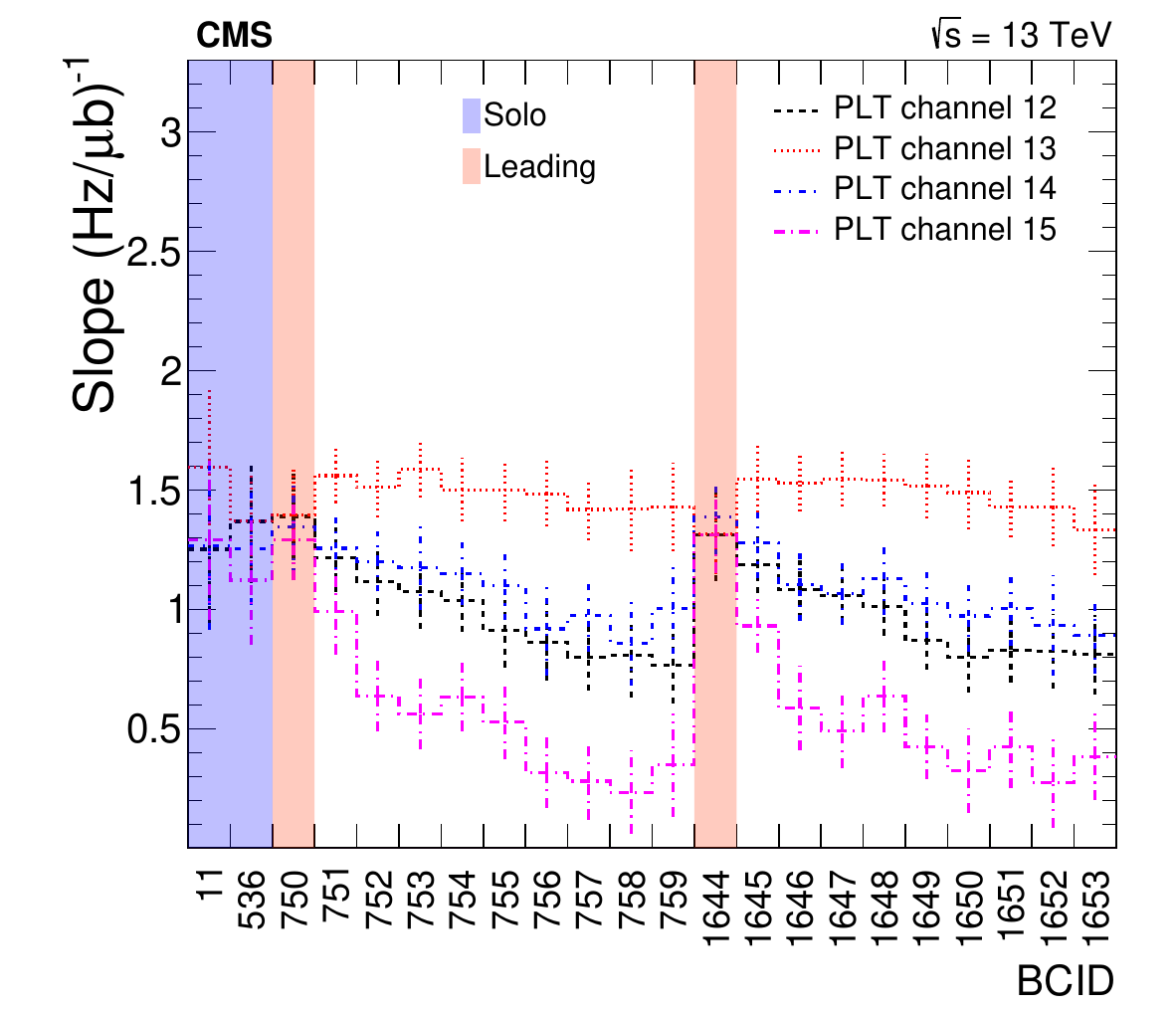}
\caption{Measured slope of the PLT/HFOC ratio as a function of BCID for isolated bunches (the two bunches at the left in the blue background), leading bunches (the two bunches on a light red background), and train bunches (other bunches) for PLT channels 12--15 in the high-pileup fill 7358.}
\label{fig:bcid_dependence}
\end{figure}

Figure~\ref{fig:bcid_dependence} summarizes the fitted slopes for all the colliding bunches in fill 7358 for the four channels in the $+z$ far quadrant (channels 12--15). The isolated bunches are shown by the blue highlight, and the leading bunches by the light red highlight. We can observe a similar pattern in the two bunch trains, with the slope somewhat different for the leading bunch and then gradually decreasing over the length of the bunch train. These train effects, which can also be seen in the emittance scan analysis discussed in Section~\ref{sec:emittance}, are most likely due to dynamic inefficiency in the ROC (where a hit in one BX causes a slightly decreased probability of registering a hit in the next BX).

Overall the results give us confidence that the PLT can still be used even in conditions with very high pileup, although it will be important to understand the linearity of the PLT well in order to minimize systematic uncertainties. The results in Fig.~\ref{fig:bcid_dependence} also illustrate the need for channel-by-channel linearity corrections for the PLT, as discussed in Section~\ref{sec:lin_eff}.

\subsection{Luminous region reconstruction}
\label{sec:beamspot}

By extrapolating the tracks measured in the PLT to the CMS interaction point, the position of the luminous region (``beamspot'') can be estimated. The beamspot position along the beam ($z$) axis is obtained with a least-square fit of a straight line to the locations of three clusters in the three planes of a PLT telescope, given in local telescope ($x$, $z$) and ($y$, $z$) coordinates. They include corrections for the alignment of the planes within the telescopes, as described in Section~\ref{sec:alignment}.

The positions are translated to global coordinates by applying additional global alignment corrections to the telescope positions. The global alignment of the telescopes with respect to each other was measured using a sample of events with tracks in both ends ($-z$ and $+z$) of the PLT. First, we locate the point on the $z$ axis where the average track $x$ and $y$ coordinates are minimized; the global $z$ position of the telescope is defined by aligning this point to $z=0$. Since during the 2016 run period the goal was to monitor the relative behavior of the beamspot, and because this closest-approach method is necessarily an approximation, this measurement is not necessarily comparable to the 3D measurement from the CMS tracker. This analysis is primarily a proof of concept to illustrate possibilities for future measurements with the PLT in Run 3.

\begin{figure}[htbp]
\centering
\includegraphics[width=.48\textwidth]{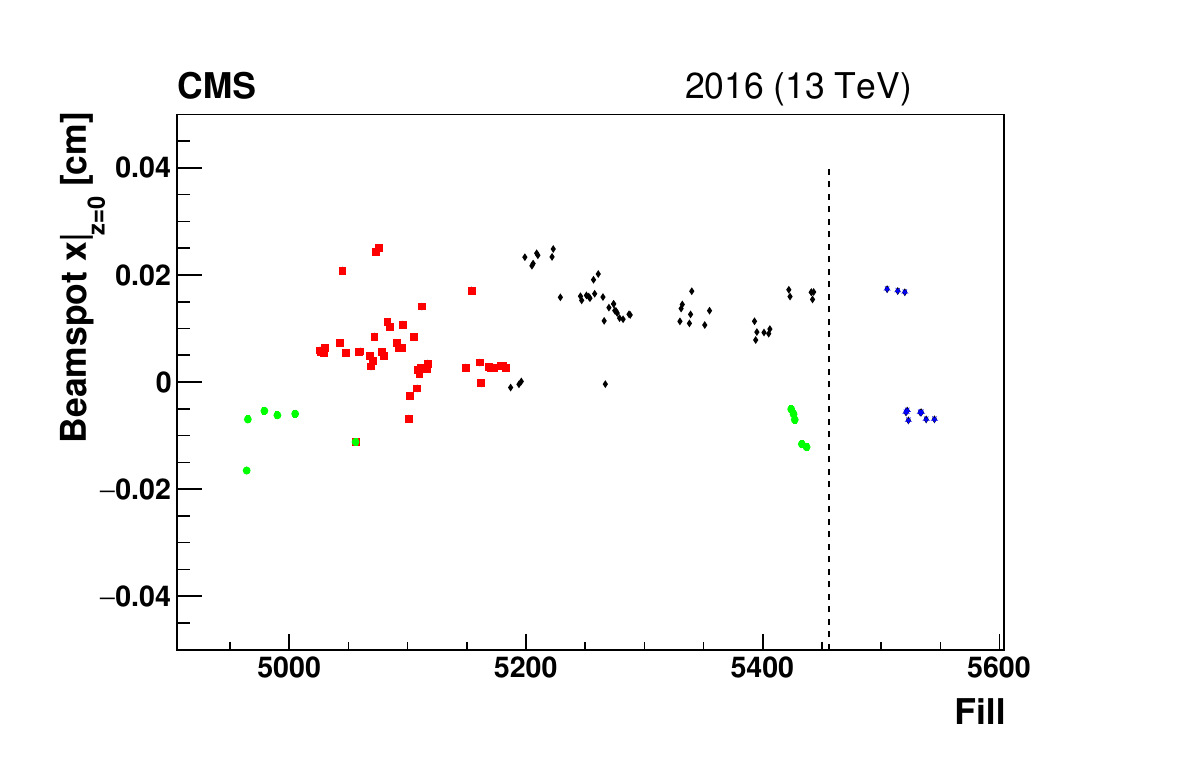}
\includegraphics[width=.48\textwidth]{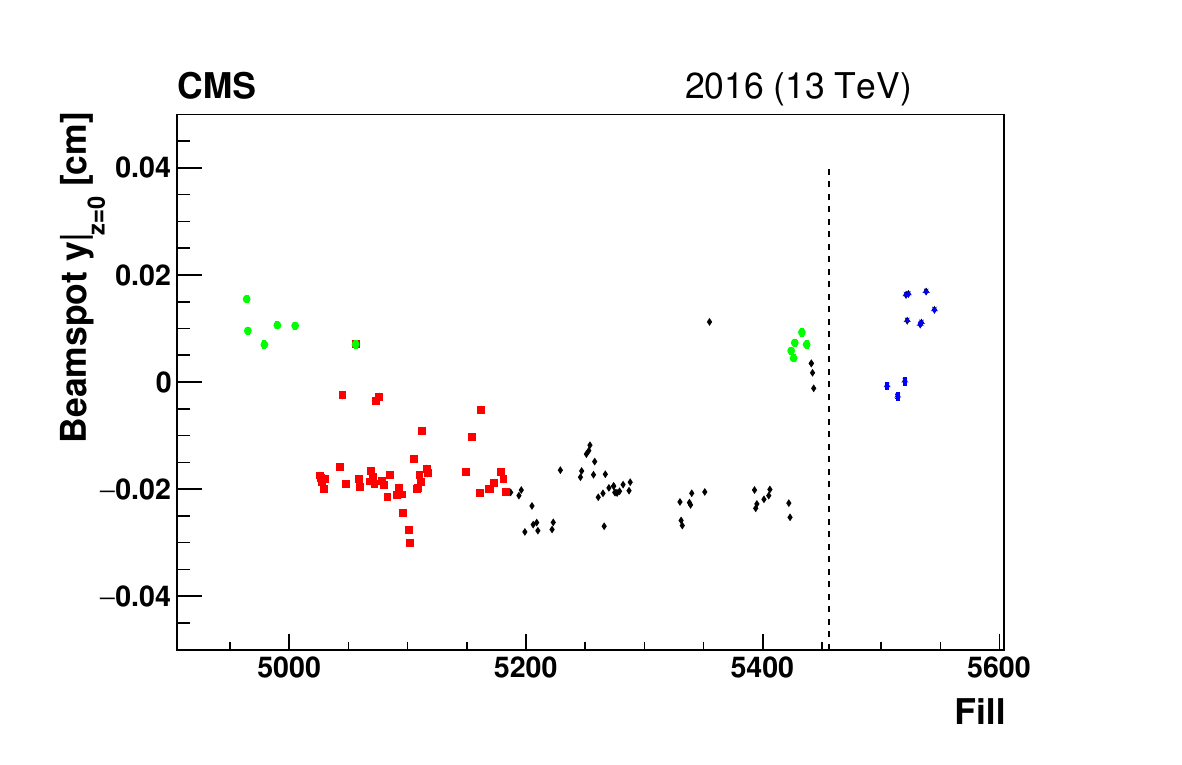}
\caption{The position of the beamspot mean in global $x$ (\cmsLeft) and $y$ (\cmsRight) coordinates vs. fill numbers. The coordinates are estimated from the straight line fits in the $x$-$z$ and $y$-$z$ projections when extrapolated to $z=0$. The distributions of the coordinates are fit to double Gaussian functions. The dashed black line indicates the start of the heavy ion run. The different marker colors and shapes indicate groups of fills for which the beamspot position is relatively constant.}
\label{fig:BSevolution} 
\end{figure}

Figure~\ref{fig:BSevolution} shows the global beamspot position in $x$ and $y$ coordinates vs. fill numbers over the course of 2016. For each fill, the first 30 minutes of data taking are skipped, and then tracks with exactly three clusters (one in each plane) are accumulated for the following 5 minutes of run time. The distributions in the $x$ and $y$ at the $z=0$ position are each fit to a double Gaussian function with a common mean. The fit is an unbinned maximum likelihood fit, performed with RooFit~\cite{Verkerke:2003ir}. In addition to the mean, the standard deviations of the two components and the relative contribution of the two components are varied. The vertical black line indicates the start of the heavy-ion run.

\begin{figure}[htbp]
\centering
\includegraphics[width=\cmsFigureWidth]{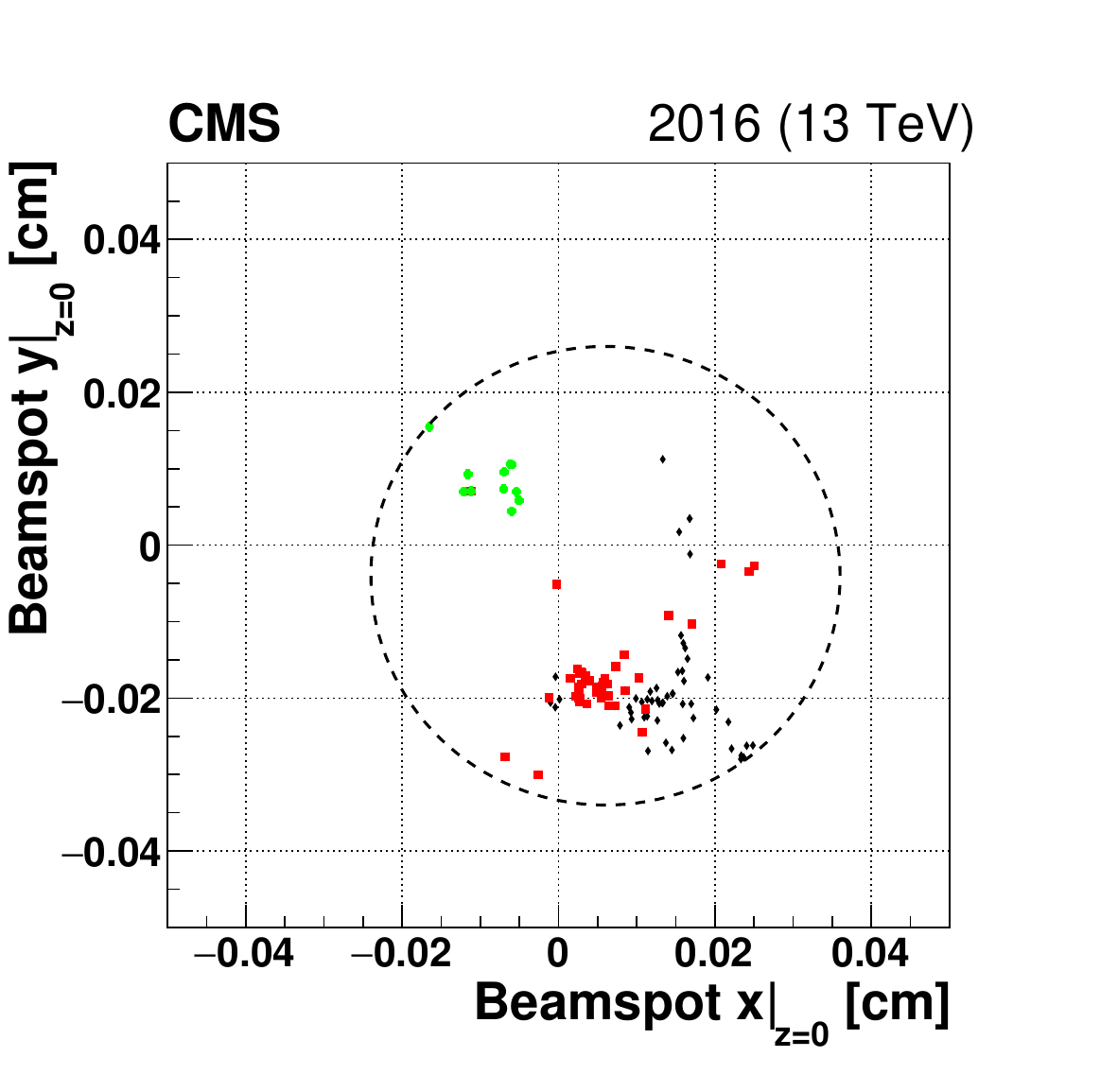}
\caption{The position of the mean beamspot in global $x$ and $y$ coordinates. The red squares indicate fills from the period of stable beamspot position, shown by the red squares in Fig.~\ref{fig:BSevolution}. The green dots indicate a secondary position that is offset from the red cluster of positions by about 150\mum in $x$ and 300\mum in $y$. Theses fills occur at the beginning and end of the 2016 $\Pp\Pp$ run period. The black diamonds correspond to other fills. The dashed circle represents the overall range of beamspot position during 2016. It is centered at $x=60\mum$ and $y=-40\mum$ and has a radius of 300\mum.}
\label{fig:BeamSpot2D}
\end{figure}

Figure~\ref{fig:BeamSpot2D} shows the position of the beamspot in the $x$-$y$ plane for each $\Pp\Pp$ fill (heavy-ion fills are excluded). The points appear in three separate groups, which correspond to different time periods during the 2016 run, shown in the same colors as in Fig.~\ref{fig:BSevolution}. The red data points are from fills in the first half of the year, and the mean $x$ positions in this cluster are well described by a Gaussian function with a width of 43\mum. The Gaussian function fit to the mean $y$ positions has a width of 66\mum; these widths give an estimate of the precision of the PLT measurement. The green points indicate the cluster of positions originating from fills at the beginning and the end of the $\Pp\Pp$ collision run. The measured beamspot positions for all $\Pp\Pp$ fills remain within a circle of radius 300\mum. At fill number 5209, after the red period, the reconstructed beamspot moves by about 0.02\unit{cm} in the positive $x$ direction, and then gradually moves back towards $x=0$. This corresponds approximately to an LHC technical stop and an increase in the number of colliding bunches to 2220.

These results show the potential for measuring the beamspot using PLT data. This both provides an intrinsic validation of the PLT alignment, and a future opportunity to compare with the tracker measurements to further improve the precision of the PLT position measurement.

\section{Luminosity measurement with the PLT}
\label{sec:lumi}

For any physics process, the rate $R$ at which the process occurs is related to the instantaneous luminosity \Linst via the fundamental relation~\cite{Sirunyan:2021qkt}
\begin{linenomath}
\begin{equation}
    R = \Linst \sigma,
\end{equation}
\end{linenomath}
where $\sigma$ is the cross section of the process in question. For a luminometer that operates by measuring the rate $R$ of a certain quantity of interest (hits, tracks, etc.), we can write:
\begin{linenomath}
\begin{equation}
    R = \Linst \sigmavis,
\label{eq:sigmavis}
\end{equation}
\end{linenomath}
where the calibration constant \sigmavis, the ``visible cross section'', is determined by the particular properties of the luminometer, such as its acceptance.

For the PLT, the principal luminosity measurement is provided by the triple coincidence rate measured using the fast-or data, and is measured bunch by bunch. If $R_i$ is the per-bunch rate of triple coincidences, we can write $R_i = \mu_i \frev$, where $\mu_i$ is the average number of triple coincidences per bunch and \frev is the LHC revolution frequency of 11.246\unit{kHz}.

To determine the value of $\mu_i$, the simplest way is to count the average number of triple coincidences per telescope per bunch. However, this method introduces systematic effects due to limitations of the fast-or readout (specifically, that multiple hits in the same double column are not counted as separate hits, and more than three hits overall are not counted). Instead, to avoid these effects, we employ a ``zero-counting'' technique. In this method, we consider collisions where no triple coincidence is observed in a given telescope (although one or two planes may be hit). If the fraction of such events is given by $f_0$, then the mean number of triple coincidences per collision $\mu$ for that telescope is given by $\mu = -\ln f_0$, assuming a Poisson distribution of the number of triple coincidences (since the Poisson probability of observing 0 events is simply $e^{-\mu}$). The main potential drawback of the zero-counting method is the ``zero starvation'' effect, when $f_0$ is so low that the uncertainties become extremely large. However, the typical PLT occupancy is on the order of 0.1--0.2 triple coincidences per telescope per colliding bunch at nominal physics luminosities for $\Pp\Pp$ running, so this is normally not a concern. The $\mu$ values are averaged over all telescopes to obtain an overall occupancy.

The determination of \sigmavis is performed using the VdM calibration procedure described below. Once \sigmavis is obtained, the per-bunch luminosity \Linsti (SBIL) can be obtained using Eq.~(\ref{eq:sigmavis}). In an ideal luminometer, this linear relation holds perfectly. In practice, however, we must apply two corrections. The first accounts for the potential loss of efficiency over time from effects such as radiation damage, and is applied to correct the measured rate, thus modifying our equation for \Linst as a function of $\mu$:
\begin{linenomath}
\begin{equation}
    \Linsti = \frac{\frev \mu_i}{\sigmavis \varepsilon},
\label{eq:sigmavis_effcorr}
\end{equation}
\end{linenomath}
where $\varepsilon$ represents the time-varying efficiency. The second accounts for potential nonlinear effects. Taking the above equation and defining $k = \frev/(\sigmavis \varepsilon)$, we can write:
\begin{linenomath}
\begin{equation}
    \Linsti = k \mu_i(1-ak\mu_i),
\label{eq:sigmavis_lincorr}
\end{equation}
\end{linenomath}
where $a$ represents the nonlinearity as a function of the (linearity uncorrected) instantaneous luminosity $k\mu_i$, typically expressed in units of \%/(\hzub). The $a$ term may also vary over time if, for example, the radiation damage affects different sensors at different rates. The procedure for deriving these $\varepsilon$ and $a$ terms is described in Section~\ref{sec:lin_eff}; since these are not necessarily the same across all channels, they are applied on a per-channel basis.

\subsection{The VdM scan method}
\label{sec:VdM}

The Van der Meer method, first developed by Simon van der Meer for luminosity measurement at the CERN Intersecting Storage Rings~\cite{vanderMeer:296752}, uses beam-separation scans (``VdM scans'') in special LHC fills to estimate the transverse size of the beam overlap region from the measured rate as a function of the beam separation. This allows us to calculate the absolute luminosity and, in conjunction with Eq.~(\ref{eq:sigmavis}), to determine \sigmavis for a given luminometer, which is then used for luminosity determination during normal physics operation.

The formula for the instantaneous luminosity for a single colliding bunch $i$, \Linsti, as a function of beam parameters is given by the following, in the case where there is no crossing angle between the beams and the beams are not separated:
\begin{linenomath}
\ifthenelse{\boolean{cms@external}}{\begin{multline}
\label{eq:beamintegral}
\Linsti = N_1^i N_2^i \frev \int{\rho^i_1(x,y)\rho^i_2(x,y)\rd x\rd y} \\
= N_1^i N_2^i \frev \int{\rho^i_{x1}(x)\rho^i_{x2}(x)\rd x}\int{\rho^i_{y1}(y)\rho^i_{y2}(y)\rd y},
\end{multline}}{\begin{equation}
\label{eq:beamintegral}
\Linsti = N_1^i N_2^i \frev \int{\rho^i_1(x,y)\rho^i_2(x,y)\rd x\rd y} =
N_1^i N_2^i \frev \int{\rho^i_{x1}(x)\rho^i_{x2}(x)\rd x}\int{\rho^i_{y1}(y)\rho^i_{y2}(y)\rd y},
\end{equation}}
\end{linenomath}
where $N_1^i$ and $N_2^i$ are the number of protons or ions in the two individual beams for the colliding bunch $i$ and $\rho^i_j$ is the normalized particle density for the bunch in beam $j$. The rightmost term of Eq.~(\ref{eq:beamintegral}) uses the assumption that $\rho^i_j$ can be factorized into independent terms in $x$ and $y$, $\rho_x(x)$ and $\rho_y(y)$, respectively.

The beam currents $N^i_j$ can be measured with high precision, but the individual bunch density functions $\rho^i_j$ cannot generally be directly measured. The VdM method determines the value of the two beam overlap integrals in Eq.~(\ref{eq:beamintegral}) by conducting a scan in which the beam separation is systematically varied and the resulting rates are measured:
\begin{linenomath}
\begin{equation}
\int{\rho_{x1}(x)\rho_{x2}(x)\rd x} = \frac{R_x(0)}{\int{R_x(\Delta)\rd\Delta}},
\end{equation}
\end{linenomath}
where $R_x(\Delta)$ is the rate measured when the two beams are separated in $x$ by a distance $\Delta$; a similar equation can be written in $y$. We then define the beam overlap width $\Sigma_x$ (and similarly $\Sigma_y$) as:
\begin{linenomath}
\begin{equation}
\label{eq:capsigma}
\Sigma_x = \frac{1}{\sqrt{2\pi}}\frac{\int{R_x(\Delta)\rd\Delta}}{R_x(0)},
\end{equation}
\end{linenomath}
yielding the final expression for luminosity:
\begin{linenomath}
\begin{equation}
\label{eq:VdMeq}
\Linsti = \frac{N_1^i N_2^i \frev}{2\pi\Sigma_x\Sigma_y}.
\end{equation}
\end{linenomath}
In practice, two separate scans in the $x$ and $y$ directions are performed to evaluate the integral in Eq.~(\ref{eq:capsigma}) in each direction. In each scan, the rate is measured (normalized by the product of the beam currents) at a certain number of separation steps, the resulting points are fit with a functional form, and the fitted function is used to determine the overall integral. Once the beam overlap widths $\Sigma_x$ and $\Sigma_y$ are determined, Eq.~(\ref{eq:sigmavis}) can
be used to obtain the overall visible cross section \sigmavis.

\subsection{Procedure for VdM scans}

The VdM scans are typically carried out under special conditions in order to maximize the precision of the VdM measurement. The luminosity of a single colliding bunch pair is significantly smaller than in regular data-taking conditions (approximately 0.05--0.1\unit{\hzub}, corresponding to a pileup of about 0.4--0.8), both to minimize any nonlinear effects in the detector and to increase the stability of the luminosity over the course of the calibration fill. The bunch size is also increased, to allow for more precise measurement of the beam overlap width. No crossing angle between the beams is used, and the number of colliding bunches is significantly reduced to ensure that each colliding bunch is well separated from any other colliding bunches. This reduces effects from long-range beam-beam interactions, as well as from detector ``afterglow'', where the signal from a single colliding bunch produces a signal in one or more following BXs.

In a typical VdM scan at the LHC, the beams are symmetrically separated from each other by a distance of $6\sigmab \approx 600\mum$ in a single plane, where \sigmab is the transverse width of each individual beam (as measured by the LHC beam monitoring systems). The separation is then varied in a sequence of 25 steps, with 30 seconds per step, until they are separated by $6\sigmab$ in the opposite direction. For studies of systematic effects, other specialized scans are also conducted with different scan procedures.

Generally, one VdM calibration fill is conducted per year during normal proton-proton ($\Pp\Pp$) running. More information about these scans and their analysis for Run 2 can be found in Refs.~\cite{Sirunyan:2021qkt,CMS-PAS-LUM-17-004,CMS-PAS-LUM-18-002}. In addition, separate calibration runs are necessary for special runs with lower energy~\cite{CMS-PAS-LUM-16-001,CMS-PAS-LUM-19-001}, with collisions using lead ions (PbPb)~\cite{CMS-PAS-LUM-18-001}, or collisions between protons and lead ions (pPb)~\cite{CMS-PAS-LUM-17-002}, as the \sigmavis for these runs will be different due to the different physics processes.

The procedure used to fit the VdM scan curve of the PLT rate as a function of separation was adjusted over time. For the 2015--17 data, the curves were fitted with the sum of two Gaussian terms with a common mean (``double Gaussian'' function). For the 2018 data, the fit quality was found to be best with a single Gaussian instead. For the 2015 and 2016 scans, the background during the VdM scan was found to be negligible and so no correction was applied. For the 2017 and 2018 scans, an independent estimation of the background was performed and this background estimate was subtracted prior to fitting. For the 2017 scans, this estimate was performed by using the measured rate in BXs in the abort gap, which are guaranteed to be empty, to determine the contribution from detector noise, and the measured rate in noncolliding BXs to determine the contribution from BIB. For the 2018 data, this estimate was performed by using a special ``super-separation'' scan, in which the two beams were separated by $6\sigmab$ along both axes, so that the contribution from collisions is negligible and the resulting rate should be due solely to background. 

\begin{figure*}[tbhp]
\centering
  \includegraphics[width=0.45\textwidth]{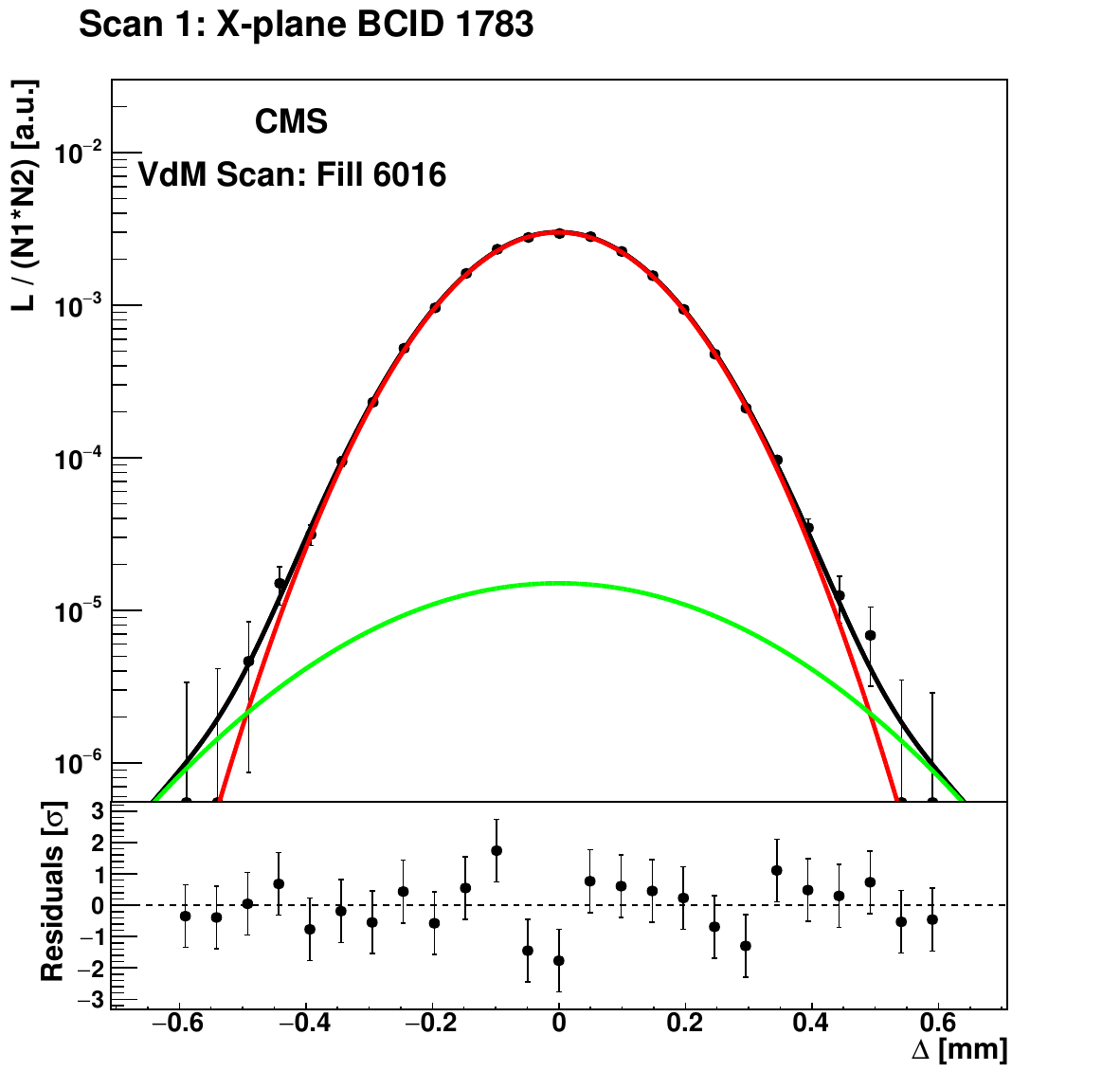}
  \includegraphics[width=0.45\textwidth]{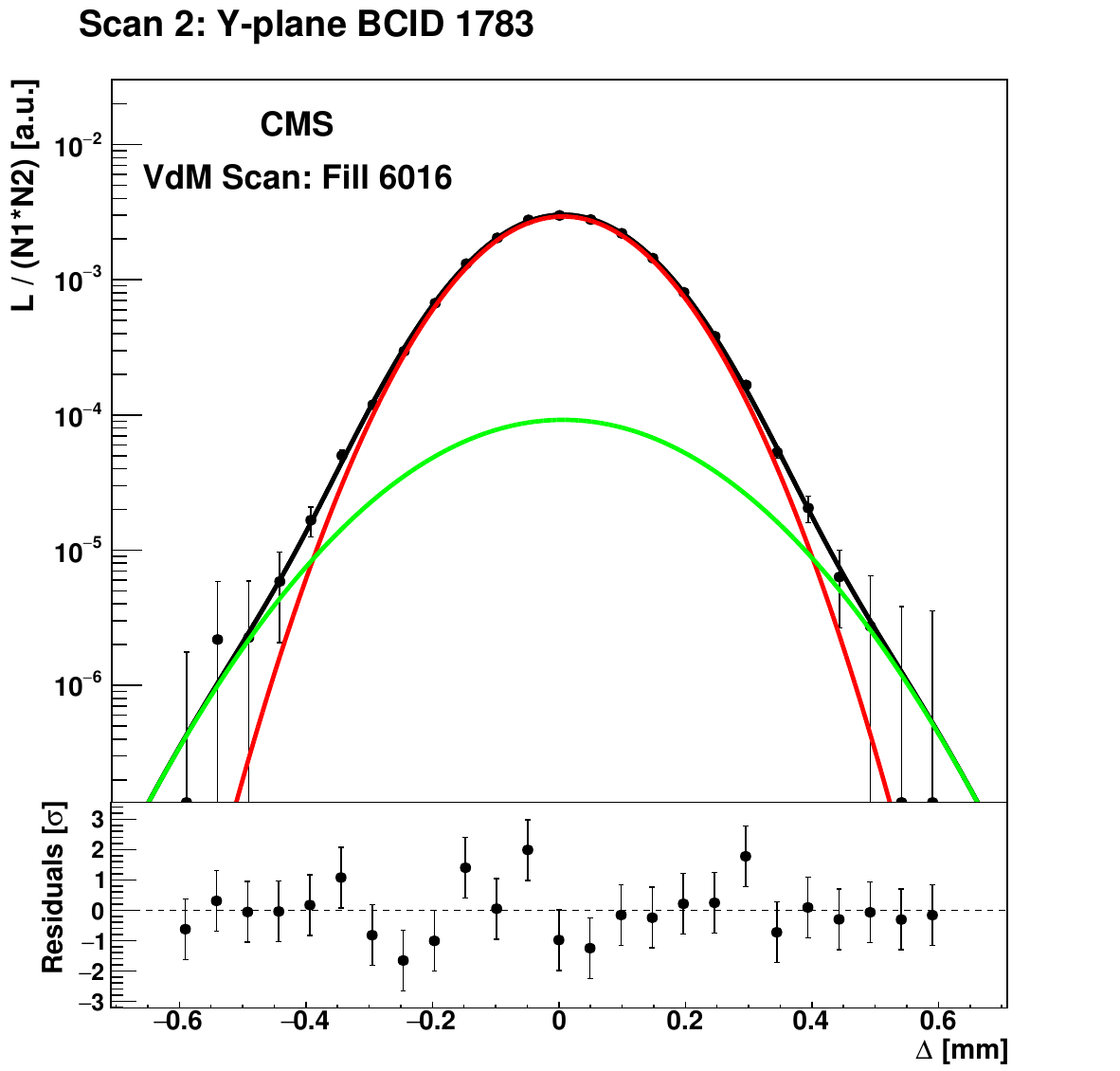}
  \includegraphics[width=0.45\textwidth]{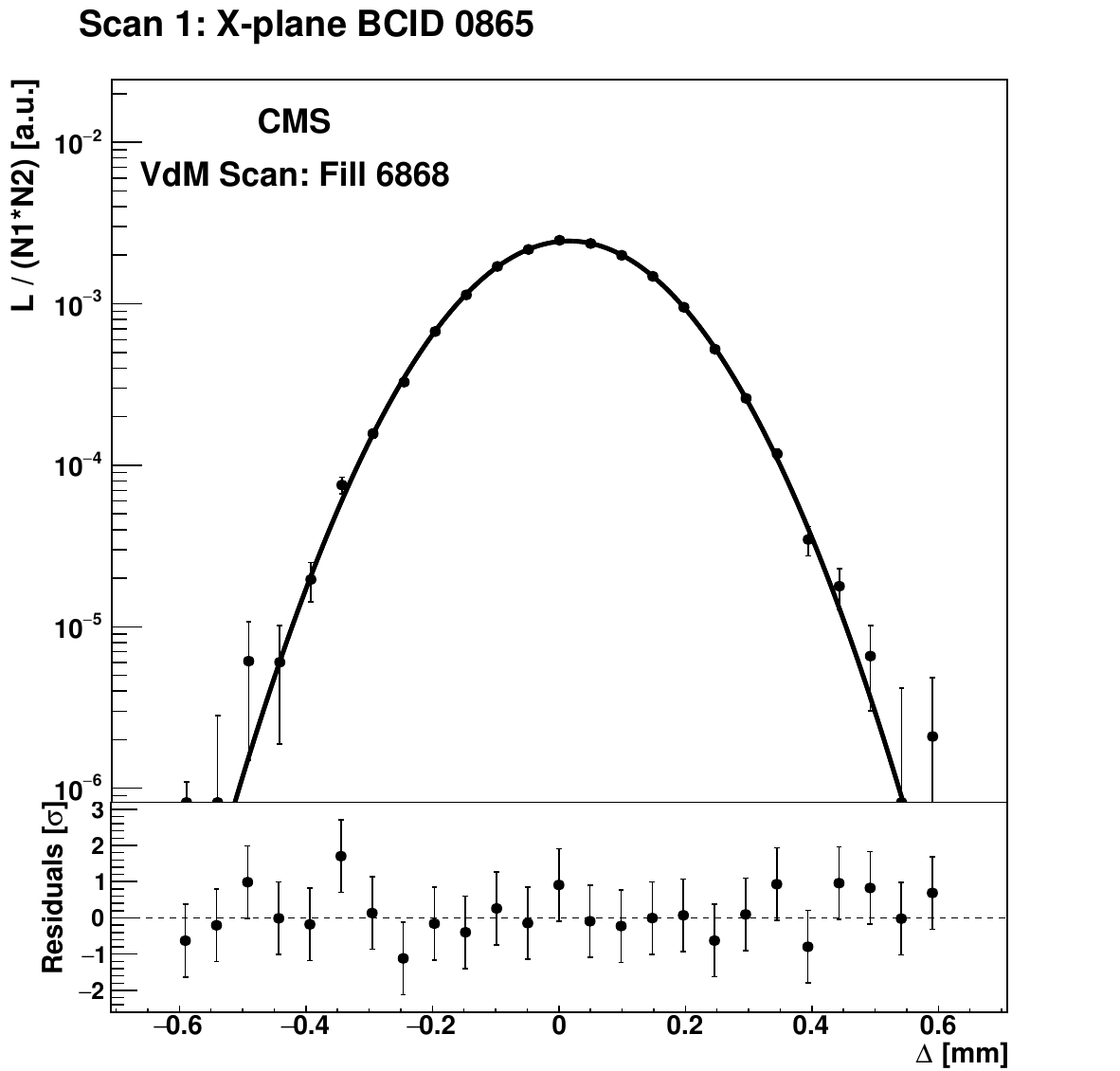}
  \includegraphics[width=0.45\textwidth]{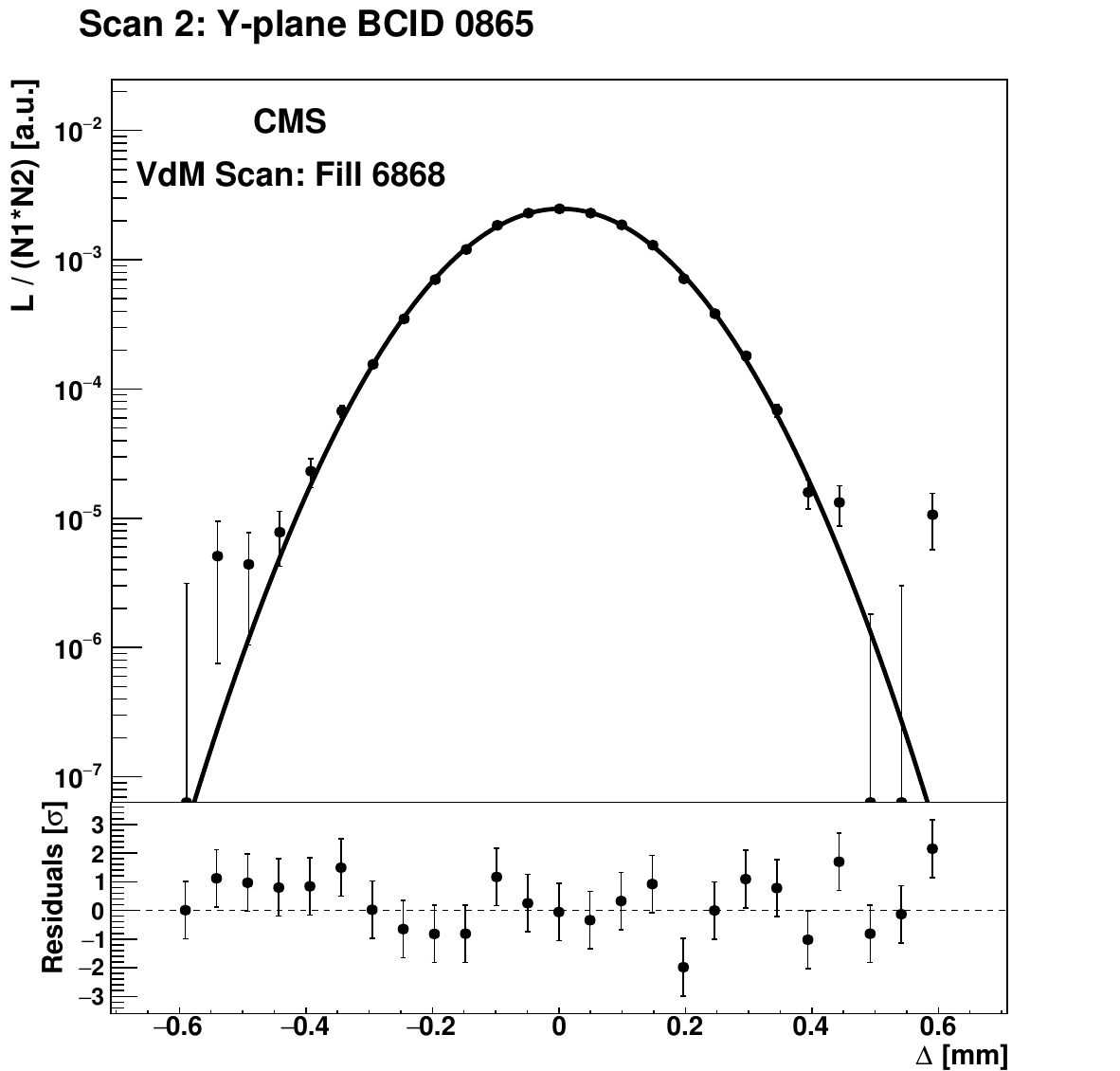}
  \caption{Normalized PLT rates (dots) and the resulting fitted Gaussian  scan curves (black curves) as a function of the beam separation ($\Delta$) for a single colliding bunch, for scans in the $x$ (left column) and $y$ (right column) direction. The top row shows results from a scan pair in the 2017 VdM program in LHC fill 6016~\cite{CMS-PAS-LUM-17-004}, using a double Gaussian fit (the two individual components are shown by the red and green curves), and the bottom row shows results using a scan pair in the 2018 VdM program in fill 6868~\cite{CMS-PAS-LUM-18-002}, using a single Gaussian fit. The background subtraction procedure described in the text has been applied to the raw data before the fit. The lower panel in each plot shows the residual difference between the fit and data, in units of the uncertainty $\sigma$. The statistical uncertainty in the $\Sigma$ values from the fit is 0.4--0.5\%.}
  \label{fig:VdMfits}
\end{figure*}

Figure~\ref{fig:VdMfits} shows some sample fits of the VdM scan curve for a single colliding bunch using the PLT data for 2017 and 2018, showing the double Gaussian fit used in 2017 and the single Gaussian fit used in 2018. While \sigmavis is an intrinsic property of the detector and thus is different for the different luminometers, the beam overlap width $\Sigma$ is a property of the beam and thus should be consistent across the different luminometers. In general good agreement is observed~\cite{Sirunyan:2021qkt,CMS-PAS-LUM-17-004,CMS-PAS-LUM-18-002}, indicating the validity of the VdM scan method. Due to the various operational changes in the PLT between years, comparison of the PLT \sigmavis values across years is generally not possible.

Several corrections must be applied to account for systematic effects in the VdM scan procedure. These include the length scale calibration, to account for the uncertainty in the actual beam separation when the LHC magnets are adjusted to produce a given beam separation; orbit drift, to account for movement of the beam from its nominal orbit position during the scan; $x$-$y$ nonfactorization, which arises from the fact that the VdM scan procedure assumes that the proton density functions can be separated into independent $x$ and $y$ terms, which does not hold perfectly in real conditions; beam-beam effects, which account for the fact that electromagnetic interactions between the two beams can result in both additional deflection of the beams and a change in the beam shape; and systematic effects in the beam current measurement. A full description of these corrections is beyond the scope of this paper, but can be found in Ref.~\cite{Sirunyan:2021qkt}.

\subsection{Stability and linearity tracking with emittance scans}
\label{sec:emittance}

During regular physics fills, the operating conditions (number of colliding bunches, beam intensity, bunch size, etc.) are significantly different from those in VdM calibration fills. Nevertheless, it is possible to perform VdM-like beam scans during normal physics fills. These scans, referred to as ``emittance scans'', were regularly performed by the LHC operators throughout Run 2, and in 2017, BRIL developed a program to automatically analyze these as VdM scans~\cite{Karacheban:2019ypt}. Emittance scans follow the same basic procedure as a regular VdM scan, but in order to minimize the loss of physics data, they are performed in a much shorter time period, with typically 7 or 9 scan points in each of the $x$ and $y$ directions, with only 10 seconds per point, so that the whole scan takes only a few minutes. In 2017--18, these were performed in as many physics fills as possible, typically with one scan at the start of the fill, and another shortly before the end of the fill (unless the beam was unexpectedly lost). The separation range covered by the scan is also smaller than in a VdM fill, covering a range of ${\pm}3 \sigma_\text{b}$. The emittance scan data can be fitted using the standard VdM methodology to extract the beam overlap widths $\Sigma_x$ and $\Sigma_y$; then, Eq.~(\ref{eq:sigmavis}) can be used to obtain the overall visible cross section \sigmavis. Because of the limited amount of data available, especially in the tails, the data are typically fitted with a single Gaussian.

There are several factors affecting the emittance scan measurement which limit the precision with which an absolute calibration can be determined. For example, the crossing angle of the beams in physics operations means that the longitudinal shape of the bunches becomes a relevant factor in determining the absolute luminosity. In physics operations, additional long-range interactions can occur because of the 25\unit{ns} separation between each colliding bunch and the next. This can result in a modification of the bunch separation during the scans. Dynamic inefficiency, where a hit in one BX causes a reduced efficiency in the following BX due to recovery time in the electronics, can also affect the emittance scan data, and nonlinear effects in the detector response (such as accidentals) become much more significant, which can result in a bias in the measured beam overlap width.

While we account for as many of these effects as can be quantified, because of the limitations they create in the measurement, we treat the \sigmavis values obtained from the emittance scan fits as a relative rather than an absolute measurement, normalizing them to the values obtained in similar emittance scans taken in regular fills near the time of the regular VdM program.

The emittance scans at the start and end of a single fill (referred to as ``early'' and ``late'' scans, respectively) typically have SBIL values differing by a factor of 2 or more, so comparing the \sigmavis values obtained can be used to measure the nonlinearity in the PLT response. Similarly, by comparing the \sigmavis values obtained in the early scans, some measure of the changing overall efficiency of the PLT with time or integrated luminosity can be obtained. Knowledge of these efficiency and nonlinearity factors can be used to correct the PLT measurements in a way that is purely intrinsic to the PLT. The same measurements can be applied to the other CMS luminometers, and the final ratio measurements between luminometers give an estimate of the uncorrelated systematic uncertainties in each calibration.

It was observed that the \sigmavis values extracted from leading bunches and train bunches were different; indeed, a structure within trains can also be observed, although at a level smaller than the current uncertainties. Therefore, the efficiency and nonlinearity corrections were computed separately for leading and train bunches. The emittance scan scan analysis was performed individually for each PLT telescope; thus, it produced efficiency and nonlinearity values for each channel for leading and train bunches separately, for each fill with at least one emittance scan.

\begin{figure}[tbhp]
\centering
  \includegraphics[width=0.45\textwidth]{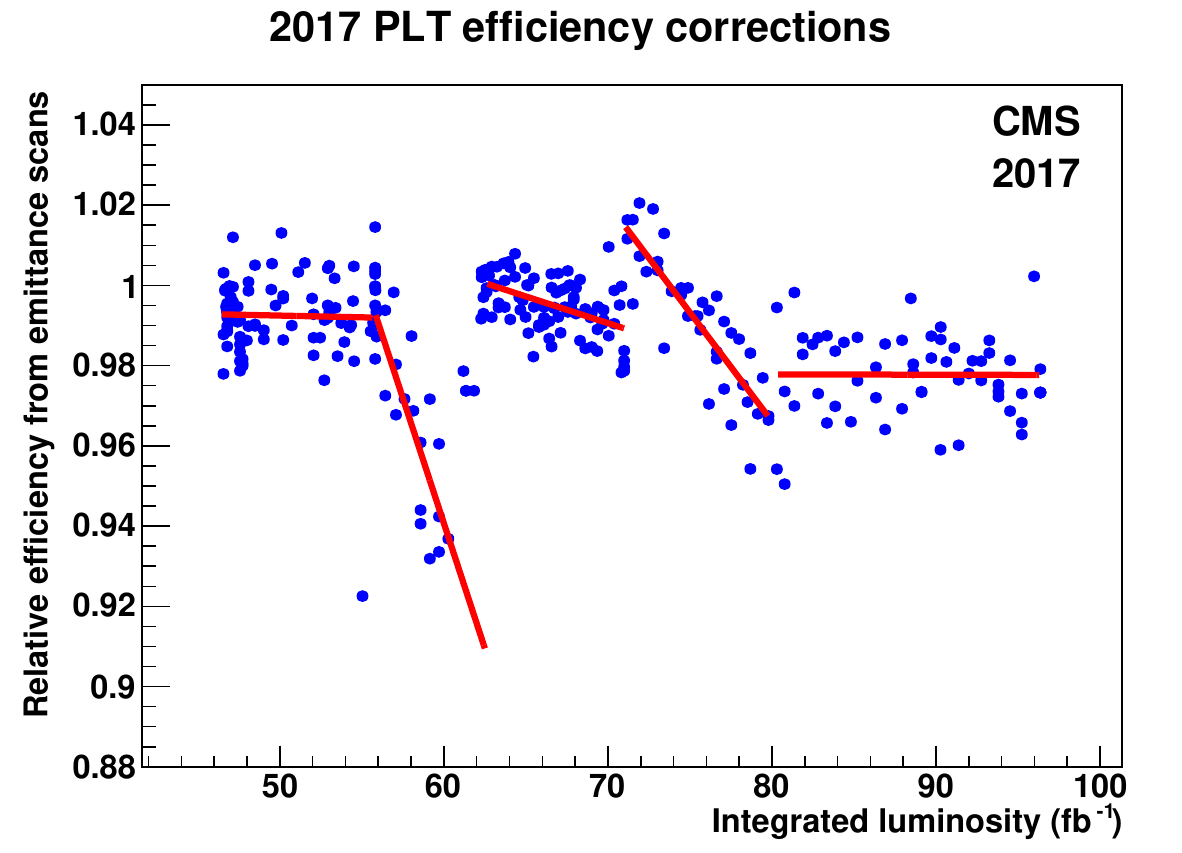}
  \includegraphics[width=0.45\textwidth]{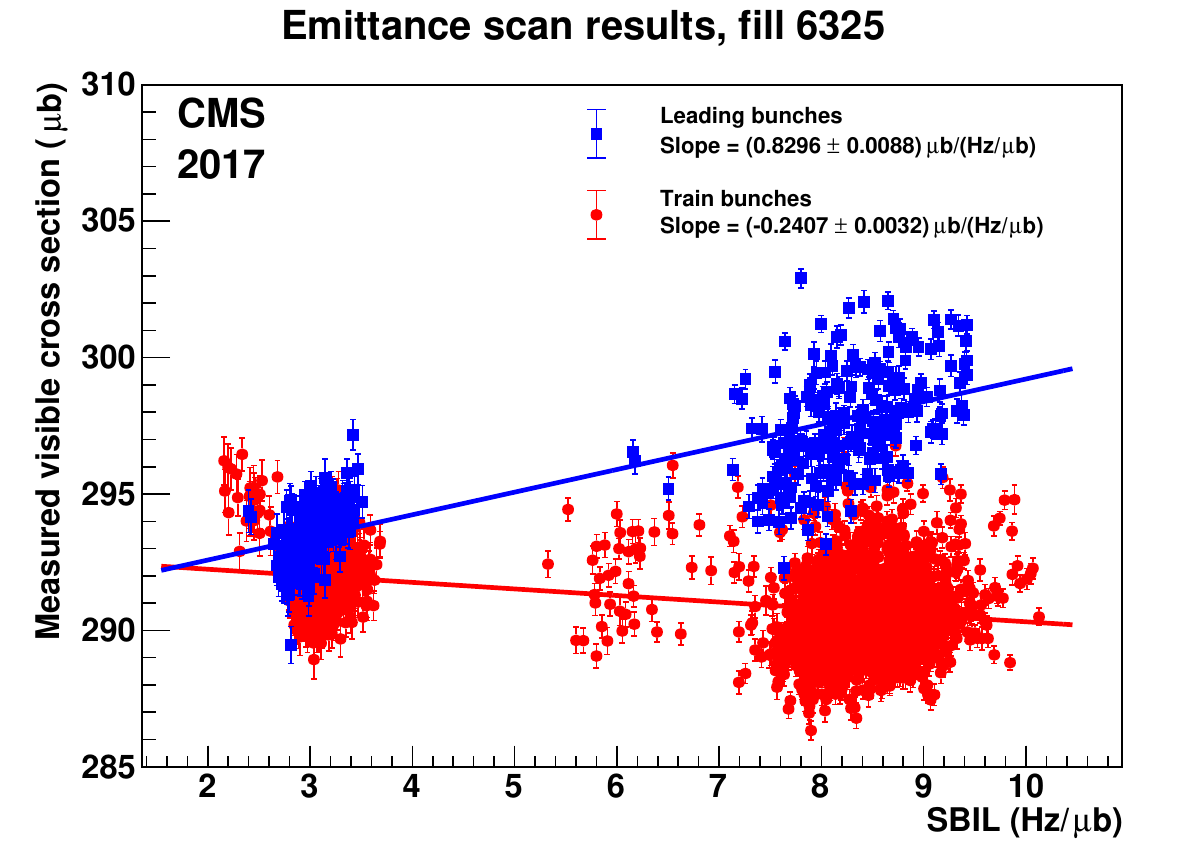}
  \caption{\cmsLeftCap: Efficiency corrections determined from the 2017 emittance scan analysis as a function of the integrated luminosity over the course of the year. \cmsRightCap: Linearity measured for a single fill (fill 6325), showing the results from emittance scans at the beginning (right side) and end (left side) of the fill for leading (blue squares) and train (red circles) bunches. The fits for each type of bunch are shown by the lines, and the resulting slopes are shown in the legend.}
  \label{fig:emittance2017}
\end{figure}

Figure~\ref{fig:emittance2017} shows some results from the emittance scan data in 2017. The \cmsLeft plot shows the efficiency, as measured by the \sigmavis relative to the measurement nearest the time of the VdM scan, and the \cmsRight plot shows the linearity for a single fill (fill 6325). The results of the scan at the beginning of the fill are shown in the points on the right (with higher SBIL), while the results of the scan at the end of the fill appear in the points on the left (with lower SBIL), so the resulting fits can be used to determine the slope for leading and train bunches separately. These per-fill slope values are then used to derive overall linearity corrections for the year.

\begin{figure}[htbp]
\centering
  \includegraphics[width=\cmsFigureWidth]{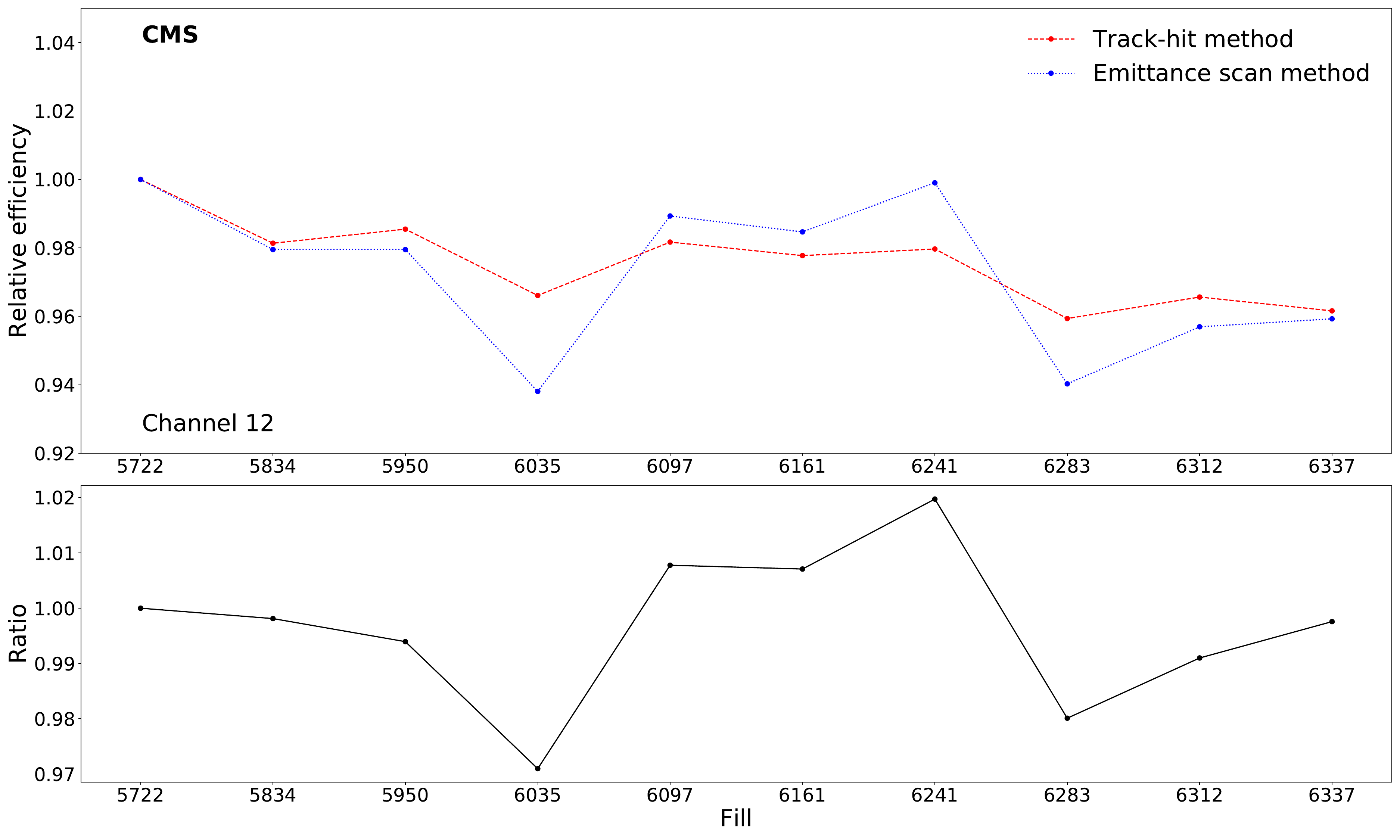}
  \caption{The average sensor efficiency for channel 12 obtained with the track-hit method (dashed red line) and the per-telescope efficiency measured from the analysis of the emittance scan data (dotted blue line), with their ratio (solid black line) shown in the lower pane, as a function of time in 2017. Both efficiencies are normalized to 1 for the first fill considered. The uncertainties in the individual values are too small to be visible.}
  \label{fig:ROCEfficiency_ComparisonEmittanceScans}
\end{figure}

The results from the emittance scan data can also be compared with those from the track reconstruction efficiency described in Section~\ref{sec:efficiency}. The final comparison is shown in Fig.~\ref{fig:ROCEfficiency_ComparisonEmittanceScans}, which shows the track-hit efficiency, the efficiency measured from the emittance scans, and their ratio over the course of 2017, where both efficiencies are normalized to 1 at the first fill considered. We observe that the relative variation over the course of the year is less than 5\%, indicating that the two efficiency measurements are generally consistent over the year.

\subsection{Measured visible cross sections}
\label{sec:sigmavis}

The final calibration of the PLT consists of two parts: the \sigmavis values derived from the VdM calibration procedure, and linearity and efficiency corrections applied to account for variations of the PLT response in normal physics running. Each of these components includes a systematic uncertainty, which determines the overall uncertainty in the PLT luminosity.

The calibrations for normal physics conditions, \ie, $\Pp\Pp$ collisions at $\sqrt{s} = 13\TeV$, as determined by the VdM scan procedure~\cite{Sirunyan:2021qkt,CMS-PAS-LUM-17-004,CMS-PAS-LUM-18-002}, are as follows:

\begin{itemize}
    \item 2015: 494\unit{$\mu$b}
    \item 2016: 329.2\unit{$\mu$b}
    \item 2017: 297.7\unit{$\mu$b}
    \item 2018: 261.6\unit{$\mu$b}
\end{itemize}

The large change from 2015 to 2016 is due to the change in active area described in Section~\ref{sec:activearea}, while the other changes are due to the gradual changes in PLT efficiency over the course of Run~2.

In addition to the calibrations described above, calibrations are also produced for various special physics runs. These calibrations are usually derived from a dedicated VdM scan during the special run, except when noted below. Because these special runs typically have very low luminosity, nonlinear effects are negligible, and the total amount of additional radiation damage during the run is very small, so the efficiency can be treated as a constant and simply included in the \sigmavis measurement. The calibrations for the special runs are as follows. Note that the proton-lead collision runs are referred to as ``$\Pp\mathrm{Pb}$'' or ``$\mathrm{Pb}\Pp$'' depending on whether the protons are in beam~1 or 2, with ``$\Pp\mathrm{A}$'' used to refer to both collectively. The factors of 82 in the $\Pp\mathrm{A}$ and PbPb runs derive from the 82 protons in a lead nucleus, and similarly for the factor of 54 in the xenon-xenon (XeXe) run.

\begin{itemize}
    \item 2015 $\Pp\Pp$ run at $\sqrt{s} = 5.02\TeV$: 355.0\unit{$\mu$b}~\cite{CMS-PAS-LUM-16-001}
    \item 2015 PbPb run at $\sqrt{s} = 5.02\TeV$/nucleon: $2.69\unit{b} = 400.1\unit{$\mu$b} \times 82 \times 82$~\cite{CMS-PAS-LUM-18-001}
    \item 2016 $\Pp\mathrm{A}$ run at $\sqrt{s} = 8.16\TeV$/nucleon: $20.7\unit{mb} = 252.8\unit{$\mu$b} \times 82$ for the Pbp period, and $19.8\unit{mb} = 241.3\unit{$\mu$b} \times 82$ for the pPb period~\cite{CMS-PAS-LUM-17-002}
    \item 2017 XeXe run at $\sqrt{s} = 5.44\TeV$/nucleon: $932\unit{mb} = 319.6\unit{$\mu$b} \times 54 \times 54$ (derived from emittance scan data)~\cite{CMS-DP-2021-002}
    \item 2017 $\Pp\Pp$ run at $\sqrt{s} = 5.02\TeV$: 192.8\unit{$\mu$b}~\cite{CMS-PAS-LUM-19-001} 
    \item 2018 $\Pp\Pp$ run at $\sqrt{s} = 900\GeV$: 162.7\unit{$\mu$b} (derived from scaling by the ratio of the theoretical inelastic cross section to that at 13\TeV)
    \item 2018 PbPb run at $\sqrt{s} = 5.02\TeV$/nucleon: $1.67\unit{b} = 249.1\unit{$\mu$b} \times 82 \times 82$~\cite{CMS-PAS-LUM-18-001}
\end{itemize}

Prior to the main 2016 $\Pp\mathrm{A}$ run, a short $\Pp\mathrm{A}$ run at $\sqrt{s} = 5.02\TeV$/nucleon was also carried out. No offline calibration was performed for this run, as the data were not used for physics.

\subsection{Linearity and efficiency corrections}
\label{sec:lin_eff}

In order to apply the VdM calibration to physics fills with significantly higher luminosity, potential nonlinear effects in the PLT response, as represented by the $a$ term in Eq.~(\ref{eq:sigmavis_lincorr}), must be measured and corrected for. In addition, changes in the detector conditions can result in changes in \sigmavis, which must also be corrected for to obtain an accurate luminosity measurement, as represented by the $\varepsilon$ term in Eq.~(\ref{eq:sigmavis_effcorr}). In this section, the following conventions are used: an efficiency value of 0.95 means that the measured efficiency is 5\% lower than the reference value, so the luminosity is corrected upward by 1/0.95. Similarly, a linearity value of 1\%/(\hzub) means that the observed luminosity exhibits an excess over the true value, so the raw data will then be corrected downwards by that amount. In all years, the efficiency is normalized to 1 for the fill containing the VdM scan program for that year.

For 2015, as the overall luminosity was low and thus the effect of radiation damage is small, no efficiency correction is included. The linearity correction is taken to be the accidental rate, whose measurement is described in Section~\ref{sec:accidentals}, and amounts to $4.76 + 2.74\cdot\text{SBIL}$ [\%], where SBIL is in \hzub. In addition, most of the 2015 PLT data were affected by an issue in the fast-or FED firmware that caused the highest signal level from the ROCs (indicating three or more hits) to be decoded as a ``0'' rather than a ``1''. The effect of this issue was measured to be $2.4 + 6.0\cdot \text{SBIL}$ [\%], with an uncertainty of 0.7\%. This issue was fixed towards the end of the 2015 run, so data from subsequent years are not affected.

For the 2016 calibration, efficiency and linearity corrections are derived by using the measurement from the RAMSES detectors~\cite{Forkel-Wirth:687619}. These are detectors located in the CMS experimental cavern, consisting of ionization chambers with an active volume of 3\unit{liters} of air at atmospheric pressure. The primary function of the RAMSES monitors is to ensure the safety of personnel in the CMS cavern; however, BRIL discovered in 2017 that the RAMSES measurement also could be used for luminosity determination~\cite{Wanczyk:2701798}. The RAMSES detectors are not fast enough to provide bunch-by-bunch luminosity measurements, and because their overall rates are significantly lower than the primary BRIL luminometers, they cannot be directly calibrated using the VdM method. However, these low rates also mean that the RAMSES measurement, although it must be integrated over a sufficiently long period, shows excellent long-term stability and linear behavior.

\begin{figure}[tbhp]
\centering
  \includegraphics[width=0.48\textwidth]{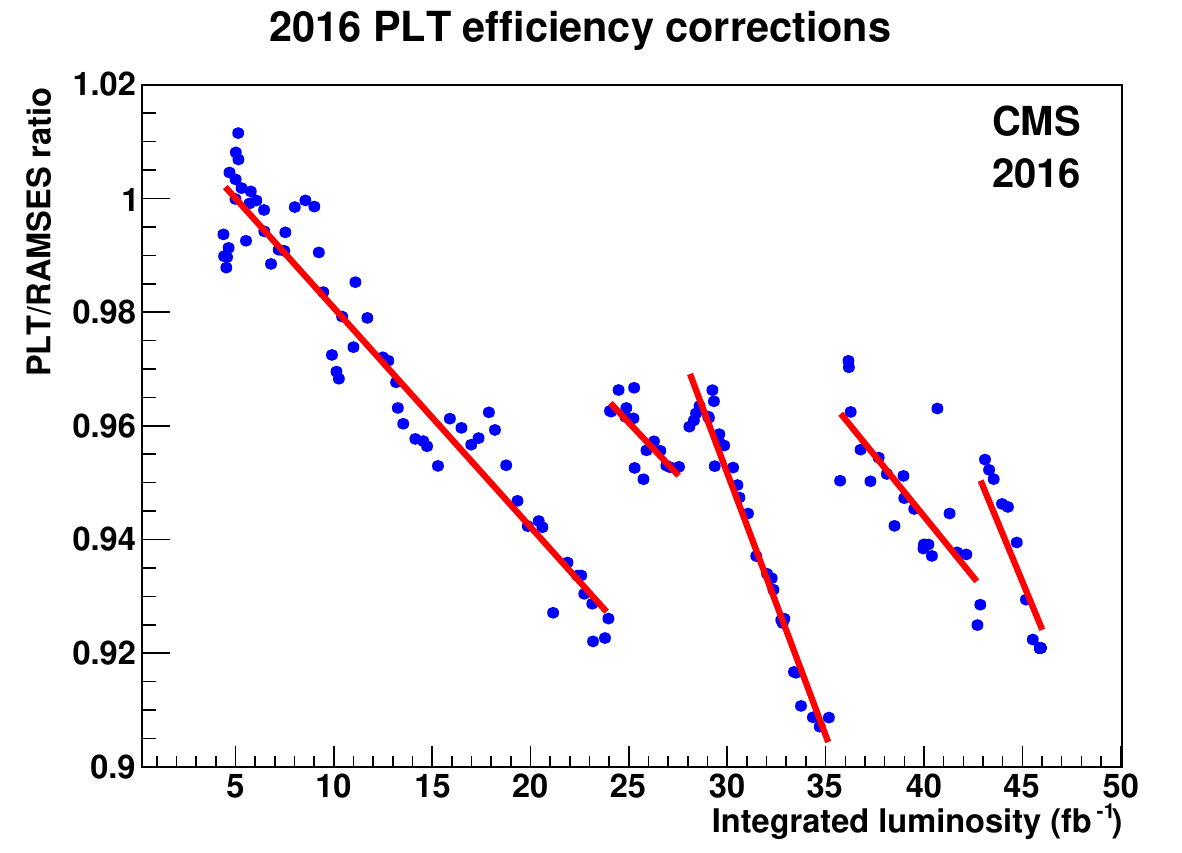}
  \includegraphics[width=0.48\textwidth]{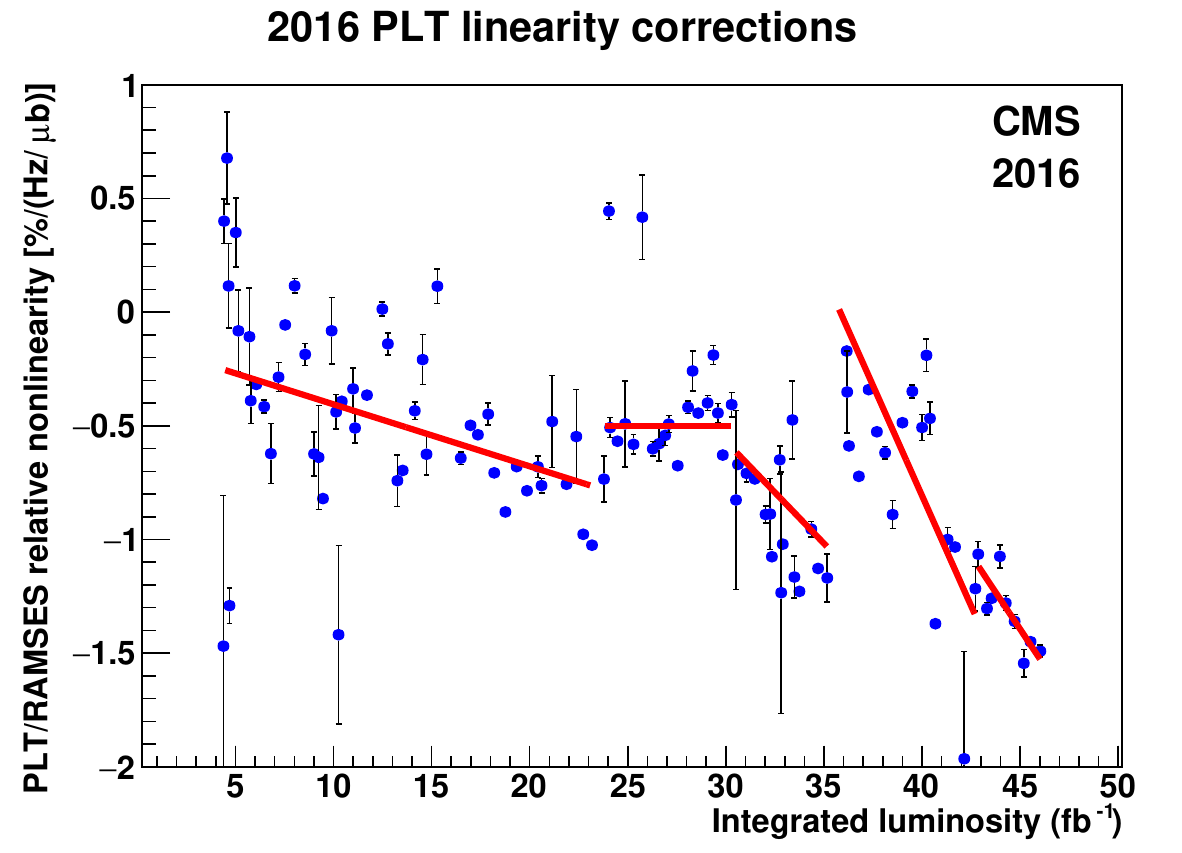}
  \caption{\cmsLeftCap: Efficiency of PLT relative to RAMSES over the course of 2016, where each point represents a single fill. \cmsRightCap: Linearity of PLT relative to RAMSES over the course of 2016, where each point indicates the fitted nonlinearity and its uncertainty for a single fill. The red lines show the fit functions that are used to obtain the final efficiency and linearity corrections for 2016. The uncertainties in the efficiency measurements are too small to be visible on the plot.}
  \label{fig:plt_vs_ramses}
\end{figure}

Consequently, for the final 2016 PLT corrections~\cite{Sirunyan:2021qkt}, stability and linearity corrections were derived using the RAMSES luminosity measurement as a baseline. This allows us to combine the excellent statistical precision of the PLT bunch-by-bunch measurement with the stability and linearity of the RAMSES measurement. The resulting corrections for the 2016 data are shown in Fig.~\ref{fig:plt_vs_ramses}. The data are divided into separate periods by examining the general trends over time, and a linear fit is used within each period to obtain the final efficiency and linearity corrections. (The efficiency and linearity corrections use a different set of five periods from each other.) The final time-dependent efficiency correction is in the range 0.90--1.00, and the time-dependent linearity correction is in the range $-0.2$ to 1.4\%/(\hzub).

In 2017 and 2018, the linearity and efficiency corrections are derived using the emittance scan data described in Section~\ref{sec:emittance}, allowing us to derive a correction using data intrinsic to the PLT. The year is divided into periods over which the PLT behavior can be observed to change linearly over time, and corrections are derived for each period separately. In 2017, the time-dependent efficiency factor is in the range 0.97--1.03, and the time-dependent linearity correction is in the range 0.2--0.9\%/(\hzub).

\begin{figure}[htbp]
\centering
\includegraphics[width=\cmsFigureWidth]{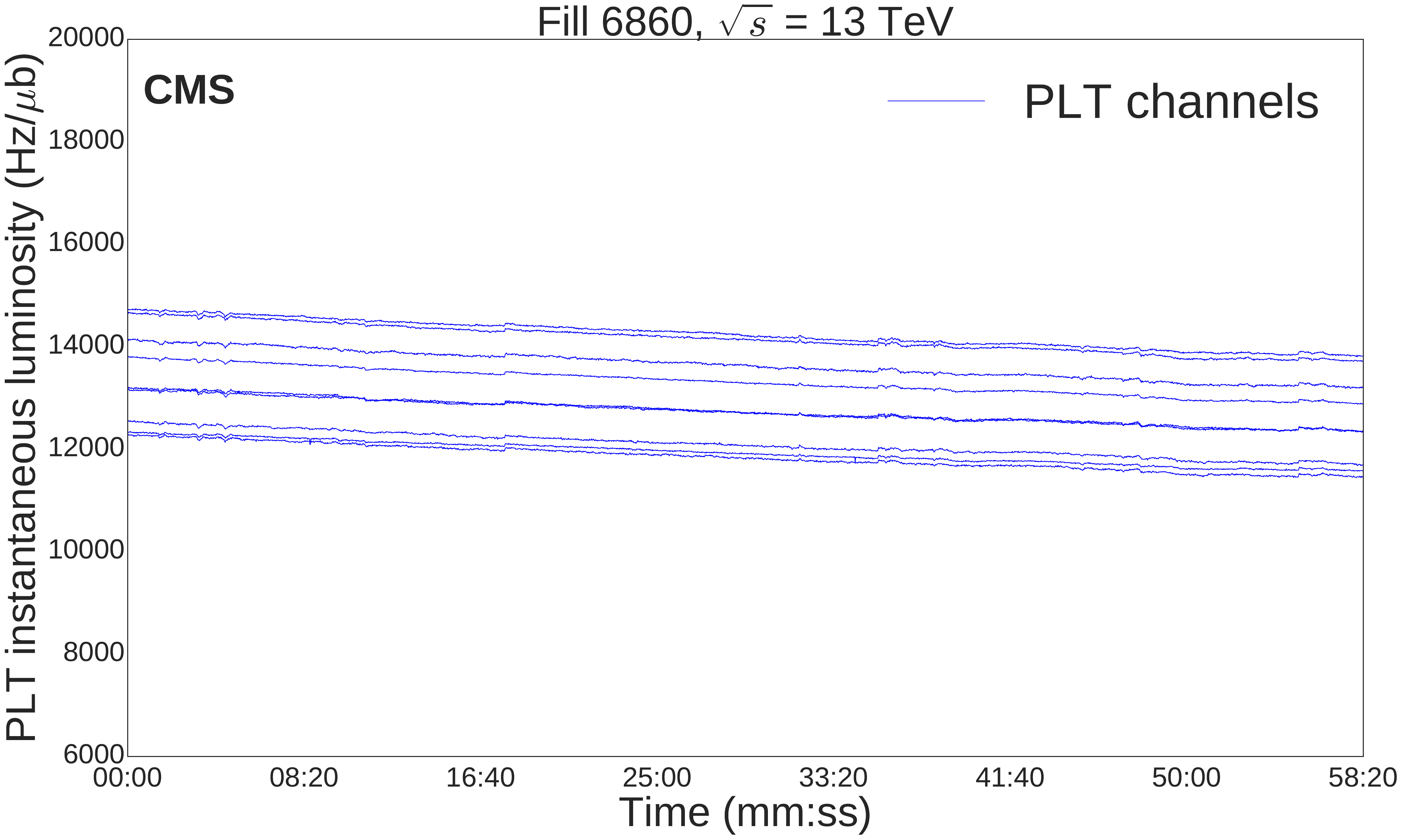}
\includegraphics[width=\cmsFigureWidth]{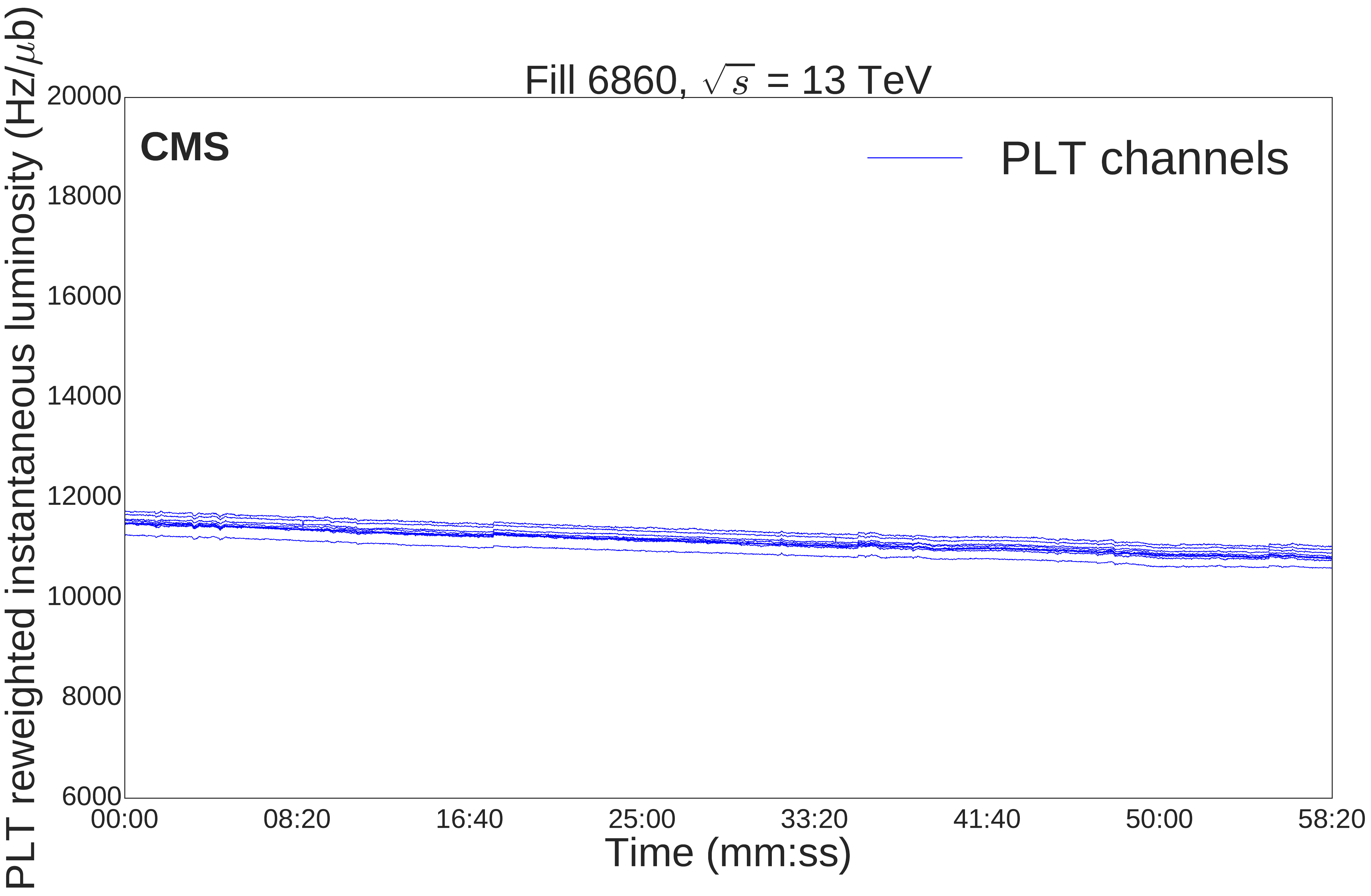}
\caption{Top: PLT per-channel luminosity values as a function of time for fill 6860 in 2018, showing the total (\ie, over all BXs) instantaneous luminosity as measured by the PLT detector. Bottom: Per-channel luminosity values for the same fill after applying the per-channel weights described in the text to correct for differing linearity and efficiency.}
\label{fig:PLTandPLTreweighted}
\end{figure}

For 2018, the corrections are applied in two steps. First, the emittance scan data are analyzed on a channel-by-channel basis, to account for different behavior in the different PLT channels. As the thresholds were adjusted in the middle of 2018, two sets of corrections are used, for the periods before and after the adjustment. The efficiency corrections used are in the range of 0.90--1.25, depending on the channel, bunch type (leading or train), and time period, and the linearity corrections are in the range 1.0--2.6\%/(\hzub). Some channels which are not well behaved over the course of the year are discarded. Figure~\ref{fig:PLTandPLTreweighted} shows the effect of the channel-by-channel corrections on the luminosity for a single fill. The per-channel corrections significantly improve the agreement between the individual channels, decreasing the relative luminosity difference between all PLT channels from about 20\% before corrections to about 5\% after the corrections are applied.

After the channel-by-channel corrections, the emittance scan analysis is repeated and a second set of corrections is applied to the overall data to account for residual effects. This includes a time-dependent efficiency term that varies from 0.96 to 1.00 over the course of the year, applied using a linear fit in two different time periods, and an additional linearity term of $-0.4\%/(\hzub)$.

\subsection{Systematic uncertainties}
\label{sec:systematics}

In order to evaluate the overall systematic uncertainty in the luminosity measurement from the PLT for $\Pp\Pp$ runs at 13\TeV, we consider two general categories:

\begin{itemize}
\item ``Normalization'' uncertainties, which result from the determination in the absolute luminosity calibration \sigmavis determined from the VdM scan procedure
\item ``Integration'' uncertainties, which result from the application of the VdM calibration to a full year of physics running, including variations in efficiency over time (stability) and nonlinearity in extrapolating to physics luminosities
\end{itemize}

The inputs to the determination of the absolute luminosity given in Eq.~(\ref{eq:VdMeq}) are affected by several systematic effects, which thus can cause biases in the \sigmavis measured in the VdM scan procedure. These effects are measured and corrected, where applicable; the uncertainty in these corrections results in an uncertainty in the final luminosity calibration. The main sources of systematic uncertainty in the VdM procedure are those associated with the nonfactorization of the spatial proton density functions, modeling of the effects on the beam position and shape due to electromagnetic interactions between the beams, and residual differences between the true beam positions and the values nominally set by LHC magnets. As these uncertainties arise from the VdM procedure itself and are not specific to the PLT, we use the uncertainties measured in Refs.~\cite{Sirunyan:2021qkt,CMS-PAS-LUM-17-004,CMS-PAS-LUM-18-002} for these. Table~\ref{tab:normCorrections} summarizes the final corrections applied for each of these sources for each year, and Table~\ref{tab:normUncertainties} shows the systematic uncertainties in these corrections, as well as some other uncertainties in the VdM procedure, considered relevant for PLT. Note that the uncertainties due to beam-beam deflection and dynamic-$\beta$ are correlated, so for all years except 2017 they are combined into a single uncertainty. (In 2017, there was no correction applied for the dynamic-$\beta$ effect, so an uncertainty was assigned to cover possible corrections.) In the 2015--16 analysis, two corrections were applied to account for beam position effects, one to account for gradual linear orbit drift and one for residual differences from the linear orbit drift. In the 2017 and 2018 analyses, only the first of these effects was considered (and was found to be negligibly small in 2017).

In 2017--18, the cross-detector consistency among the detectors in the VdM fill is evaluated by computing the integrated luminosity during stable periods in the VdM fill (\ie, when a scan is not in progress). In 2017, a correction was applied to each luminometer to bring it to the average, and the largest such correction taken as the uncertainty due to cross-detector consistency. In 2018, the individual luminometers were not corrected, but the largest deviation from average was taken as the systematic. The correction for PLT in 2017 was $<$0.1\%, and the uncertainty due to cross-detector consistency was 0.6 and 0.5\% in 2017 and 2018, respectively.

\begin{table}[htbp]
\centering
\topcaption{Summary of corrections to \sigmavis (in \%) in the VdM scan procedure for each effect considered. Entries for which the correction is either negligibly small, or no correction is applied but the effect is taken into account in the uncertainty, are marked with \NA.}
\cmsTable{
\begin{tabular}{ccccc}
\hline
\multirow{2}{*}{Systematic} & \multicolumn{4}{c}{Correction (\%)}  \\
& 2015 & 2016 & 2017 & 2018 \\
\hline
Length scale & $-0.4$ & $-1.3$ & $-0.9$ & $-0.8$ \\
Orbit drift & +0.6 to +1.0 & +0.2 to +1.0 &  \multirow{2}{*}{\NA} & \multirow{2}{*}{+0.2} \\
Residual beam position & $-0.6$ to 0.4 & $-0.5$ to $-0.2$ & & \\
$x$-$y$ nonfactorization & +0.8 to +1.3 & +0.6 & +0.8 & \NA \\
Beam-beam deflection & \multirow{2}{*}{+0.6} & \multirow{2}{*}{+0.4} & +1.6 & +1.5 \\
Dynamic-$\beta$ &  &  & \NA & $-0.5$ \\
Ghosts and satellites & +0.2 & +0.3 & \NA & +0.4 \\
Background subtraction & \NA & \NA & \NA & +0.3 \\
\hline
\end{tabular}
}
\label{tab:normCorrections}
\end{table}

\begin{table}[htbp]
\centering
\topcaption{Summary of contributions to the relative systematic uncertainty in \sigmavis (in \%) in the VdM scan procedure. Each correction in Table~\ref{tab:normCorrections} has an associated uncertainty, and additional uncertainties are assigned for variation between the individual scans and bunches considered in the VdM analysis.}
\cmsTable{
\begin{tabular}{ccccc}
\hline
\multirow{2}{*}{Systematic} & \multicolumn{4}{c}{Uncertainty (\%)}  \\
& 2015 & 2016 & 2017 & 2018 \\
\hline
Length scale & 0.2 & 0.3 & 0.3 & 0.2 \\
Orbit drift & 0.2 & 0.1 &  0.2 & 0.1 \\
Residual beam position differences & 0.8 & 0.5 & \NA & \NA \\
$x$-$y$ nonfactorization & 0.5 & 0.5 & 0.8 & 2.0 \\
Beam-beam deflection & \multirow{2}{*}{0.5} & \multirow{2}{*}{0.5} & 0.4 & \multirow{2}{*}{0.2} \\
Dynamic-$\beta$ &  &  & 0.5 & \\
Beam current calibration & 0.2 & 0.2 & 0.3 & 0.2 \\
Ghosts and satellites & 0.1 & 0.1 & 0.1 & 0.1 \\
Scan to scan variation & \multirow{2}{*}{0.6} & \multirow{2}{*}{0.3} & 0.9 & 0.3 \\
Bunch to bunch variation &  &  & 0.1 & 0.1 \\
Background subtraction & \NA & \NA & \NA & 0.1 \\[\cmsTabSkip]
Total & 1.3 & 1.0 & 1.5 & 2.1 \\
\hline
\end{tabular}
}
\label{tab:normUncertainties}
\end{table}

\begin{figure*}[tbhp]
\centering
  \includegraphics[width=0.3\textwidth]{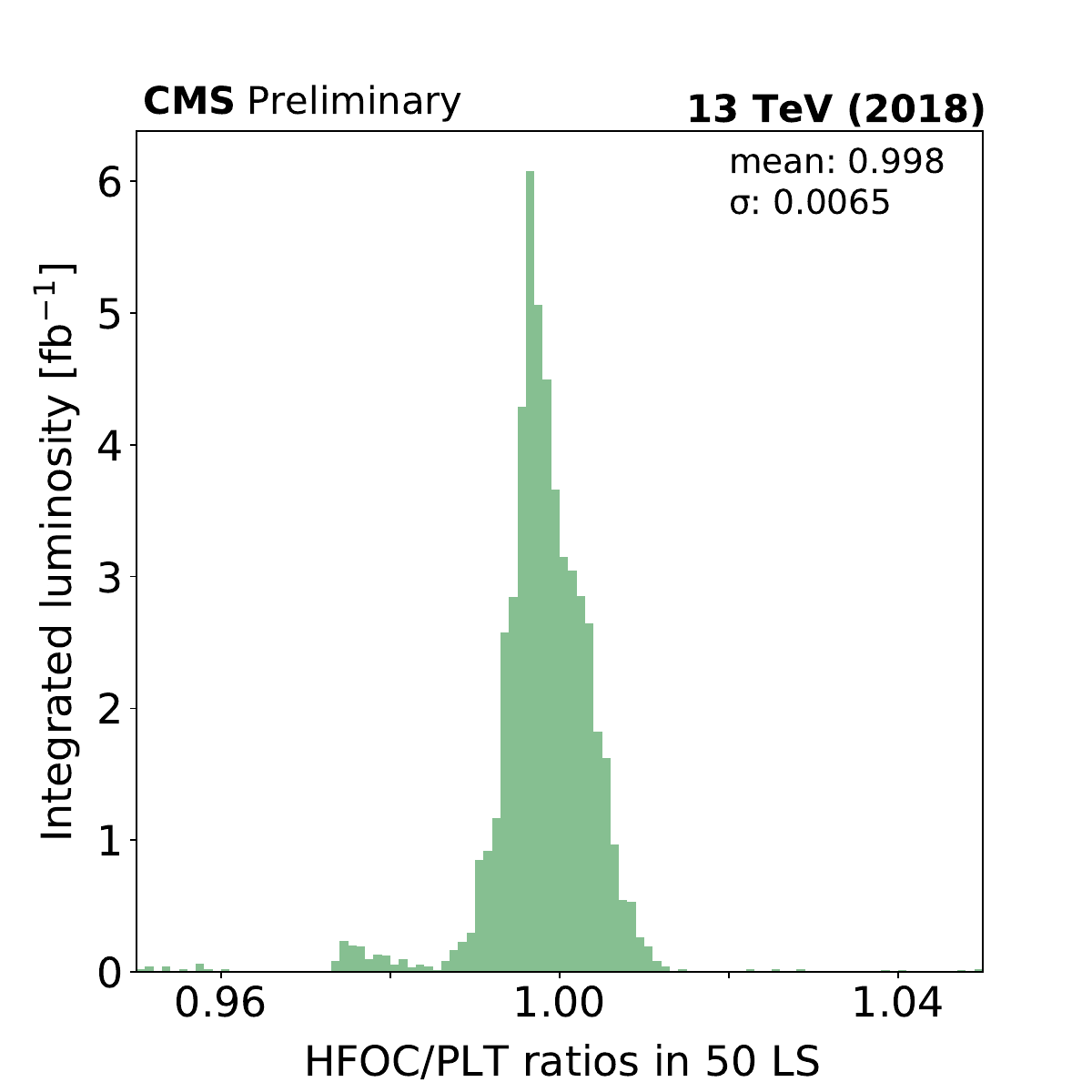}
  \includegraphics[width=0.3\textwidth]{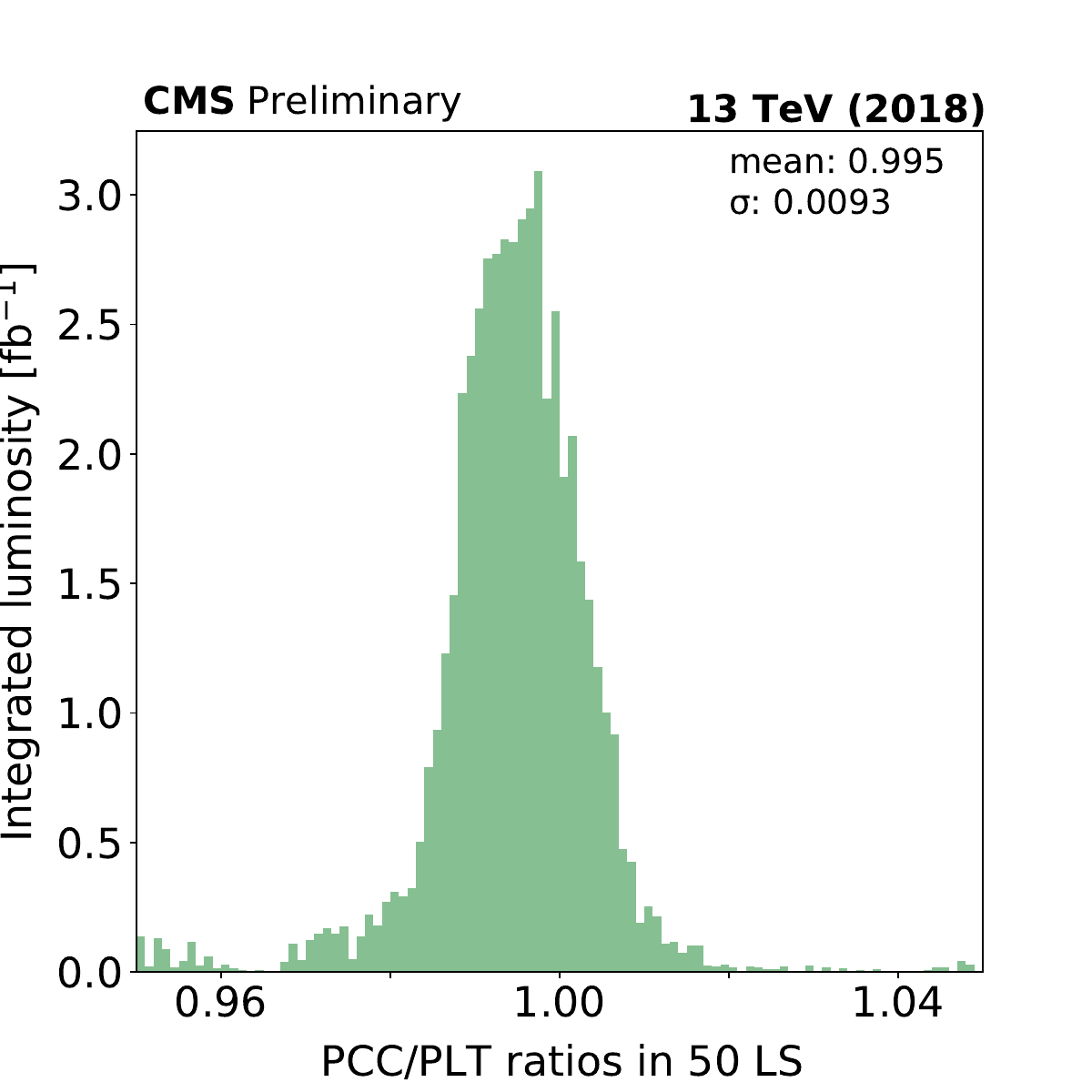}
  \includegraphics[width=0.3\textwidth]{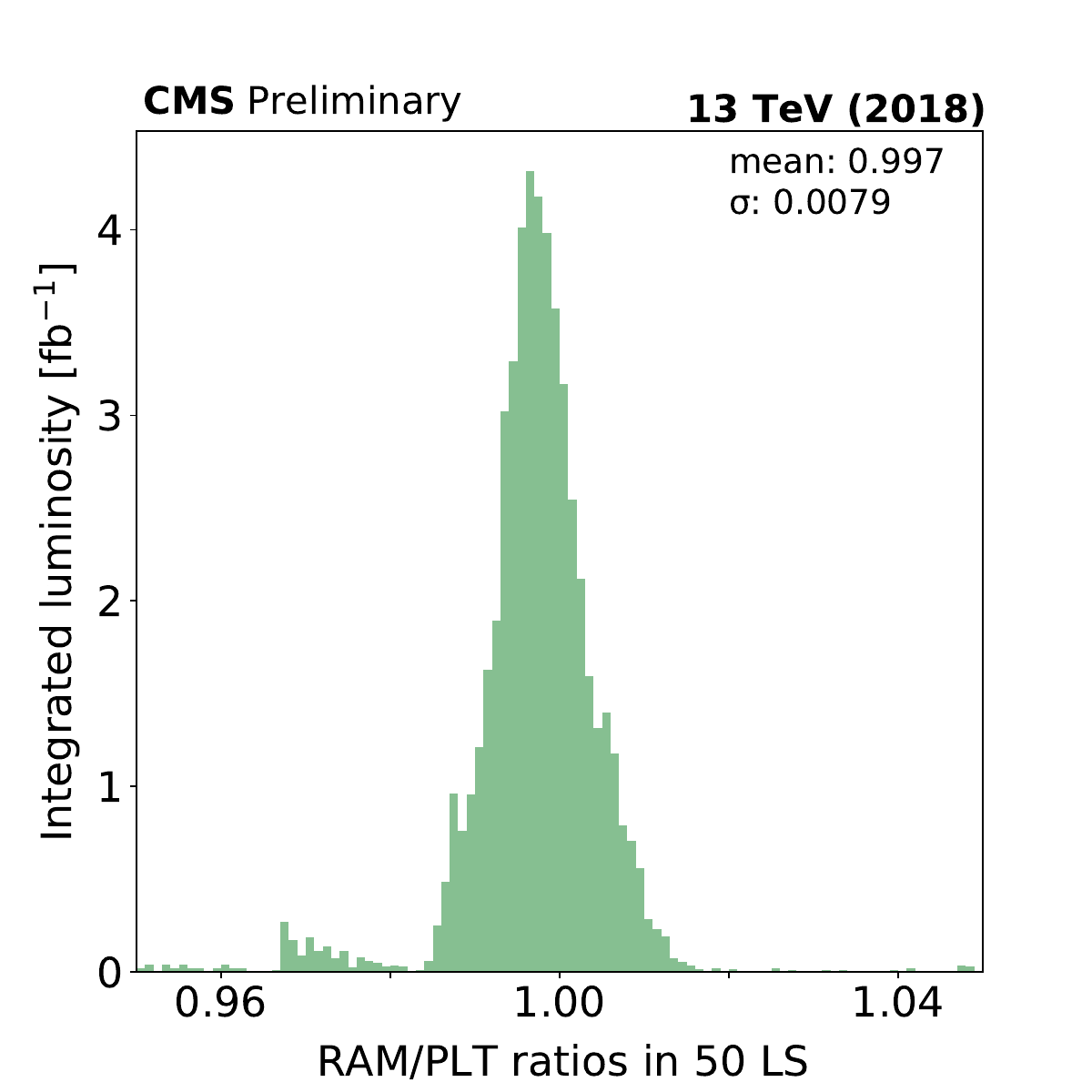}
  \caption{Ratio histograms for different luminometer pairs during 2018. Each entry represents a period of 50 lumi sections, weighted by the luminosity in that period. Left: HFOC/PLT; middle: PCC/PLT; right: RAMSES/PLT.}
  \label{fig:ratio2018}
\end{figure*}

Once the linearity and efficiency corrections described in Section~\ref{sec:lin_eff} have been applied, any remaining nonlinearity and efficiency effects are measured through comparison to other luminometers. The overall consistency with respect to other luminometers can be quantified by taking the ratio of the reported PLT luminosity to that of another luminometer (integrated into 50 LS bins). The uncertainty due to stability can be calculated by binning the resulting ratios into a histogram weighted by luminosity. The relative stability is then given by the standard deviation of the distribution; the potential bias from the difference of the mean of the distribution from unity is always a subdominant effect. Figure~\ref{fig:ratio2018} shows the luminosity-weighted ratio distributions for HFOC/PLT, PCC/PLT, and RAMSES/PLT in the 2018 data.

\begin{figure*}[tbhp]
\centering
  \includegraphics[width=0.3\textwidth]{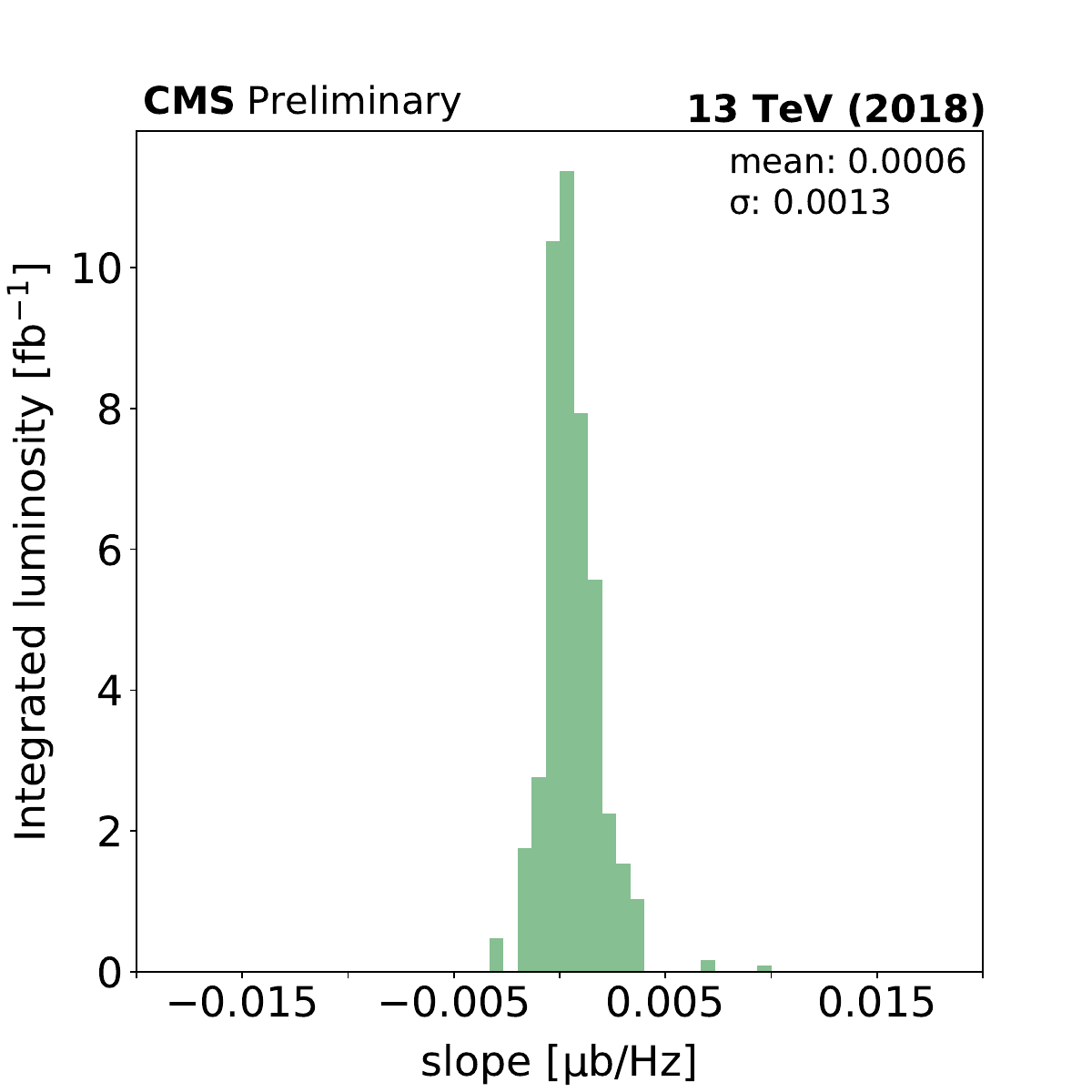}
  \includegraphics[width=0.3\textwidth]{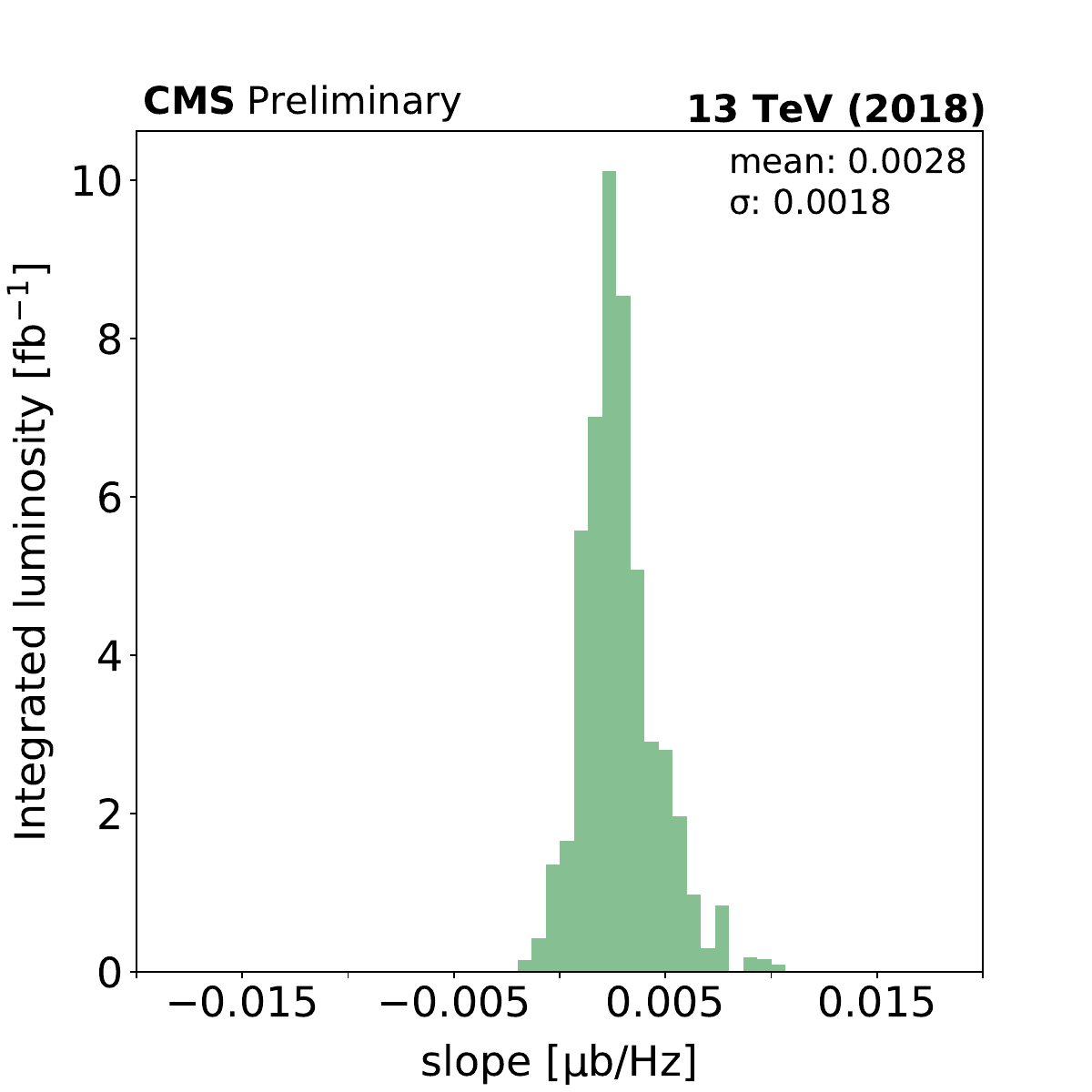}
  \includegraphics[width=0.3\textwidth]{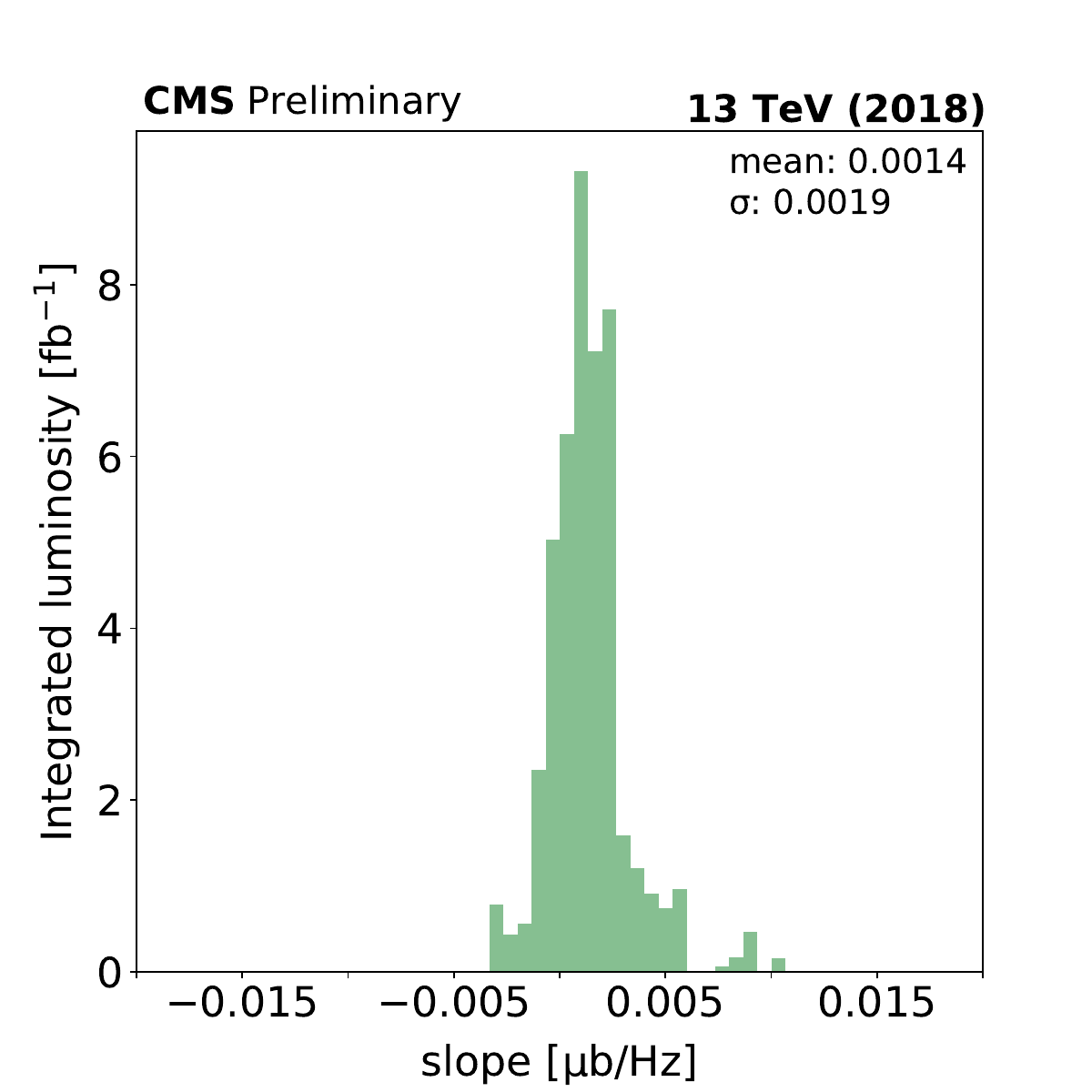}
  \caption{Slope distribution measuring the relative nonlinearity between different luminometer pairs during 2018: (left) HFOC/PLT, (middle) PCC/PLT, (right) RAMSES/PLT.}
  \label{fig:slope2018}
\end{figure*}

A systematic uncertainty in the linearity is assigned by comparing the linearity response to that of other luminometers. In order to evaluate this uncertainty for two given luminometers, the ratio of luminosity values as a function of average SBIL is fitted with a line for each fill. The resulting slope of the fitted line is taken as the relative nonlinearity for those two luminometers for that fill. The resulting slopes are plotted as a function of integrated luminosity and binned into a histogram. Figure~\ref{fig:slope2018} shows the slope distribution during 2018 for PLT compared to HFOC, RAMSES, and PCC.

The uncertainty in the cross-detector stability is determined by selecting the maximum standard deviation of the ratio distribution among the available luminometer pairs. For the uncertainty in the cross-detector linearity, the largest of the mean and standard deviation of the slope distributions among the different detector pairs is taken. As this gives us the uncertainty in \%/(\hzub), it is then scaled by the average SBIL for each year to obtain the overall uncertainty.

In 2015, the cross-detector linearity comparisons are not available, so the uncertainty is taken instead from the uncertainties in the accidental and firmware corrections described in Section~\ref{sec:lin_eff}.

Luminometers such as the PLT measure the instantaneous luminosity delivered by the LHC to CMS. However, the quantity of interest to CMS physics analysis is the amount of luminosity actually recorded by CMS; these quantities are related by the CMS deadtime, so the uncertainty in this measurement also affects the uncertainty in the recorded luminosity.

Table~\ref{tab:totalUncertainties} summarizes the final systematic uncertainties considered. Throughout all years, the $x$-$y$ nonfactorization of the proton bunch density functions, measurement of the beam position, and the modeling of beam-beam interactions are the dominant sources of uncertainty in the normalization. The total normalization uncertainty in the luminosity calibration ranges from 1.0 to 2.1\%. When including the integration uncertainties, the total uncertainty is in the range 1.6--3.8\%.

\begin{table}[htbp]
\centering
\topcaption{Summary of total uncertainty in the PLT luminosity measurement, including the total normalization uncertainty, individual sources of integration uncertainty, and total integration uncertainty.}
\cmsTable{
\begin{tabular}{ccccc}
\hline
\multirow{2}{*}{Systematic} & \multicolumn{4}{c}{Uncertainty (\%)}  \\
& 2015 & 2016 & 2017 & 2018 \\
\hline
Total normalization uncertainty & 1.3 & 1.0 & 1.5 & 2.1 \\[\cmsTabSkip]
Cross-detector stability & 3.0 & 0.9 & 1.3 & 1.0 \\
Cross-detector linearity & 1.8 & 0.8 & 1.4 & 1.5 \\
CMS deadtime & 0.5 & $<$0.1 & $<$0.1 & $<$0.1 \\ [\cmsTabSkip]
Total integration uncertainty & 3.6 & 1.2 & 2.0 & 1.8 \\ [\cmsTabSkip]
Total uncertainty & 3.8 & 1.6 & 2.4 & 2.8 \\
\hline
\end{tabular}
}
\label{tab:totalUncertainties}
\end{table}

We can combine the data from the different years, treating the stability uncertainty as uncorrelated between years, the linearity uncertainty as correlated, and the normalization uncertainties following the scheme described in Ref.~\cite{Sirunyan:2021qkt}, to obtain a total systematic uncertainty of 2.2\% in the PLT luminosity measurement on the Run 2 $\Pp\Pp$ data set at $\sqrt{s} = 13\TeV$.

The PLT-specific uncertainty is not evaluated for the various special runs mentioned in Section~\ref{sec:sigmavis}. In general, however, because these runs feature very low instantaneous luminosity, the effects of the stability and linearity uncertainties are significantly reduced, and the overall uncertainty is dominated by the normalization uncertainties, which are discussed in Refs.~\cite{CMS-PAS-LUM-16-001,CMS-PAS-LUM-19-001,CMS-PAS-LUM-18-001,CMS-PAS-LUM-17-002,CMS-DP-2021-002}.

\subsection{Luminosity using track data}
\label{sec:track_lumi}

While the fast-or luminosity measurement is the primary luminosity output from the PLT, reconstructed tracks using the pixel data can also be used to produce a luminosity measurement. The principal advantage of such a measurement is that the track reconstruction should reduce the contribution from accidentals and other noncollision sources, producing a more linear measurement than the fast-or triple coincidence measurement. Because of the low trigger rate used to gather the pixel data, the statistical precision of this measurement is lower than that of the fast-or method; in order to obtain a reasonable precision, the track data are aggregated over all BXs in 5-minute intervals. As a result, this measurement is not suitable for online operations; however, it can provide a valuable offline crosscheck of the stability and linearity of the fast-or technique.

The method to obtain the luminosity from reconstructed tracks in normal physics fills is as follows. First, tracks are reconstructed with the pixel data, as described in Section~\ref{sec:alignment}. Then, accidentals are rejected using the procedure described in Section~\ref{sec:accidentals}. For this study, a tighter selection of $2\sigma$ is used for rejection; while this will result in rejecting some good tracks, it ensures an event sample of high-quality tracks, crucial to the track counting technique.

In order to avoid the difficulties of reconstructing multiple tracks in a single telescope, a zero-counting method is used, as for the fast-or luminosity; a telescope is considered to have a track if any of the possible combinations of hits in the telescope form a track passing the accidental rejection. For each channel, the luminosity (integrated over all BXs) is calculated separately, and then the channels are averaged to obtain a final luminosity measurement.

For this study, ten fills in 2016 were chosen (as the pixel data quality was generally better in 2016 than in 2017--18), spaced throughout the year and with no known operational issues in PLT.

\begin{figure*}[tbhp]
\centering
  \includegraphics[width=0.45\textwidth]{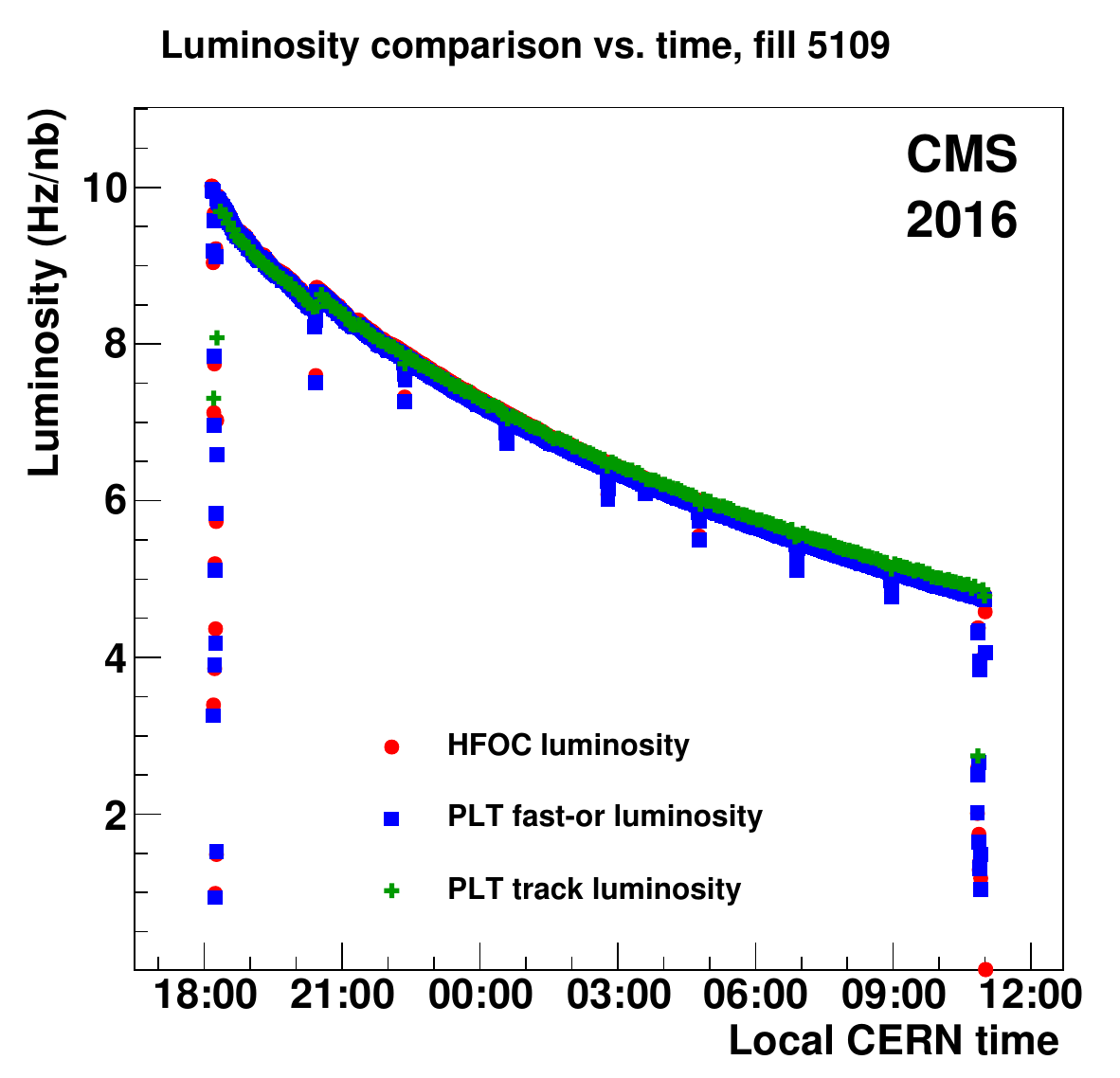}
  \includegraphics[width=0.45\textwidth]{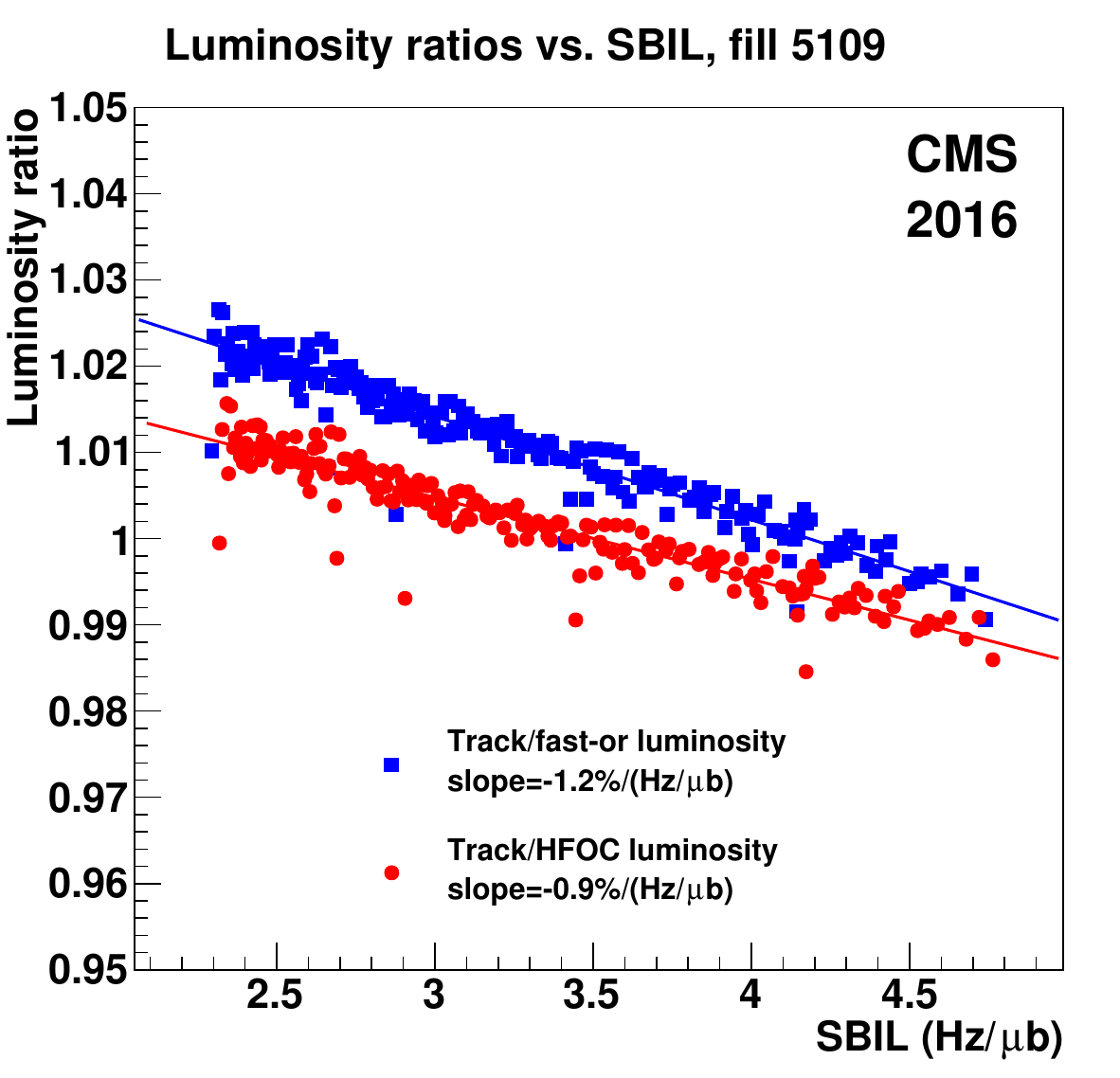}
  \caption{Left: Luminosity obtained from track reconstruction (green crosses) vs. PLT fast-or luminosity (blue squares) and forward hadron calorimeter luminosity (HFOC, red circles) for fill 5109 as a function of time. The track luminosity is cross-calibrated to the HFOC luminosity at the beginning of the fill. Right: Ratios of the track luminosity to the fast-or and HFOC luminosities as a function of SBIL measured by the luminometer in the denominator of the ratio.}
  \label{fig:tracklumi_fill_good}
\end{figure*}

Figure~\ref{fig:tracklumi_fill_good} shows the results for fill 5109 in 2016. The left plot shows the luminosity from track counting compared to the luminosity from the regular fast-or method and the luminosity from the forward hadron calorimeter (HFOC). The fast-or and HFOC luminosities are independently calibrated and fully corrected using their final 2016 corrections~\cite{Sirunyan:2021qkt}. The track counting luminosity is cross-calibrated to the HFOC luminosity (\ie, normalized to the HFOC value) at the beginning of the fill, but otherwise no additional corrections are applied. The right plot shows the ratio of the track luminosity to the HFOC and fast-or luminosities as a function of the SBIL, fitted with a linear function to determine the overall slope. Overall, we see good agreement between the track luminosity and the other two luminometers, although there is a residual nonlinearity of approximately 1\%/(\hzub). This suggests that an additional correction will still be needed for the track luminosity technique, although the magnitude of this correction is less than the correction applied to the fast-or luminosity measurement.

However, looking at fills over the course of a year, we observe that many of the fills are affected by cases where issues in the readout hardware (presumably caused by an SEU) caused some loss in the pixel data without affecting the fast-or readout, and so these problems were not immediately noticed. While an automatic algorithm was developed to find and correct for these, there is also significant fill-to-fill variation in the observed normalization of the track luminosity measurement, due to unobserved changes in the performance of the pixel readout. This suggests that, while the track luminosity measurement shows promise as a complementary measurement to the fast-or luminosity, the data-taking conditions for the pixel data need to be considerably more stable in order for this method to produce reliable results. For Run 3, it is thus important to improve our monitoring of the pixel data quality, and implement procedures to quickly recover from any observed problems. This is discussed further in Section~\ref{sec:run3}.

During normal physics fills, the trigger rate at which the pixel data are recorded is too low to make bunch-by-bunch analysis possible on short timescales. However, for the VdM fills in 2016--18, a special high-rate trigger was employed, which allows for the possibility of performing the VdM analysis as described in Section~\ref{sec:VdM} for the reconstructed track data.  The high-rate trigger includes two changes from the regular trigger. First, the trigger only selects a small number of BXs (mostly colliding bunches, with a few noncolliding and empty BXs also included), since the vast majority of BXs in a VdM fill are empty. Second, the overall rate of the trigger is increased. This study uses the data from one of the 2017 VdM scans, where the trigger rate used was approximately 9.7\unit{kHz}.

The results shown here come from the fourth $x$-$y$ scan pair in the 2017 VdM scan program. In contrast to the results discussed above, these measurement use the regular $5\sigma$ criterion to reject accidental tracks, as the overall track rate in the VdM scan is extremely low and so we want to avoid unnecessarily rejecting good tracks. However, even with this looser criterion, the track rate in the noncolliding BXs was still either exactly zero or very close to zero, thus indicating that there is no constant background term necessary in the VdM fits.

\begin{figure*}[tbhp]
\centering
  \includegraphics[width=0.45\textwidth]{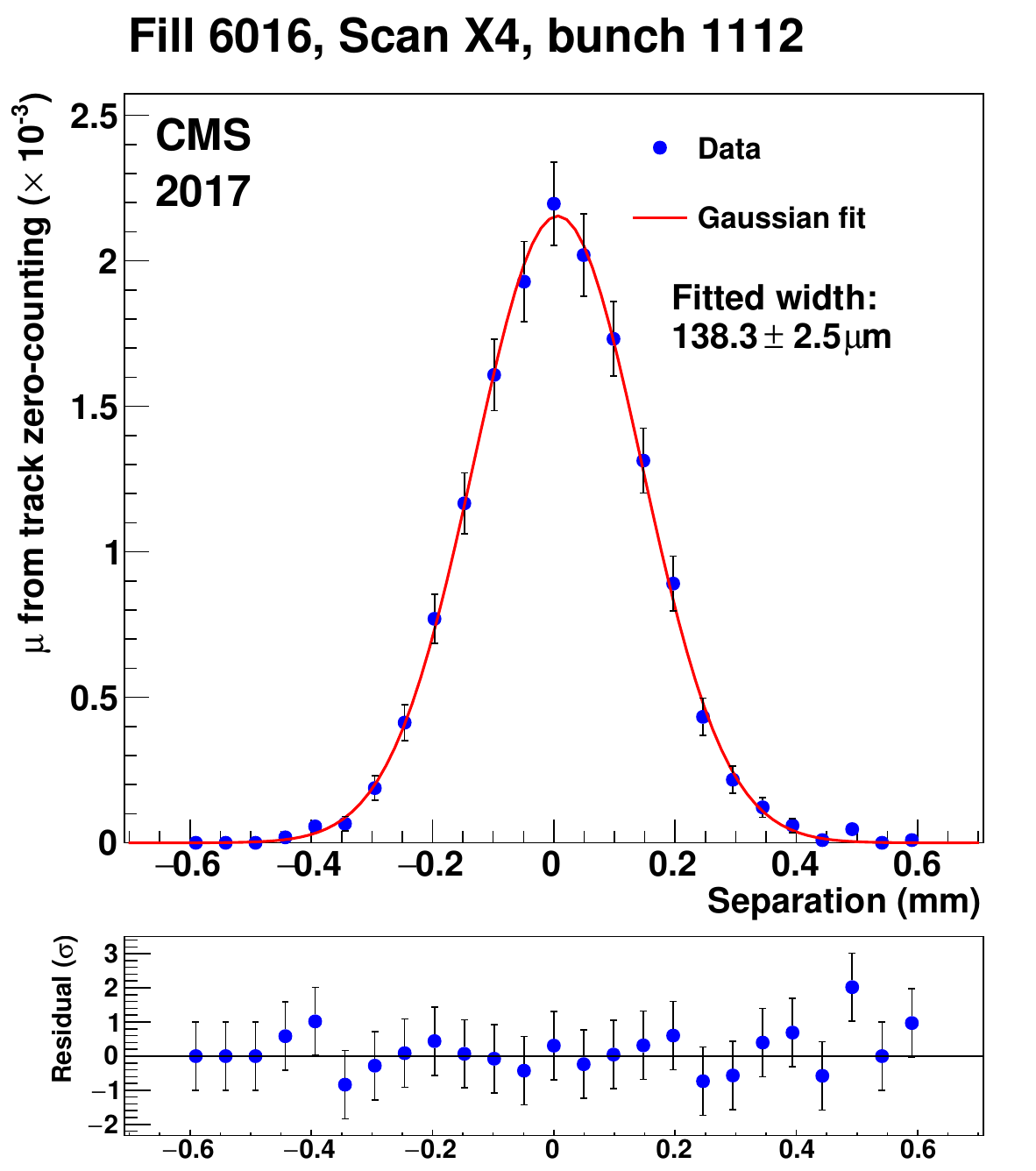}
  \includegraphics[width=0.45\textwidth]{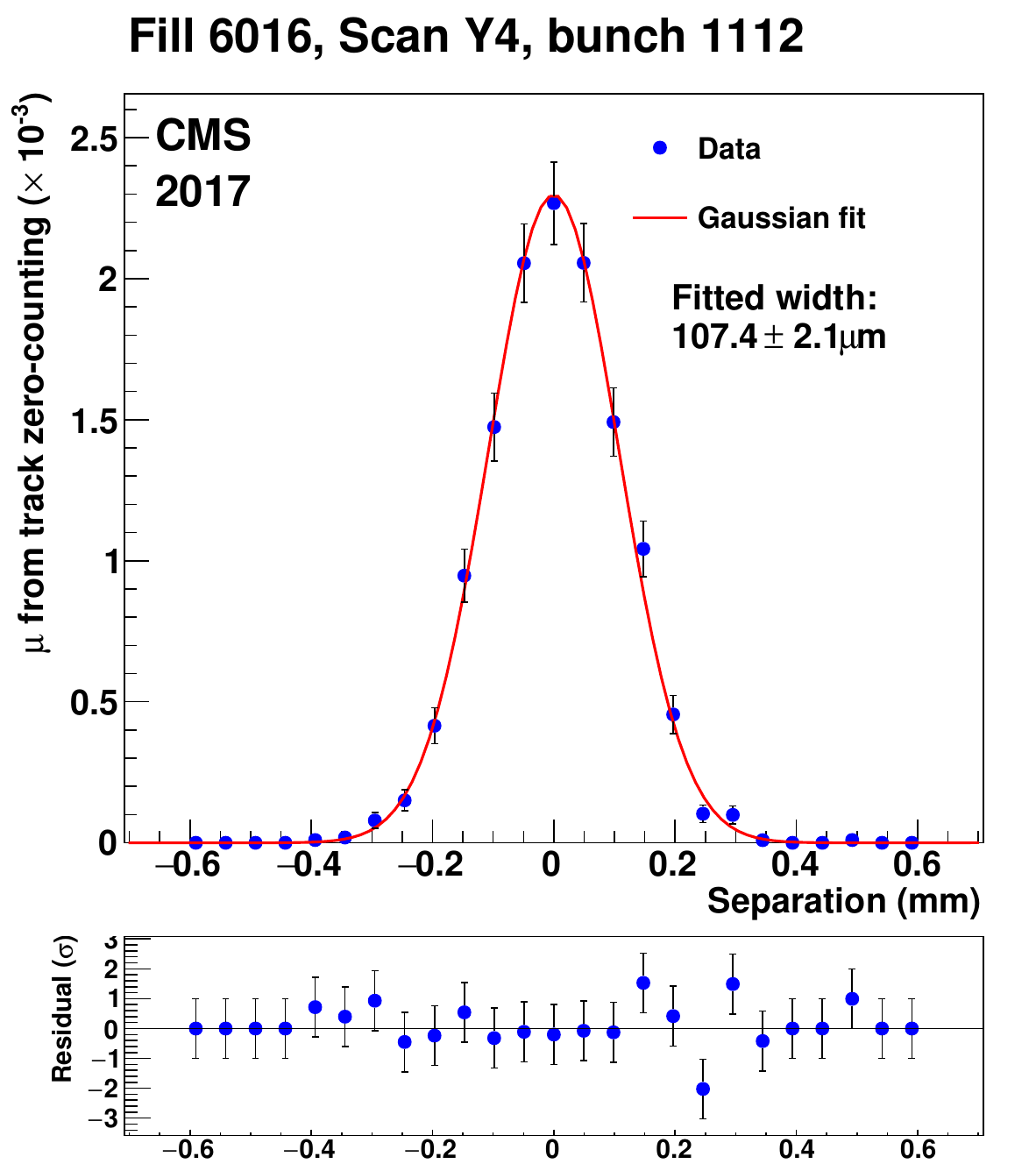}
  \caption{Scan curves using the track luminosity data during the fourth VdM scan pair in the 2017 VdM fill (fill 6016) for a single colliding bunch (BCID 1112) in the $x$ (left) and $y$ (right) directions. The extracted $\Sigma$ and its statistical uncertainty are also shown.}
  \label{fig:tracklumi_VdMfits}
\end{figure*}

Figure~\ref{fig:tracklumi_VdMfits} shows an example of the VdM fit to the track luminosity for a single colliding bunch (BCID 1112) for the average of all channels, including the fitted $\Sigma_x$ and $\Sigma_y$ values. The fit function is a single Gaussian, since there is not enough data in the tails for a second Gaussian component to be well determined, and the background is negligible. Because of the limited number of triggers per colliding bunch, the resulting statistical precision on the measured $\Sigma$ is approximately 2\%.

\begin{figure*}[tbhp]
\centering
  \includegraphics[width=0.8\textwidth]{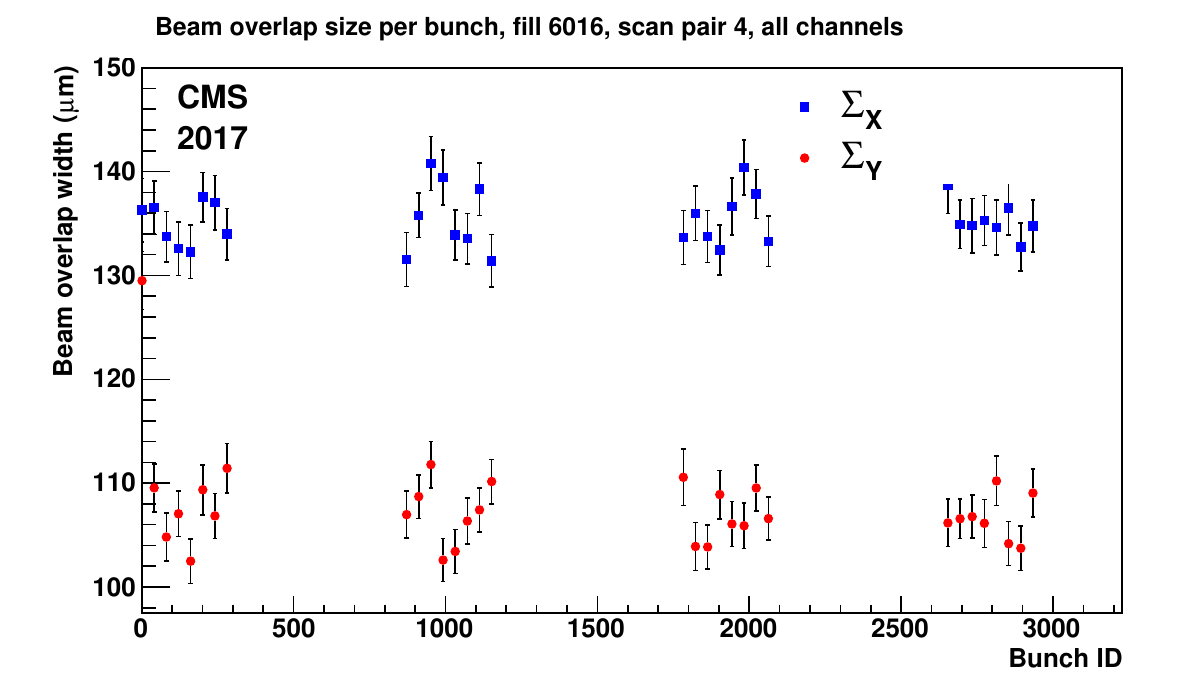}
  \includegraphics[width=0.8\textwidth]{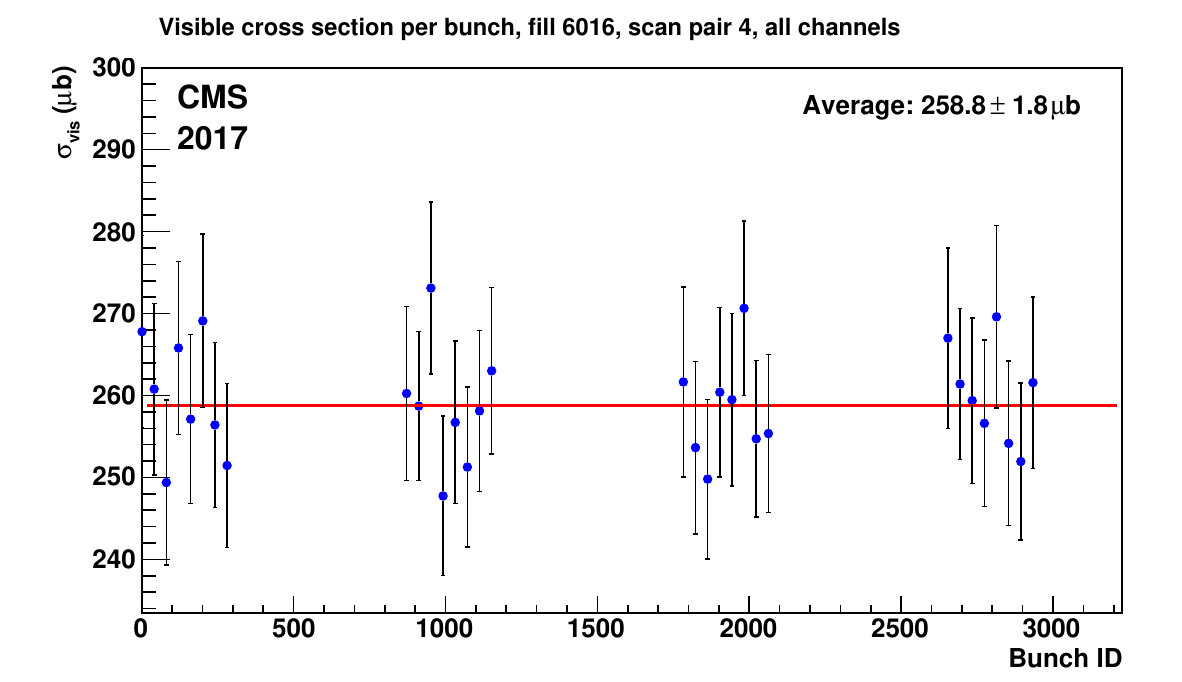}
  \caption{Top: Measured $\Sigma_x$ (blue squares) and $\Sigma_y$ (red circles) values as a function of BCID for the track luminosity measurement. Bottom: Measured \sigmavis value as a function of BCID. The red line indicates the fitted average over all bunches.}
  \label{fig:tracklumi_VdMresults}
\end{figure*}

The final measured $\Sigma$ and \sigmavis values for each bunch are shown in Fig.~\ref{fig:tracklumi_VdMresults}. The measured $\Sigma$ values are of course expected to show some bunch-by-bunch variation, and they agree with the measured values from the other detectors in Ref.~\cite{CMS-PAS-LUM-17-004} (including the high $\Sigma_y$ value in BCID 1). The measured \sigmavis values show good consistency across all bunches. Note that the systematic corrections described in Section~\ref{sec:VdM} have not yet been applied here, so we can compare this to the uncorrected value of \sigmavis of the fast-or measurement for the 2017 VdM scan of $292.8 \pm 1.8\unit{$\mu$b}$. As we expect, the \sigmavis for the track luminosity measurement of $260.7 \pm 1.8\unit{$\mu$b}$ is somewhat lower than for the fast-or measurement, as the track reconstruction requirement results in lower efficiency overall.

The analysis can also be performed on a channel-by-channel basis. However, in this case the statistical precision is significantly lower, and in particular, points in the tails of the VdM scan curves will often have zero counts, which causes the resulting width to be systematically underestimated. As a consequence, the results in this case are not reliable.

In conclusion, these results show that a successful VdM analysis can be performed using the track luminosity data, and so the track luminosity can be independently calibrated without having to rely on a cross-calibration to another luminometer. However, because of the limited trigger bandwidth, care must be taken in deriving a trigger scheme in order to obtain useful results. In particular, studies should be performed to find the maximum sustainable trigger rate without risking data loss or corruption. It may also be desirable to consider a trigger scheme in which only a small subset of colliding bunches are triggered, rather than attempting to take data for all colliding bunches, to ensure an adequate trigger rate for the bunches considered. With these improvements, it may even be possible to do a channel-by-channel VdM analysis for the track luminosity in Run~3.

\section{Preparations for Run 3}
\label{sec:run3}

The LS2 period was used to rebuild PLT, with one full copy completed and installed in July 2021 for the start of Run 3, and a second copy currently in production to be used as a spare in case full or partial replacement is required; this is foreseen to be likely because of the radiation damage expected in Run 3. This required the preparation and comprehensive testing of each new component, assembly of each independent quadrant, and a period of stress testing each assembled quadrant, with the quadrant in an as ready-to-install a configuration as possible.

Most of the newly produced hardware components have no changes in their design from Run 2, with the exception of the slow hub chips on the OMBs, which are responsible for distributing the control signals. In the Run~2 PLT, two physical chips were used for this functionality, but the Run~3 design uses a single combined chip. Operational experience from Run 2 strongly suggests that several components---including slow hub chips, ALT chips, and LCDS chips---are sensitive to thermal effects. Thus, a crucial part of the stress testing involves an extensive thermal cycling program.

One other change in the PLT copy installed for the beginning of Run 3 is that one of the telescopes (channel 7) was constructed using prototype sensors for the Phase-2 CMS tracker~\cite{Steinbruck:2020mnb}. These prototype sensors have the same $150{\times}100\mum$ pixel size as the sensors used in the rest of the PLT (although the final design is expected to have smaller pixel sizes). It uses an n-in-p design with an overall thickness of 150\mum, so it is expected that these sensors should need less bias voltage to reach maximum efficiency. The installation of these sensors will allow us to collect valuable data on their performance during Run 3, while not affecting the overall performance of the PLT.

The source testing of PSI46 sensor planes is ongoing as they become available, as they are being produced concurrently at the Paul Scherrer Institute (PSI). Each plane is tested individually and graded. The HDIs are tested both without and with high voltage applied. New port cards are monitored closely and tested frequently since the LCDS chips have shown a significant rate of failure under operational and lab conditions, as well as cases in new port cards where the LCDS chips are dead upon initial installation. The new OMBs are tested to verify functionality, with an emphasis on the ALT driver chips and the newly designed combined slow hub chips, since in Run 2, both exhibited indications of partial or total failure under lab and operational conditions. The slow hub chips are especially critical since their failure leads to the loss of an entire quadrant.

Based on the experience gained from PLT operations during Run 2, a number of potential challenges have been identified that need to be addressed for Run 3. As previously mentioned, the new hardware needs to be vetted via stress testing in order to validate its reliability under operational conditions. Assembled quadrants will be subjected to continuous running with thermal cycles and periodic source testing in order to identify and replace specific components which could otherwise fail after installation inside the detector volume.

In addition, one of the most significant challenges during operations is the monitoring and optimization of the performance and efficiency of the detector with accumulated radiation dose. The most consequential lessons from Run 2 operations can be summarized as follows:
\begin{enumerate}
    \item The sensor depletion voltage must be monitored very regularly and the HV set points must be maintained above the  measured depletion voltage.
    \item The ROC thresholds must be frequently reoptimized to allow the efficient operation of the detector, especially once the HV set points are adjusted.
    \item All monitoring of detector performance and efficiency must be done independently for each channel, since their behavior can vary significantly.
\end{enumerate}

During Run 2, increases to the operational HV set points were found to be the most effective and simplest way to compensate for the gradual reduction in detector performance. In order to streamline this process, automated software has been developed to execute HV scans and log the results in a consistent way. In addition, analysis software has been developed to process the logged results and determine the depletion voltage for each channel. Based on Run 2 experience, HV scans should be run at least once a month.

In conjunction with HV monitoring and adjustment, the threshold settings of the chips should also be monitored and updated in order to retain good efficiency for reconstructing hits. As changing the thresholds can affect other aspects of the ROC performance, such as timing, developing a comprehensive program to ensure that these can be updated easily is necessary.

As illustrated in Section~\ref{sec:emittance}, indicators of detector performance, such as efficiency, cannot be assumed to be uniform for all channels. Thus, all measurements should be done separately for all channels. This involves some modification to the analysis code and implementation of automation, so that conclusive results are available as soon as possible after the completion of a fill. A comparison in performance between channels is one of the most effective tools in finding and troubleshooting issues in one or multiple channels, especially when evaluated as a function of time and/or integrated luminosity. Thus, an automated end-of-fill performance summary is planned for Run 3 in order to publish the accumulated results as promptly as possible.

Finally, while the system deployed in Run 2 performed well in detecting operational issues affecting the fast-or luminosity readout and automatically recovering from these issues, the pixel data readout was not similarly monitored, so issues which affected the pixel data but not the fast-or data could go undetected. In order to ensure that the analyses taking advantage of pixel data can work effectively, algorithms are currently being developed for more comprehensive detection of problems in the pixel data and automatically fixing these problems.

\section{Conclusions}
\label{sec:summary}

In Run 2, the PLT accomplished its goal of providing high-precision per-bunch luminosity in all LHC operating modes and beam energies. The PLT was especially valuable as an online luminometer, as it was capable of providing fast feedback to the LHC for machine operations such as beam optimization, with a statistical uncertainty of $\approx$0.5\% per bunch or $\approx$0.01\% total for an integration period of 1.4\unit{s} under normal operating conditions. We also demonstrated that, even when the PLT was calibrated and its data corrected entirely without reference to other luminometers, it produced results that were highly consistent (within 0.5\%) with the other luminosity detectors operated by BRIL, making it valuable for calibration and cross-detector comparisons.

The principal challenges in providing a good calibration were changes in the efficiency over time (up to 10\% over the course of a year), primarily due to radiation damage effects, and nonlinear effects as a function of instantaneous luminosity, which ranged up to approximately 2\%/(\hzub). The emittance scan analysis described in Section~\ref{sec:emittance} provided a powerful technique to measure and correct for these effects over time, but given the uncertainty inherent in these corrections (as seen in Table~\ref{tab:totalUncertainties}), a goal for Run 3 is to improve the inherent stability of the PLT. This will require a sustained effort on several fronts, including the provisioning of the spare PLT, close and prompt monitoring of the efficiency, and regular adjustments of HV and threshold settings.

In addition to the triple-coincidence luminosity measurement that is the primary deliverable from the PLT, there are many other quantities of interest that can be measured from the PLT data, such as the beam-induced background, accidental rate, beam spot position, and the luminosity with track data. A proof of concept for these promising analyses has been presented here, but additional work will be necessary to develop these for Run 3. In particular, these efforts will depend on the previously mentioned efforts to ensure consistent pixel data quality and per-channel analysis.

The replacement PLT was installed in July 2021, successfully operated during the LHC beam tests in October 2021, and is currently undergoing calibration and commissioning in preparation for the start of Run~3; production of the spare PLT quadrants is well underway and they should be available soon in case a partial or total replacement of the PLT is necessary and possible during Run 3. For Phase 2, the upgraded tracker is expected to occupy the current location of the PLT~\cite{briltdr}, so a PLT-like detector is not planned for Phase 2, but the experience gained from PLT operations in Runs 2 and 3 will be invaluable in further planning and operations of BRIL luminometers; the results from the Phase-2 sensors will also be of use to the CMS Phase-2 tracker project.

\begin{acknowledgments}
We congratulate our colleagues in the CERN accelerator departments for the excellent performance of the LHC and thank the technical and administrative staffs at CERN and at CMS institutes worldwide for their contributions to the success of the CMS effort. We thank the CERN bonding laboratory, in particular Florentina Manolescu and Ian McGill; the LHC operations team, in particular Michi Hostettler and Jorg Wenninger; the CMS technical coordination team, for their help during operations; the CMS tracker project for their support with the services; the CMS cooling team for the maintenance of the cooling system and quality control on the cooling connections; and the CMS engineering and integration office for their work in routing and supervision of the installation of the cooling services in LS1. We acknowledge the enduring support for the construction and operation of the CMS BRIL detectors by the following institutes and funding agencies: CERN; the Secretariat for Higher Education, Science, Technology and Innovation (Ecuador); the Federal Ministry of Education and Research (BMBF), Deutsche Forschungsgemeinschaft (DFG), and Helmholtz-Gemeinschaft Deutscher Forschungszentren (HGF) (Germany); the Hungarian Academy of Sciences (MTA) and the National Research, Development and Innovation Office (NKFIH) (Hungary); the Istituto Nazionale di Fisica Nucleare (INFN) (Italy); the Mexican National Council for Science and Technology (CONACYT) (Mexico); the Ministry of Business, Innovation and Employment (New Zealand); the US CMS operations program, the US National Science Foundation (NSF), and the US Department of Energy (DOE) (USA).

Individuals have received support from NKFIH research grants K 124845, K 128713, K 143460, and TKP2021-NKTA-64 (Hungary), the US NSF research grants NSF-2121686, PHY-1945366, PHY-2111554, and PHY-2209460, and the US DOE Office of High Energy Physics awards DE-AC02-07CH11359, DE-SC0011845, and DE-SC0020267 (USA).
\end{acknowledgments}

\bibstyle{cms_unsrt}
\bibliography{DN-21-008}

\providecommand{\href}[2]{#2}\begingroup\raggedright\begin{thebibliography}{10}%
\makeatletter
\providecommand{\hrefCMSnoop }[0]{\@secondoftwo}%
\makeatother
\providecommand{\doi}{\texttt{doi:}\begingroup \urlstyle{tt}\Url}

\bibitem{Kornmayer:2016wkz}
\hrefCMSnoop {}{{A. Kornmayer on behalf of the CMS Collaboration}, ``The {CMS}
  pixel luminosity telescope'',} \textit{ Nucl. Instrum. Meth. A} \textbf{ 824}
  (2016) 304,
\href{http://dx.doi.org/10.1016/j.nima.2015.09.104}{\doi{10.1016/j.nima.2015.09.104}}.

\bibitem{Lujan:2017kvh}
\hrefCMSnoop {}{{P. Lujan on behalf of the CMS Collaboration}, ``Performance of
  the {Pixel Luminosity Telescope} for luminosity measurement at {CMS} during
  {Run 2}'',} \textit{ PoS} \textbf{ 314} (2017) 504,
\href{http://dx.doi.org/10.22323/1.314.0504}{\doi{10.22323/1.314.0504}}.

\bibitem{eps2021}
\hrefCMSnoop {}{{P. Lujan on behalf of the CMS Collaboration}, ``The {Pixel
  Luminosity Telescope}: a silicon sensor detector for luminosity measurement
  at {CMS}'',} \textit{ PoS} \textbf{ EPS-HEP2021} (2022) 820,
  \href{http://dx.doi.org/10.22323/1.398.0820}{\doi{10.22323/1.398.0820}}.

\bibitem{Dabrowski:2016rqm}
\hrefCMSnoop {}{{CMS} Collaboration, ``Upgrade of the {CMS} instrumentation for
  luminosity and machine induced background measurements'',} \textit{ Nucl.
  Part. Phys. Proc.} \textbf{ 273-275} (2016) 1147,
\href{http://dx.doi.org/10.1016/j.nuclphysbps.2015.09.180}{\doi{10.1016/j.nuclphysbps.2015.09.180}}.

\bibitem{Chatrchyan:2008zzk}
\hrefCMSnoop {}{{CMS} Collaboration, ``The {CMS} experiment at the {CERN}
  {LHC}'',} \textit{ JINST} \textbf{ 3} (2008) S08004,
  \href{http://dx.doi.org/10.1088/1748-0221/3/08/S08004}{\doi{10.1088/1748-0221/3/08/S08004}}.

\bibitem{Bolla:2002my}
G.~Bolla\hrefCMSnoop {}{ {et~al.}, ``Sensor development for the {CMS} pixel
  detector'',} \textit{ Nucl. Instrum. Meth. A} \textbf{ 485} (2002) 89,
\href{http://dx.doi.org/10.1016/S0168-9002(02)00537-5}{\doi{10.1016/S0168-9002(02)00537-5}}.

\bibitem{Allkofer:2007ek}
\hrefCMSnoop {}{Y.~Allkofer {et~al.}, ``Design and performance of the silicon
  sensors for the {CMS} barrel pixel detector'',} \textit{ Nucl. Instrum. Meth.
  A} \textbf{ 584} (2008) 25,
  \href{http://dx.doi.org/10.1016/j.nima.2007.08.151}{\doi{10.1016/j.nima.2007.08.151}},
\href{http://www.arXiv.org/abs/physics/0702092}{\texttt{arXiv:physics/0702092}}.

\bibitem{Kastli:2005jj}
H.~C. K{\"a}stli\hrefCMSnoop {}{ {et~al.}, ``Design and performance of the
  {CMS} pixel detector readout chip'',} \textit{ Nucl. Instrum. Meth. A}
  \textbf{ 565} (2006) 188,
  \href{http://dx.doi.org/10.1016/j.nima.2006.05.038}{\doi{10.1016/j.nima.2006.05.038}},
\href{http://www.arXiv.org/abs/physics/0511166}{\texttt{arXiv:physics/0511166}}.

\bibitem{Barbero:467141}
\href {https://cds.cern.ch/record/467141}{M.~Barbero, ``Development of a
  radiation-hard pixel read out chip with trigger capability''}.
\newblock PhD thesis, Basel University, 2003.

\bibitem{Leonard:2014gpa}
\hrefCMSnoop {}{J.~L. Leonard {et~al.}, ``Fast beam condition monitor for
  {CMS}: performance and upgrade'',} \textit{ Nucl. Instrum. Meth. A} \textbf{
  765} (2014) 235,
  \href{http://dx.doi.org/10.1016/j.nima.2014.05.008}{\doi{10.1016/j.nima.2014.05.008}},
\href{http://www.arXiv.org/abs/1405.1926}{\texttt{arXiv:1405.1926}}.

\bibitem{Hempel:2017nvn}
\hrefCMSnoop {}{M.~Hempel, ``Development of a novel diamond based detector for
  machine induced background and luminosity measurements''}.
\newblock PhD thesis, DESY, Hamburg, 2017.
\newblock
\href{http://dx.doi.org/10.3204/PUBDB-2017-06875}{\doi{10.3204/PUBDB-2017-06875}}.

\bibitem{Forkel-Wirth:687619}
D.~Forkel-Wirth\href {http://cds.cern.ch/record/687619}{ {et~al.}, ``Radiation
  monitoring system for the environment and safety project'',} ST Note
  2002-006-MA, 2002.

\bibitem{Evans:2008zzb}
\hrefCMSnoop {}{L.~Evans and P.~Bryant, ``{LHC} machine'',} \textit{ JINST}
  \textbf{ 3} (2008) S08001,
\href{http://dx.doi.org/10.1088/1748-0221/3/08/S08001}{\doi{10.1088/1748-0221/3/08/S08001}}.

\bibitem{BartzTBM}
\hrefCMSnoop {}{E.~Bartz, ``The 0.25$\mu$m token bit manager chip for the {CMS}
  pixel readout'',} in \textit{ Proceedings, 11$^\text{th}$ Workshop on
  Electronics for LHC and Future Experiments}.
\newblock 2005.
\newblock
  \href{http://dx.doi.org/10.5170/CERN-2005-011.153}{\doi{10.5170/CERN-2005-011.153}}.

\bibitem{Friedl:2004ji}
\hrefCMSnoop {}{{CMS Tracker} Collaboration, ``Analog optohybrids for the
  readout of the {CMS} silicon tracker'',} \textit{ Nucl. Instrum. Meth. A}
  \textbf{ 518} (2004) 515,
\href{http://dx.doi.org/10.1016/j.nima.2003.11.073}{\doi{10.1016/j.nima.2003.11.073}}.

\bibitem{Benotto:2008zz}
F.~Benotto\href {http://cds.cern.ch/record/1112032}{ {et~al.}, ``Design and
  test of the digital opto hybrid module for the {CMS} tracker inner barrel and
  disks'',} CMS Note 2008/013, 2008.

\bibitem{Furtado:920425}
\hrefCMSnoop {}{H.~Furtado, ``Delay25 an {ASIC} for timing adjustment in
  {LHC}'',} in \textit{ Proceedings, 11$^\text{th}$ Workshop on Electronics for
  LHC and Future Experiments}, p.~148.
\newblock 2005.
\newblock
  \href{http://dx.doi.org/10.5170/CERN-2005-011.148}{\doi{10.5170/CERN-2005-011.148}}.

\bibitem{Mueller:1319599}
\href {http://cds.cern.ch/record/1319599}{S.~M{\"u}ller, ``The {Beam Condition
  Monitor} 2 and the radiation environment of the {CMS} detector at the
  {LHC}''}.
\newblock PhD thesis, Karlsruhe University, 2011.
\newblock CERN-THESIS-2011-085.

\bibitem{Guthoff:1977429}
\href {https://cds.cern.ch/record/1977429}{M.~Guthoff, ``Radiation damage to
  the diamond-based Beam Condition Monitor of the CMS detector at the LHC''}.
\newblock PhD thesis, Karlsruhe Institute of Technology, 2014.
\newblock
CERN-THESIS-2014-216, CMS-TS-2014-043.

\bibitem{Kassel:2271160}
\href {http://cds.cern.ch/record/2271160}{F.~R. Kassel, ``The rate dependent
  radiation induced signal degradation of diamond detectors''}.
\newblock PhD thesis, Karlsruhe Institute of Technology, 2017.
\newblock
CERN-THESIS-2017-071.

\bibitem{Pernicka:1091743}
M.~Pernicka\hrefCMSnoop {}{ {et~al.}, ``The {CMS} pixel {FED}'',} in \textit{
  Proceedings, Topical Workshop on Electronics for Particle Physics (TWEPP07),
  Prague, Czech Republic, September 3--7, 2007}, p.~487.
\newblock 2008.
\newblock
  \href{http://dx.doi.org/10.5170/CERN-2007-007.487}{\doi{10.5170/CERN-2007-007.487}}.

\bibitem{vanderBij:1997rc}
\hrefCMSnoop {}{E.~van~der Bij, R.~A. McLaren, O.~Boyle, and G.~Rubin,
  ``{S-LINK}, a data link interface specification for the {LHC} era'',}
  \textit{ IEEE Trans. Nucl. Sci.} \textbf{ 44} (1997) 398,
  \href{http://dx.doi.org/10.1109/23.603679}{\doi{10.1109/23.603679}}.

\bibitem{Hegeman:2016hlt}
\hrefCMSnoop {}{J.~Hegeman {et~al.}, ``The {CMS} timing and control
  distribution system'',} in \textit{ {Proceedings, 2015 IEEE Nuclear Science
  Symposium and Medical Imaging Conference (NSS/MIC 2015), San Diego,
  California, United States}}, p.~7581984.
\newblock 2016.
\newblock
\href{http://dx.doi.org/10.1109/NSSMIC.2015.7581984}{\doi{10.1109/NSSMIC.2015.7581984}}.

\bibitem{bib:DIP}
\hrefCMSnoop {}{B.~Copy, E.~Mandilara, I.~Prieto~Barreiro, and F.~Varela,
  ``Monitoring of {CERN's} data interchange protocol {(DIP)} system'',} in
  \textit{ {Proceedings, 16th International Conference on Accelerator and Large
  Experimental Physics Control Systems (ICALEPCS 2017), Barcelona, Spain,
  October 8--13, 2017}}, p.~THPHA162.
\newblock 2018.
\newblock
\href{http://dx.doi.org/10.18429/JACoW-ICALEPCS2017-THPHA162}{\doi{10.18429/JACoW-ICALEPCS2017-THPHA162}}.

\bibitem{HallWilton:2011zz}
\hrefCMSnoop {}{R.~Hall-Wilton {et~al.}, ``Results from a beam test of a
  prototype {PLT} diamond pixel telescope'',} \textit{ Nucl. Instrum. Meth. A}
  \textbf{ 636} (2011) S130,
\href{http://dx.doi.org/10.1016/j.nima.2010.04.097}{\doi{10.1016/j.nima.2010.04.097}}.

\bibitem{Bugg:2011zz}
\hrefCMSnoop {}{W.~Bugg {et~al.}, ``Studies of mono-crystalline {CVD} diamond
  pixel detectors'',} \textit{ Nucl. Instrum. Meth. A} \textbf{ 650} (2011) 50,
\href{http://dx.doi.org/10.1016/j.nima.2010.12.161}{\doi{10.1016/j.nima.2010.12.161}}.

\bibitem{Gan:2015zya}
\hrefCMSnoop {}{{RD42} Collaboration, ``Diamond particle detectors systems in
  high energy physics'',} \textit{ PoS} \textbf{ TIPP2014} (2015) 103,
\href{http://dx.doi.org/10.22323/1.213.0103}{\doi{10.22323/1.213.0103}}.

\bibitem{Pompeo:2017kig}
\href {http://tesi.cab.unipd.it/57497/}{G.~Pompeo, ``Study of the performance
  of the {CMS} Pixel Luminosity Telescope''}.
\newblock Laurea thesis, University of Padua, 2017.

\bibitem{kmeans}
\href
  {https://projecteuclid.org/ebooks/berkeley-symposium-on-mathematical-statistics-and-probability/Proceedings%20of%20the%20Fifth%20Berkeley%20Symposium%20on%20Mathematical%20Statistics%20and%20Probability,%20Volume%201:%20Statistics/chapter/Some%20methods%20for%20classification%20and%20analysis%20of%20multivariate%20observations/bsmsp/1200512992}{J.~MacQueen,
  ``Some methods for classification and analysis of multivariate
  observations'',} in \textit{ Proceedings of the {Fifth Berkeley Symposium on
  Mathematical Statistics and Probability}, Volume 1: Statistics}, p.~281.
\newblock 1967.

\bibitem{CMS-PAS-LUM-18-002}
\href {https://cds.cern.ch/record/2676164/}{{{CMS}} Collaboration, ``{CMS}
  luminosity measurement for the 2018 data-taking period at $\sqrt{s}$ = 13
  {TeV}'',} CMS Physics Analysis Summary CMS-PAS-LUM-18-002, 2019.

\bibitem{Verkerke:2003ir}
\hrefCMSnoop {}{W.~Verkerke and D.~P. Kirkby, ``The {RooFit} toolkit for data
  modeling'',} in \textit{ {Proceedings, 13th International Conference on
  Computing in High-Energy and Nuclear Physics (CHEP 2003), La Jolla,
  California, March 24--28, 2003}}, p.~MOLT007.
\newblock 2003.
\newblock
  \href{http://www.arXiv.org/abs/physics/0306116}{\texttt{arXiv:physics/0306116}}.

\bibitem{Sirunyan:2021qkt}
\hrefCMSnoop {}{{CMS} Collaboration, ``Precision luminosity measurement in
  proton-proton collisions at $\sqrt{s} =$ 13 {TeV} in 2015 and 2016 at
  {CMS}'',} \textit{ Eur. Phys. J. C} \textbf{ 81} (2021) 800,
  \href{http://dx.doi.org/10.1140/epjc/s10052-021-09538-2}{\doi{10.1140/epjc/s10052-021-09538-2}},
  \href{http://www.arXiv.org/abs/2104.01927}{\texttt{arXiv:2104.01927}}.

\bibitem{vanderMeer:296752}
\href {https://cds.cern.ch/record/296752}{S.~van~der Meer, ``Calibration of the
  effective beam height in the {ISR}'',} Technical Report CERN-ISR-PO-68-31,
  1968.

\bibitem{CMS-PAS-LUM-17-004}
\href {https://cds.cern.ch/record/2621960/}{{{CMS}} Collaboration, ``{CMS}
  luminosity measurement for the 2017 data-taking period at $\sqrt{s}$ = 13
  {TeV}'',} CMS Physics Analysis Summary CMS-PAS-LUM-17-004, 2018.

\bibitem{CMS-PAS-LUM-16-001}
\href {http://cds.cern.ch/record/2235781}{{CMS Collaboration}, ``{CMS}
  luminosity calibration for the pp reference run at
  {$\sqrt{s}=5.02~\mathrm{TeV}$}'',} CMS Physics Analysis Summary
  CMS-PAS-LUM-16-001, 2016.

\bibitem{CMS-PAS-LUM-19-001}
\href {http://cds.cern.ch/record/2765655}{{CMS Collaboration}, ``Measurement of
  the integrated luminosity for proton-proton collisions at 5.02 {TeV} recorded
  by the {CMS} experiment in 2017'',} CMS Physics Analysis Summary
  CMS-PAS-LUM-19-001, 2020.

\bibitem{CMS-PAS-LUM-18-001}
\href {http://cds.cern.ch/record/2809613}{{CMS} Collaboration, ``{CMS}
  luminosity measurement using nucleus-nucleus collisions at
  $\sqrt{s_\text{NN}}$ = 5.02 {TeV} in 2018'',} CMS Physics Analysis Summary
  CMS-PAS-LUM-18-001, 2022.

\bibitem{CMS-PAS-LUM-17-002}
\href {https://cds.cern.ch/record/2628652}{{CMS} Collaboration, ``{CMS}
  luminosity measurement using 2016 proton-nucleus collisions at
  $\sqrt{s_\text{NN}}$ = 8 {TeV}'',} CMS Physics Analysis Summary
  CMS-PAS-LUM-17-002, 2018.

\bibitem{Karacheban:2019ypt}
\hrefCMSnoop {}{{CMS} Collaboration, ``{Emittance scans for CMS luminosity
  calibration in Run 2}'',} \textit{ PoS} \textbf{ EPS-HEP2019} (2020) 193,
  \href{http://dx.doi.org/10.22323/1.364.0193}{\doi{10.22323/1.364.0193}}.

\bibitem{CMS-DP-2021-002}
\href {https://cds.cern.ch/record/2751564}{{{CMS}} Collaboration, ``{CMS}
  luminosity measurement using nucleus-nucleus collisions in {Run 2}'',} CMS
  Detector Performance Summary CMS-DP-2021-002, 2021.

\bibitem{Wanczyk:2701798}
\href {http://cds.cern.ch/record/2701798}{J.~Wanczyk, ``Measurements and
  estimates of the radiation levels in the {CMS} experimental cavern using
  {Medipix} and {RAMSES} monitors, and the {FLUKA} {Monte Carlo} code'',}
  Master's thesis, AGH University of Science and Technology, 2019.

\bibitem{Steinbruck:2020mnb}
\hrefCMSnoop {}{{CMS Tracker Group} Collaboration, ``Development of planar
  pixel sensors for the {CMS} inner tracker at the high-luminosity {LHC}'',}
  \textit{ Nucl. Instrum. Meth. A} \textbf{ 978} (2020) 164438,
  \href{http://dx.doi.org/10.1016/j.nima.2020.164438}{\doi{10.1016/j.nima.2020.164438}}.

\bibitem{briltdr}
\href {https://cds.cern.ch/record/2759074}{{{CMS}} Collaboration, ``The
  {Phase-2} upgrade of the {CMS} beam radiation, instrumentation, and
  luminosity detectors'',} Technical Design Report CERN-LHCC-2021-008,
  CMS-TDR-023, 2021.

\end{thebibliography}\endgroup

\clearpage
\appendix
\section{The CMS BRIL Collaboration}
\cmsinstitute{Escuela Politecnica Nacional, Quito, Ecuador}
{\tolerance=6000
E.~Ayala\cmsorcid{0000-0002-0363-9198}
\par}
\cmsinstitute{Universidad San Francisco de Quito, Quito, Ecuador}
{\tolerance=6000
E.~Carrera~Jarrin\cmsorcid{0000-0002-0857-8507}
\par}
\cmsinstitute{National Institute of Chemical Physics and Biophysics, Tallinn, Estonia}
{\tolerance=6000
I.~Ahmed
\par}
\cmsinstitute{Deutsches Elektronen-Synchrotron, Hamburg, Germany}
{\tolerance=6000
A.~Campbell\cmsorcid{0000-0003-4439-5748}, V.~Danilov, L.I.~Estevez~Banos\cmsorcid{0000-0001-6195-3102}, A.~Giraldi\cmsorcid{0000-0003-4423-2631}, M.~Guthoff\cmsorcid{0000-0002-3974-589X}, M.~Hempel, H.~Henschel, J.~Knolle\cmsAuthorMark{1}\cmsorcid{0000-0002-4781-5704}, W.~Lange, J.~Leonard\cmsorcid{0000-0003-1761-8221}, W.~Lohmann\cmsAuthorMark{2}\cmsorcid{0000-0002-8705-0857}, A.B.~Meyer\cmsorcid{0000-0001-8532-2356}, V.~Myronenko\cmsorcid{0000-0002-3984-4732}, M.~Penno, B.~Ribeiro~Lopes\cmsorcid{0000-0003-0823-447X}, J.~R\"{u}benach, A.~Saggio\cmsorcid{0000-0002-7385-3317}, V.~Scheurer\cmsAuthorMark{3}, R.E.~Sosa~Ricardo\cmsorcid{0000-0002-2240-6699}, O.~Turkot\cmsorcid{0000-0001-5352-7744}, D.~Walter\cmsorcid{0000-0001-8584-9705}
% ...
\par}

$^{1}$Now at Ghent University, Ghent, Belgium\\
$^{2}$Also at Brandenburg University of Technology, Cottbus, Germany\\
$^{3}$Now at Purdue University, West Lafayette, Indiana, USA\\
% ...

\end{document}